\documentclass[fleqn,usenatbib]{mnras}

\usepackage{newtxtext,newtxmath}

\usepackage[T1]{fontenc}

\usepackage{tikz} 
\usetikzlibrary{shapes.geometric, arrows}
\tikzstyle{arrow} = [thick,->,>=stealth]
\usepackage{dirtytalk}
\DeclareRobustCommand{\VAN}[3]{#2}
\let\VANthebibliography\thebibliography
\def\thebibliography{\DeclareRobustCommand{\VAN}[3]{##3}\VANthebibliography}

\usepackage{graphicx}	
\usepackage{amsmath}	
\usepackage[italian]{}
\usepackage{lscape}
\usepackage{booktabs}
\usepackage{longtable}
\usepackage{supertabular}
\usepackage{subcaption}

\usepackage{ulem}

\title[Viscous + wind disc evolution with dead zones present]{A question of personalities: evolution of viscous and wind-driven protoplanetary discs in the presence of dead zones}

\author[S. Tong, R. Alexander and G. Rosotti]{
Simin Tong,$^{1}$\thanks{E-mail: st547@leicester.ac.uk, astro.stong@gmail.com}
Richard Alexander,$^{1}$
and Giovanni Rosotti$^{2}$
\\
$^{1}$School of Physics \& Astronomy, University of Leicester, University Road, Leicester, LE1 7RH, UK\\
$^{2}$Dipartimento di Fisica, Università degli Studi di Milano, Via Celoria, 16, 20133, Milano, Italy\\
}

\date{Accepted XXX. Received YYY; in original form ZZZ}

\pubyear{2024}

\begin{document}
\label{firstpage}
\pagerange{\pageref{firstpage}--\pageref{lastpage}}
\maketitle

\begin{abstract} 
Whether the angular momentum of protoplanetary discs is redistributed by viscosity or extracted by magnetised winds is a long-standing question. Demographic indicators, such as gas disc sizes and stellar accretion rates, have been proposed as ways of distinguishing between these two mechanisms. In this paper, we implement one-dimensional gas simulations to study the evolution of ``hybrid'' protoplanetary discs simultaneously driven by viscosity and magnetised winds, with dead zones present. We explore how the variations of disc properties, including initial disc sizes, dead zone sizes and angular momentum transport efficiency, affect stellar accretion rates, disc surface density profiles, disc sizes, disc lifetimes, and cumulative mass loss by different processes. Our models show that the expansion of the gas disc size can be sustained when the majority of angular momentum is removed by the magnetised wind for individual protoplanetary discs. However, when we can only observe discs via demographic screenshots, the variation of disc sizes with time is possibly diminished by the disc \say{personalities}, by which we mean the variations of initial disc properties among different discs. Our \say{hybrid} models re-assess association of the two demographic indicators with mechanisms responsible for angular momentum transport and suggest additional diagnostics are required to assist the differentiation.
\end{abstract}




\section{Introduction}
Protoplanetary discs are by-products of star formation due to the conservation of angular momentum. These discs of dust and gas fuel materials to the central star and provide the necessary components for planet formation. Therefore, understanding protoplanetary discs and how they evolve is fundamental to the study of planetary systems.

\par 
Material in the disc must lose angular momentum in order to be accreted from the protoplanetary disc to the central star. Two scenarios have been suggested to address where the angular momentum has gone: redistribution of the angular momentum by turbulence, and extraction of the angular momentum by magneto-hydrodynamic (MHD) winds. Turbulence can arise from gravitational instabilities \citep{1987MNRAS.225..607L, 2016ARA&A..54..271K}, hydrodynamical instabilities, such as vertical shear instabilities \citep{1998MNRAS.294..399U, 2003A&A...404..397U, 2013MNRAS.435.2610N}, and magneto-hydrodynamical instabilities, such as the magneto-rotational instability (MRI) \citep{1991ApJ...376..214B, 1995ApJ...440..742H}, which has long been thought to be the main driver of disc turbulence. When MRI-induced ``viscous'' turbulence redistributes the angular momentum in the disc, the small fraction of the outer disc carrying a large quantity of the angular momentum moves outwards and gives rise to an increasing gas disc size over time \citep{1974MNRAS.168..603L, 1998ApJ...495..385H}. 

\par The MRI is sensitive to the degree of ionization. It can be sustained when the disc is sufficiently ionized by the thermal and non-thermal processes \citep{1981PASJ...33..617U, 2009ApJ...690...69U, 2013ApJ...772....5C} to couple with magnetic fields. These conditions are usually fulfilled in the very inner and possibly outer discs, indicating a large fraction of the disc remains MRI-quenched. These regions are called \say{dead zones} \citep{1996ApJ...457..355G}. MHD winds also rely on charged particles and the magnetic field. When charged particles in protoplanetary discs are coupled to the magnetic field and are lifted from the disc to launch the magnetised wind, the tail of the ionized gas exerts torques on the disc and takes away angular momentum from discs \citep{1997A&A...319..340F}. Several studies utilising different methods have shown that this wind is capable of driving the observed stellar accretion rate \citep[e.g.,][]{ 2013ApJ...769...76B, 2023A&A...674A.165W} and inducing gas disc size shrinking while the characteristic radius remains unchanged \citep{2022MNRAS.512.2290T, 2022ApJ...926...61T}, making it a viable alternative mechanism to viscous accretion. 

\par Advances in observational techniques in the last decade have now enabled us to characterise protoplanetary discs properties, such as disc sizes and stellar accretion rates, systematically and statistically. The Atacama Large Millimeter/submillimeter Array (ALMA) reveals substructures in the dust and gas discs \citep[e.g.,][]{2013Sci...340.1199V, 2015ApJ...808L...3A, 2018MNRAS.475.5296D,2018ApJ...869L..41A, 
2020ApJ...895L..18B, 2021ApJS..257....1O}, and provides radial intensity profiles of dust \citep[e.g.,][]{2016ApJ...818L..16Z, 2018ApJ...869L..42H} and molecular emission \citep[e.g.,][]{2016ApJ...823L..18H, 2021ApJS..257....3L, 2021AJ....161...38O} from discs in the nearby star-forming regions. These observations, along with surveys dedicated to specific star-forming regions \citep[e.g.][]{2017ApJ...851...85B, 2018ApJ...859...21A,2019MNRAS.482..698C}, help us to quantify the disc sizes subject to some observational limitations, such as sensitivity, resolution and sample selection biases. The X-shooter instrument mounted on the Very Large Telescope (VLT) has measured accretion rates for hundreds of discs in nearby star-forming regions, including Lupus \citep{2014A&A...561A...2A, 2017A&A...600A..20A}, Chamaeleon I \citep{2016A&A...591L...3M, 2017A&A...604A.127M}, $\eta$-Chamaeleon \citep{2018A&A...609A..70R}, TW Hydrae association \citep{2019A&A...632A..46V} and Upper Scorpius \citep{2020A&A...639A..58M}. 

\par 
Recent studies built on the established theories and ample observations have attempted to discern whether viscosity or the MHD wind drives disc evolution.

\par \citet{2018ApJ...864..168N} collated gas sizes of Class I and Class II discs characterised by different tracers. They found that the Class I gas discs are typically smaller than those of Class II, implying that gas discs spread in the T Tauri phase and supporting the viscous picture. \citet{2022ApJ...931....6L} adopted a similar approach but with larger samples (44 discs) that were consistently traced by spatially resolved $\mathrm{{}^{12}CO}$ ($J=2-1$). They found no correlations between $\mathrm{{}^{12}CO}$ disc sizes and stellar ages (see their Figure 5f), and attributed this to the large uncertainties in the stellar ages.

\par
\citet{2022MNRAS.512.2290T} presented an analytical solution of the magnetised wind, whose efficiency in removing angular momentum is parametrized by an $\alpha_\mathrm{DW}$, equivalent to the $\alpha_\mathrm{SS}$ for viscosity in \citet{1973A&A....24..337S}. This solution facilitates the study of disc evolution by incorporating the wind component into 1-D evolution models.  \citet{2022MNRAS.512L..74T} reproduced the correlation between disc masses and accretion rates observed in Lupus by generating a population of inviscid wind-driven discs. This indicates that a wind-only model can possibly explain some observations. \citet{2023MNRAS.524.3948A} generated populations of purely wind-driven and viscous discs, respectively, and discussed the possibility of distinguishing two scenarios by the distribution of stellar accretion rates. They suggested that a slightly larger sample size than we currently have is required to answer the question. 

\par
\citet{2022ApJ...926...61T} integrated the $\alpha_\mathrm{DW}$-prescribed pure wind model into the thermochemical model \texttt{DALI} \citep{2012A&A...541A..91B, 2013A&A...559A..46B} and showed that the gas disc sizes in Lupus and Upper Sco are reproducible without the inclusion of viscosity. \citet{2023ApJ...954...41T} expanded applications of thermochemical models to viscosity and/or wind discs. The inferred disc characteristic radii decreasing from younger to older clusters are inconsistent with either of the two scenarios, hinting other physical mechanisms (such as external photoevaporation) could also play a role.

\par 1-D gas+dust evolution incorporating viscosity and winds was investigated in \citet{2022MNRAS.514.1088Z}. Their model reproduced the observed dust sizes (in 0.89-mm observations) in Lupus, Chamaeleon I and Upper Sco, irrespective of the relative strength of the two mechanisms, implying that current dust sizes limited by observational sensitivitiy \citep{2019MNRAS.486.4829R} are not reliable differentiators of wind and viscosity scenarios, and alternative tracers should be considered \citep{2023A&A...672L..15Z}.

\par 
However, all of these prior studies presume $\alpha_\mathrm{DW}$ and $\alpha_\mathrm{SS}$ are constant at all disc radii, and overlook the presence of dead zones, which can alter the spatial distribution of $\alpha_\mathrm{SS}$. Likewise, $\alpha_\mathrm{DW}$ should also be treated as a radial variable in addition to its time variations due to changes in the configuration of magnetic fields with time. Comparison of disc sizes and stellar accretion rates in previous research is also based on the implicit assumption that all the discs share the same $\alpha_\mathrm{SS}$ and/or $\alpha_\mathrm{DW}$. But discs formed in different environments might possess very different properties. Those around massive stars tend to have a larger dead zone inner edges \citep{2016ApJ...827..144F}. The ambient thermal and non-thermal radiation can also impact the outer edge of dead zones. External radiation \citep{2018ApJ...860...77E, 2018MNRAS.478.2700W, 2022MNRAS.514.2315C}, along with the local stellar density \citep{2016ApJ...828...48V, 2021ApJ...923..221O}, potentially truncates the outer disc, leading to smaller disc sizes than their counterparts growing in a more \say{friendly} environment \citep[][Anania et al., \textit{in prep.}]{2020A&A...640A...5T}. Therefore, in this work, we propose to explore how these various parameters influence the evolution of more realistic discs driven by viscosity and winds prescribed by radially varying $\alpha$, and attempt to determine whether the two main observable diagnostics, gas disc sizes and stellar accretion rates, are still valid discriminators between the two mechanisms when the \say{personalities} of discs -- fundamental initial disc properties, such as disc masses and disc sizes, varying among individuals -- are considered. We limit our models to isolated discs, so effects directly caused by environments, such as external photoevaporation and dynamical encounters, are beyond the scope of this study.

\par
This paper is structured as follows. In Section \ref{sec:method}, we introduce the disc evolution and the dead zone models. Section \ref{sec:fiducial} shows how the disc evolution is altered with the dead zone present. We explore the effect of parameters, including $\alpha_\mathrm{SS}(R)$, $\alpha_\mathrm{DW}(R)$, initial characteristic radii $R_\mathrm{c,0}$ and dead zone outer edges $R_\mathrm{dz,out}$, on the disc evolution from perspectives of stellar accretion rates, surface densities, disc sizes, lifetimes and cumulative mass loss in Section \ref{sec:explor}. The discussion of our models and selection of some parameters are presented in Section \ref{sec:discussion}. Based on the aforementioned studies, we perform two small scale population syntheses and show the results in Section \ref{sec:disc_pop}. Then we discuss observational implications and limitations of this work in Section \ref{sec:impli_and_limit}, and summarise our results in Section \ref{sec:conclusion}.

\section{Method}\label{sec:method}
\subsection{Disc evolution model}\label{sec:method:evo}
In our model, we consider geometrically thin protoplanetary discs regulated by viscosity, MHD winds and internal photoevaporation to assist the rapid clearing at the end of evolution. The gas surface density ($\Sigma_g$) of a viscous disc can be expressed as \citep{1974MNRAS.168..603L}
\begin{equation}
    \frac{\partial \Sigma_g}{\partial t}= \frac{3}{R}\frac{\partial}{\partial R}\Bigl[R^{1/2}\frac{\partial }{\partial R
    }(\nu \Sigma_g R^{1/2})\Bigr],
\end{equation}
where $\nu$ is the viscosity and can be quantified by $\nu = \alpha_\mathrm{SS}c_sH$ \citep{1973A&A....24..337S}. Here, $\alpha_\mathrm{SS}$ is a dimensionless parameter, measuring the efficiency of angular momentum redistribution by turbulence, $c_s$ is the sound speed, and $H$ is the disc scale height. We adopt the prescription for MHD winds developed in \citet{2022MNRAS.512.2290T}, where they use an $\alpha_\mathrm{SS}$-equivalent parameter $\alpha_\mathrm{DW}$ along with the magnetic lever arm parameter $\lambda$ \citep{1982MNRAS.199..883B} to characterise the efficiency of angular momentum removal by winds. We incorporate the analytical model of photoevaporation from \citet{2012ApJ...757L..29A} to account for the rapid disc clearing at late evolutionary stages. Combinations of above mechanisms give the master equation of our model
\begin{multline}\label{eq:master}
    \frac{\partial \Sigma_g}{\partial t}= \frac{3}{R}\frac{\partial}{\partial R}\Bigl[R^{1/2}\frac{\partial }{\partial R
    }(\nu \Sigma_g R^{1/2})\Bigr]\\
    +\frac{3}{2R}\frac{\partial}{\partial R}\Bigl(\frac{\alpha_\mathrm{DW}\Sigma_g c_s^2}{\Omega}\Bigr)
    -\frac{3\alpha_{DW}\Sigma_g c_s^2}{4(\lambda-1)R^2\Omega}-\dot{\Sigma}_w(R,t),
\end{multline}
where $\Omega=\sqrt{GM_*/R^3}$ is the Keplerian orbital frequency at radius $R$ around a central star of $1~M_*$. The first term on the right hand side is the viscous diffusion term. The second and third terms are for the advection term and mass extraction term by the magnetised wind, respectively. The last term is the sink term by internal photoevaporation, which is prescribed as  
\begin{equation}\label{eq:pe_diff}
    \dot{\Sigma}_w(R) = \frac{\dot{M}_\mathrm{thick}}{4\pi R_\mathrm{crit}^2}\Bigl(\frac{R}{R_\mathrm{crit}}\Bigr)^{-5/2},\quad R \geqslant R_\mathrm{crit},
\end{equation}
when the disc within $R_\mathrm{crit}\simeq 0.2GM_*/c_s^2$ is optically thick. If the inner disc becomes optically thin, then the mass-loss rate is modelled as 
\begin{equation} \label{eq:pe_dir}
    \dot{\Sigma}_w(R) = \frac{\dot{M}_\mathrm{thin}}{4\pi R_\mathrm{in}^2}\Bigl(\frac{R}{R_\mathrm{in}}\Bigr)^{-5/2}\Bigl(\frac{R}{2R_\mathrm{crit}}\Bigr)^{1/2},\quad R \geqslant R_\mathrm{in}.
\end{equation}
Here $\dot{M}_\mathrm{thick}$ and $\dot{M}_\mathrm{thin}$ are measurements of the mass-loss rate in \say{diffuse radiation field} and \say{direct radiation field} defined in \citet{2012ApJ...757L..29A}. $R_\mathrm{in}$ is the innermost radius where the surface density is optically thin.

\par We replace variables in Eq. \ref{eq:master} and then solve the equation using an explicit first-order integrator following \citet{1981MNRAS.194..967B}. We evaluate diffusion and advection terms in two steps and impose different boundary conditions on each. For the diffusion term, we impose zero-torque boundary conditions for both the inner and outer boundaries. For the advection term, a zero-torque is only applied for the outer boundary and we replace the inner boundary with a constant power law condition.

\subsection{Dead/wind zone model}\label{sec:model_dz}
\par
\begin{figure}
    \centering
    \includegraphics[width=0.44\textwidth]{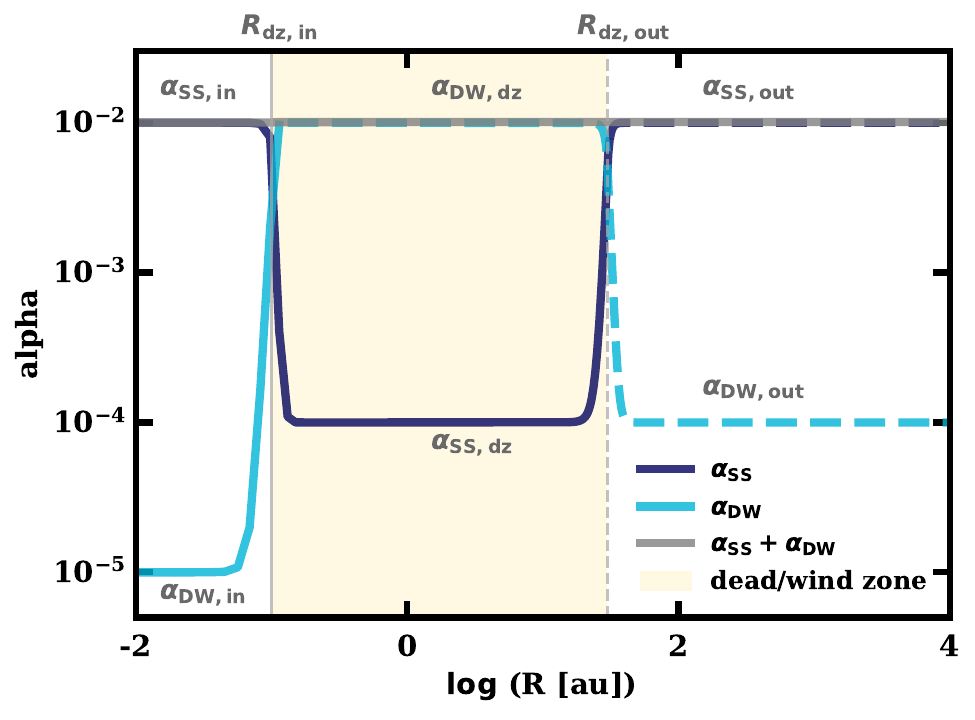}
    \caption{Transition profiles of $\alpha_\mathrm{SS}$ (dark blue lines) and $\alpha_\mathrm{DW}$ (light blue lines): three regions for each are defined in our model following observations. $\alpha_\mathrm{SS,in}=10^{-2}$, $\alpha_\mathrm{SS,dz}=10^{-4}$, $\alpha_\mathrm{DW,in}=10^{-5}$ and the inner edge of the ``dead/wind zone'' $R_\mathrm{dz,in}=0.1~$au are fixed (shown in solid lines). $\alpha_\mathrm{DW,dz}$, $\alpha_\mathrm{DW,out}$, $\alpha_\mathrm{SS,out}$ and $R_\mathrm{dz,out}$ are left as variables in our study (shown in dashed lines). The yellow shaded region is the ``dead/wind zone'' in our model.} 
    \label{fig:alpha_trans}
\end{figure}

 Following \citet{2012MNRAS.420.2851M}, \citet{2019ApJ...871...53G} and \citet{2021A&A...655A..18G}, we adopt a ``three-zone'' model for both viscosity $\alpha_\mathrm{SS}$ and the disc wind $\alpha_\mathrm{DW}$, with two-step transitions between zones. The disc is therefore modelled as a dead zone sandwiched between MRI-active regions. The radial variation of $\alpha_\mathrm{SS}$ is specified by 
\begin{multline}\label{eq:alpha_ss}
    \alpha_\mathrm{SS}(R)=
    \begin{cases}
        \alpha_\mathrm{SS,in}+(\alpha_\mathrm{SS, dz}-\alpha_\mathrm{SS,in})\\ \;\; \quad \cdot
        \begin{cases}
            1/2\exp{\bigl( \frac{R-R_\mathrm{dz,in}}{w_\mathrm{in}}\bigr) } & R<R_\mathrm{dz,in}\\
            \Bigl[ 1-1/2\exp{\bigl( \frac{R_\mathrm{dz,in}-R}{w_\mathrm{in}}\bigr) } \Bigr] & R_\mathrm{dz,in}\leq R<R_\mathrm{m}\\
        \end{cases}\\
        \alpha_\mathrm{SS,dz}+(\alpha_\mathrm{SS, out}-\alpha_\mathrm{SS,dz}) \\ \;\; \quad \cdot
        \begin{cases}
            1/2\exp{\bigl( \frac{R-R_\mathrm{dz,out}}{w_\mathrm{out}}\bigr) } & R_\mathrm{m}<R\leq R_\mathrm{dz,out}\\
            \bigl[1-1/2\exp{\bigl( \frac{R_\mathrm{dz,out}-R}{w_\mathrm{dz,out}}\bigr) }\bigr] & R>R_\mathrm{dz,out}\\
        \end{cases}\\
    \end{cases},
\end{multline}
where $R_\mathrm{dz,in}$ and $R_\mathrm{dz,out}$ delineate the boundary between the inner MRI-active region and the dead zone, and the boundary between the dead zone and the outer MRI-active region, respectively. $R_\mathrm{m}$ is the middle point between $R_\mathrm{dz,in}$ and $R_\mathrm{dz,out}$. $w_\mathrm{in}=R_\mathrm{dz,in}/20$ and $w_\mathrm{out}=R_\mathrm{dz,out}/20$ are adopted to achieve a sharp but continuously differentiable transition between regions. A sharp transition comes from the dead zone model in \citet{2016A&A...596A..81P}, where they start with a slow transition, which evolves to a sharp one at the late time. For simplicity, we assume MHD winds take over the removal of angular momentum in regions covered by the dead zone and describe $\alpha_\mathrm{DW}$ in a similar way as Eq. \ref{eq:alpha_ss}, i.e. replacing $\alpha_\mathrm{SS}$ with $\alpha_\mathrm{DW}$ correspondingly would yield the description of $\alpha_\mathrm{DW}(R)$. We keep boundaries transiting to each region and the width of transition the same for $\alpha_\mathrm{SS}(R)$ and $\alpha_\mathrm{DW}(R)$. We fix $\alpha_\mathrm{DW,in}=10^{-5}$, $\alpha_\mathrm{SS,in}=10^{-2}$ and $\alpha_\mathrm{SS,dz}=10^{-4}$, and explore how the variations of $\alpha_\mathrm{DW,dz}$, $\alpha_\mathrm{DW,out}$ and $\alpha_\mathrm{SS,out}$ affect the disc evolution and observable disc properties. In our dead zone model, though the dead zone is inactive to the MRI, it is active to the MHD wind and can be renamed as \say{dead/wind zone}. An illustration of the \say{dead/wind zone} model is shown in Figure \ref{fig:alpha_trans}. Solid and dashed lines label fixed and free parameters, respectively. Two vertical grey lines indicate locations of the \say{dead/wind zone} inner and outer boundaries, respectively. We fix the inner boundary $R_\mathrm{dz,in}=0.1~$au and vary the outer boundary $R_\mathrm{dz,out}$, which is not well constrained by observations, to investigate the impact of \say{dead/wind zone} sizes on the disc evolution. 

\subsection{Simulation Setup}\label{sec:ch1_setup}
We adopt a time-independent temperature $T(R)\propto R^{-1/2}$ \citep{1987ApJ...323..714K, 1997ApJ...490..368C}, which is a standard temperature for the \say{flaring disc} model and results in viscosity proportional to the radius. We set the aspect ratio $H/R$ to be $0.05$ at $1~$AU, corresponding to a local temperature of $\sim600~$K. We assume an initial disc mass $M_d=0.01~M_\odot$ disc with a characteristic radius $R_\mathrm{c,0}=60~$au surrounding an $M_*=1~M_\odot$ star. The initial gas surface density profile is described by a \say{cutoff} power-law function
\begin{equation}\label{eq:visc_ini_slp}
    \Sigma_g(R)=\frac{M_d}{2\pi R_\mathrm{c,0}^2}\Bigl(\frac{R}{R_\mathrm{c,0}}\Bigr)^{-1}\exp\Bigl(-\frac{R}{R_\mathrm{c,0}}\Bigr), 
\end{equation}
distributed among 8000 cells equispaced in $R^{1/2}$ between $0.0056~$au and $40,000~$au, which is sufficiently large to allow discs with a large $\alpha_\mathrm{SS,out}$ to continuously expand during the entire evolution. Eq. \ref{eq:visc_ini_slp} is not a self-similar solution when the wind component is also taken into account (see Eq. \ref{eq:sigma_profile}), though the impact is probably small. We assume the magnetic field evolves in a way more slowly than that of the gas surface density and conforms to $\alpha_\mathrm{DW}(R,t)\propto \Sigma_c(R,t)^{-\omega}$, where $\Sigma_c=M_d(t)/2\pi R_c(t)^2$, with $\omega$ between $0$ and $1$ \citep{2016A&A...596A..74S, 2022MNRAS.512.2290T}. Aside from the strength of the magnetic field, $\alpha_\mathrm{SS}$ is also sensitive to the degree of ionization \citep[e.g.,][]{2018ApJ...865...10S}, which is not depicted in our simple 1-D model. Therefore, we leave it as a constant with time for a given radius in this work as most work based on 1-D models and 2-D hydrodynamical simulations. Accurately tracing $R_c(t)$ in simulations is challenging, as substructures and disc winds make the disc surface density profile deviate from the original one (see Section \ref{sec:para_exp_spread_fixed} and Appendix \ref{sec:appendix_r_c}). We instead use $\alpha_\mathrm{DW}(R,t)\propto M_d(t)^{-\omega}$, to avoid computing $R_c(t)$ \say{on the fly}. The latter is equivalent to the former when the disc is purely evolved under MHD winds, but underestimates $\alpha_\mathrm{DW}$ when $R_c$ continuously increases, which is true for models studied here. We adopt $\omega = 0.5$ throughout the paper. As $\alpha_\mathrm{DW}$ is a time-varying parameter, if it is not otherwise specified, the value of $\alpha_\mathrm{DW}$ assigned in this paper refers to its initial value at $t=0$. We set $\lambda = 3$, following previous theoretical studies \citep{2022ApJ...926...61T, 2022MNRAS.514.1088Z} and observations of disc winds from Class II objects ($\lambda=1.6$-$2.3$) \citep{2018A&A...618A.120L, 2021ApJS..257...16B}. We also discuss the selection of $\lambda$ in Section \ref{sec:discussion_leverarm}.
$\dot{M}_\mathrm{thick}$ in Eq. \ref{eq:pe_diff} and $\dot{M}_\mathrm{thin}$ in Eq. \ref{eq:pe_dir} are fixed to representative values of $10^{-10}~M_\odot~\mathrm{yr^{-1}}$ and $10^{-9}~M_\odot~\mathrm{yr^{-1}}$ \citep[e.g.,][]{2004ApJ...607..890F,2021MNRAS.508.1675E}, respectively, noting that the stronger photoevaporative wind is only triggered when the surface density in the inner disc becomes optically thin at late times. We impose a maximum evolution time of $12~$Myr, when simulations are automatically terminated regardless of remaining mass in the disc. It is worth noting that $t=0$ in our simulations represents the time when the envelope infall rate is smaller than the stellar accretion rate instead of the initial time when the disc is formed. The stage studied in this work is close to the Class II disc defined from the infrared excess.

\subsection{Code Testing}
To test our code against analytical solutions provided in \citet{2022MNRAS.512.2290T}, we implement four simulations with constant $\alpha$ along the radius by activating 1) only the viscosity component; 2) only the wind component with no magnetic field evolution; 3) viscosity+MHD wind with no magnetic field evolution; and 4) only the wind component with magnetic field evolution. When testing cases involving MHD winds, we modify the slope of the initial surface density profile by adding $\xi$
\begin{equation}\label{eq:sigma_profile}
    \Sigma_g(R)=\frac{M_d}{2\pi R_\mathrm{c,0}^2}\Bigl(\frac{R}{R_\mathrm{c,0}}\Bigr)^{-1+\xi}\exp\Bigl(-\frac{R}{R_\mathrm{c,0}}\Bigr) 
\end{equation}
to ensure they have the same initial surface density profile as analytical solutions. $\xi$ is the mass ejection index and can be quantified by $\psi=\alpha_\mathrm{DW}/\alpha_\mathrm{SS}$ and the lever arm $\lambda$ as \citep{2022MNRAS.512.2290T}
\begin{equation}\label{eq:wind_slope}
    \xi=\frac{1}{4}(\psi+1)\Bigg[\sqrt{1+\frac{4\psi}{(\lambda-1)(\psi+1)^2}}-1  \Bigg].
\end{equation}
Our numerical method recovers the analytic solutions well: the comparison between the numerical and analytical solutions is shown in Appendix \ref{sec:appendix_test}. 

\section{Fiducial Model}\label{sec:fiducial}
We initiate our study by building a fiducial model adopting a total $\alpha_\mathrm{tot}(R)=\alpha_\mathrm{SS}(R)+\alpha_\mathrm{DW}(R)\simeq 10^{-2}$, to simulate a disc with an almost constant total $\alpha$ throughout the whole disc and investigate the roles that MHD winds and \say{dead/wind zones} play in comparison to a fundamental viscous disc with a constant $\alpha_\mathrm{SS}=10^{-2}$ facilitated by photoevaporative winds. We assume the \say{dead/wind zone} (see Section \ref{sec:method}) spanning from $0.1~$au to $30~$au, within which $\alpha_\mathrm{DW,dz}=10^{-2}$. Viscosity dominates over the magnetised wind in the outer disc, where $\alpha_\mathrm{SS,out}=10^{-2}$ and $\alpha_\mathrm{DW,out}=10^{-4}$. Other parameters are fixed as specified in Section \ref{sec:model_dz}. This transition profile is equivalent to the one depicted in Figure \ref{fig:alpha_trans}. 

\par
\begin{figure*}
    \centering
    \includegraphics[width=\textwidth]{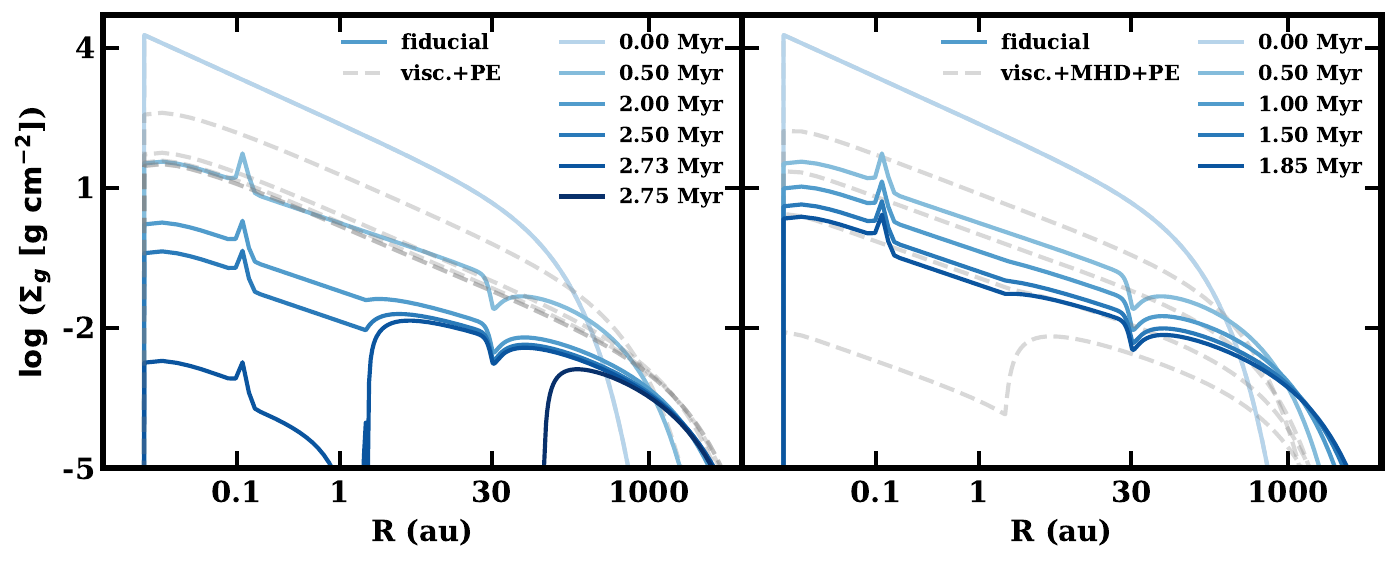}
    \caption{Evolution of the gas surface density for the fiducial disc ($R_{c,0}=60~$au, $R_\mathrm{dz,out}=30~$au and $\alpha$ as illustrated in Figure \ref{fig:alpha_trans}) compared with two discs without dead zone models shown in dashed grey lines in the left and right panels. Time of surface density profiles indicated on the upper right corner of each panel is different for two panels. In the left panel, the fiducial model is compared with a viscosity-only model ($\alpha_\mathrm{SS}=10^{-2}$) considering photoevaporation. In the right panel, the fiducial model is in comparison with a simple ``hybrid'' disc considering photoevaporation ($\alpha_\mathrm{SS}=5\times10^{-3}$ and $\alpha_\mathrm{DW}=5\times10^{-3}$). The inclusion of magnetised winds can efficiently accelerate the evolution and the inclusion of dead zones can create substructures in gas discs.}
    \label{fig:fid_l3}
\end{figure*}

\begin{figure}
    \centering
    \includegraphics[width=0.46\textwidth]{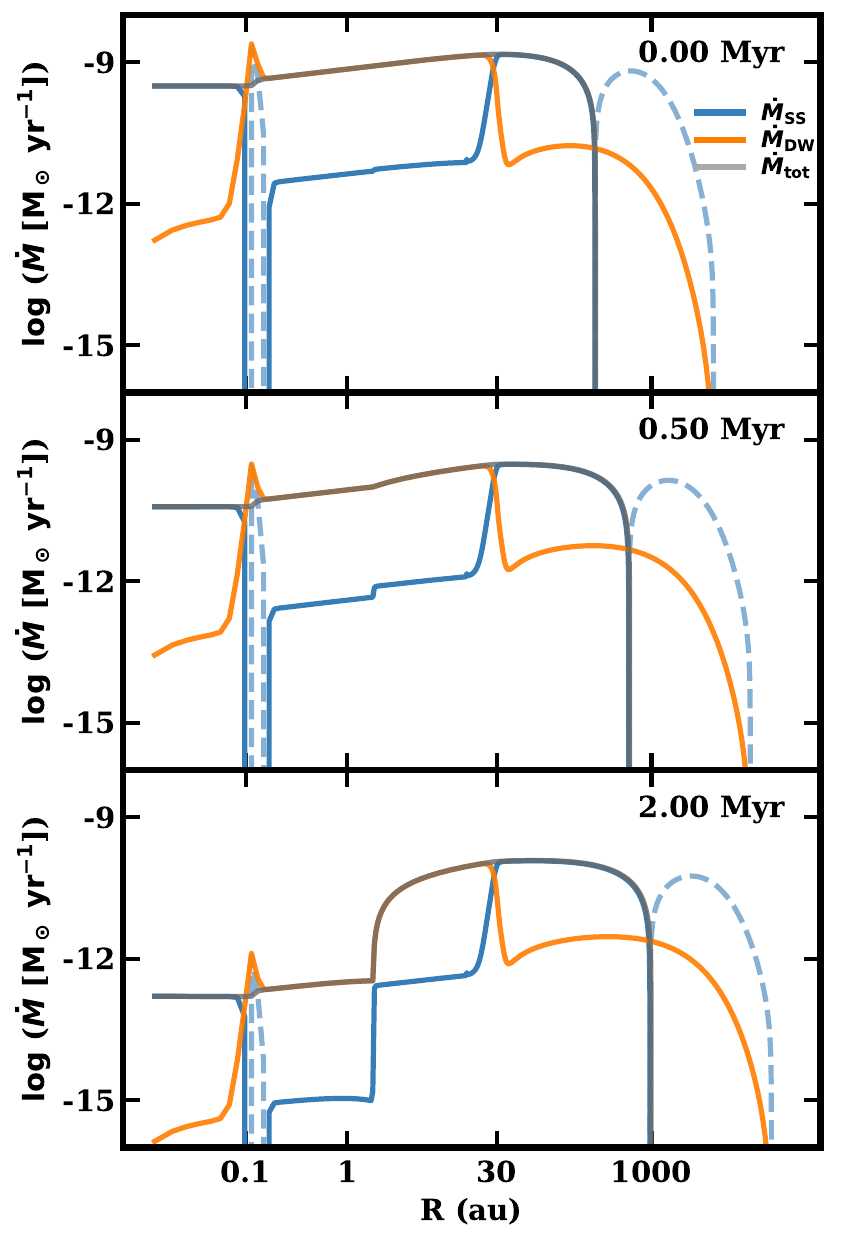}
    \caption{Local accretion rates driven by viscosity (blue lines), by MHD winds (orange lines) and by both (grey lines) at $t=0$ Myr (top panel), $0.5$ Myr (middle panel) and $2$ Myr (bottom panel) for a fiducial model. $\dot{M}(R)>0$ indicates a flow moving towards the host star (accretion, solid lines) while $\dot{M}(R)<0$ is for flows moving towards the outer disc (``decretion'', blue dashed lines).}
    \label{fig:fid_l3_mdot}
\end{figure}

 Figure \ref{fig:fid_l3} illustrates how the gas surface density evolves in the fiducial model compared concurrently with pure viscosity+photoevaporation model in the left panel and \say{hybrid} (viscosity+wind) +photoevaporation model in the right panel. It is evident that winds flatten the slope of $\Sigma_g$ and accelerate disc evolution by extracting mass from the disc and subtly aiding the stellar accretion (Figure \ref{fig:fid_l3_mdot}). At $\sim2.5~$Myr, the viscous disc (in the left panel) still has relatively high surface densities, while ``hybrid'' discs have lost a great proportion of mass before $\sim 2~$Myr.

\par The inclusion of the \say{dead/wind zone} can alter the smooth gas surface density profiles to ones with substructures formed around the inner and outer edges of the \say{dead/wind zone}. More detailed discussion on these substructures can be found in Section \ref{sec:para_exp_surf_cat}.
In these edges, $\dot{M}(R)$ for viscosity and for MHD winds change substantially due to the sharp transition of $\alpha_\mathrm{DW}(R)$ and $\alpha_\mathrm{SS}(R)$, and bring up additional mass accumulated or removed locally. The underlying physics is well illustrated by Figure \ref{fig:fid_l3_mdot}, and the analytical solutions of mass accretion rates by viscosity and winds are given in Eq. \ref{eq:ch1_mdot_ss} and Eq. \ref{eq:ch1_mdot_dw}.

\begin{equation}\label{eq:ch1_mdot_ss}
    \dot{M}_\mathrm{SS}(R) = \frac{6\pi}{R\Omega}\frac{\partial}{\partial R}\Bigl(\Sigma_g c_s^2\alpha_\mathrm{SS}R^2\Bigr),
\end{equation} 
and by MHD winds 
\begin{equation}\label{eq:ch1_mdot_dw}
    \dot{M}_\mathrm{DW}(R)=\frac{3\pi \Sigma_g c_s^2\alpha_\mathrm{DW}}{\Omega}.
\end{equation}
\noindent The wind responds to the change of $\alpha$ in a distinct way from viscosity. Accretion rates driven by winds change proportionally to $\alpha_\mathrm{DW}$ as $\dot{M}_\mathrm{DW}(R)\propto \alpha_\mathrm{DW}c_sH\Sigma_g$, while $\dot{M}_\mathrm{SS}(R)$ varies with the gradient of the product in parentheses in Eq. \ref{eq:ch1_mdot_ss}. When $\alpha_\mathrm{DW}$ decreases abruptly, $\dot{M}_\mathrm{DW}(R)$ drops significantly, leaving sufficient gas piled up in the \say{dead/wind zone}. In contrast, the decrease in $\alpha_\mathrm{SS}$ results in a positive velocity. Therefore, gas flows outwards to smooth out the gas accumulation. This is clearly shown by the dashed blue lines at $\sim0.1~$au in Figure \ref{fig:fid_l3_mdot}. This mass outflow persists for a majority of the disc lifetime. Similar behaviour of $\dot{M}_\mathrm{SS}$ has also been found by other studies that incorporate the dead zone model, such as \citet{2012MNRAS.420.2851M} (their Figure 3) and \citet{2021A&A...655A..18G} (their Figure 4). Although significant changes exist in both $\dot{M}_\mathrm{SS}$ and $\dot{M}_\mathrm{DW}$ when the \say{dead/wind zone} is taken into account, the total mass accretion rate $\dot{M}_\mathrm{tot}$ remains smooth in the fiducial case (solid grey lines in Figure \ref{fig:fid_l3_mdot}), where $\alpha_\mathrm{tot}$ is nearly constant along the radius, as expected from a purely viscous disc with constant $\alpha_\mathrm{SS}$, until rapid clearing is switched on at late stages (the bottom panel of Figure \ref{fig:fid_l3_mdot}).

\par In addition to substructures created by the incorporation of \say{dead/wind zones}, the strong photoevaporation triggered at a later stage when $\Sigma_g(R\simeq R_\mathrm{crit})$ becomes optically thin takes away gas and then opens a gap around the critical radius $R_\mathrm{crit}$ (see Section \ref{sec:method:evo}). The gap further becomes an inner cavity when the disc interior to $R_\mathrm{crit}$ is fully accreted on to the star and the gas from the outer disc cannot fuel the inner disc, due to the photoevaporative mass-loss rate exceeding the local accretion rate \citep{2001MNRAS.328..485C}.

\section{Parameter Exploration}\label{sec:explor}

Following the fiducial model, we expand the three free parameters ($\alpha_\mathrm{DW,dz}$, $\alpha_\mathrm{DW,out}$ and $\alpha_\mathrm{SS,out}$, see Figure \ref{fig:alpha_trans}) to broader parameter space (see Table \ref{tb:ch1_para} for specific values) to study how variations of $\alpha_\mathrm{SS}$ and $\alpha_\mathrm{DW}$ affect stellar accretion rates, surface density profiles, gas disc sizes, lifetimes, and cumulative mass loss by different physical processes. We then further extend our investigation to impacts of the \say{dead/wind zone} size and the initial disc characteristic radius $R_{c,0}$ on disc evolution. Accompanying these \say{hybrid} models are two \say{naive} models designed to compare and illustrate how the inclusion of \say{dead/wind zones} makes disc behaviours differ from what we expect for a commonly assumed constant-$\alpha$ disc. One of the \say{naive} models is a viscous disc ($\alpha_\mathrm{SS}=10^{-3}$) with internal photoevaporation (as introduced in Section \ref{sec:method:evo}); the other is a wind-only disc ($\alpha_\mathrm{DW}=10^{-3}$) incorporating a magnetic field evolved in the same way as that in \say{hybrid} models. Parameters that we examine in the following sections are listed in Table \ref{tb:ch1_para} above the dividing line, below which we also provide parameters that are fixed in simulations. We adopt a small $\alpha_\mathrm{DW,out}$ as non-ideal MHD simulations show the accretion rate in the outer disc is dominated by the MRI-driven accretion caused by FUV-induced ionization in upper layers \citep{2013ApJ...772...96B, 2013ApJ...775...73S, 2015ApJ...798...84B}.

\par We conduct 92 simulations in two separate groups. First, we run 27 simulations with all combinations of varying $\alpha$ in Table \ref{tb:ch1_para} for discs with fixed $R_{c,0}=60~$au and $R_\mathrm{dz,out}=30~$au. Among them, we select 13 representative combinations of $\alpha$ to study the disc size problem. We stretch the initial characteristic radius $R_{c,0}$ from $60$ to $120~$au to examine how the disc size affects disc evolution. As the \say{dead/wind zone} outer edge $R_\mathrm{dz,out}$, fixed in the first group of simulations, is also not well determined by observations and simulations, we vary it from $30~$au to $75~$au, and to $135~$au. These $13\times(6-1)=65$ simulations constitute the second group of simulations. Results for two groups of simulations can be found in Table \ref{tb:parameters}, which is also visualised in Figure \ref{fig:cum_mass_loss} and Figure \ref{fig:table_visual} to assist reading.

\begin{table}
\caption{Summary of parameters explored (above the dividing line) and fixed in our models (below the dividing line). Values shown in bold are those adopted in the fiducial model.}\label{tb:ch1_para}
\centering
\begin{tabular}{c@{\hskip 0.7in}c}
\hline
\hline
Parameter & Values \\ 
\hline
$\alpha_\mathrm{DW,dz}$ & $\mathbf{10^{-2}}$, $10^{-3}$, $5\times10^{-4}$ \\
$\alpha_\mathrm{DW,out}$ & $10^{-3}$, $\mathbf{10^{-4}}$, $10^{-5}$ \\
$\alpha_\mathrm{SS,out}$ & $\mathbf{10^{-2}}$, $10^{-3}$, $3\times10^{-4}$ \\
Characteristic radius $R_{c,0}$ (au) & $\mathbf{60}$, $120$ \\
Dead zone outer edge $R_\mathrm{dz,out}$ (au) & $\mathbf{30}$, $75$, $135$\\
\hline
Disc mass $M_d$ ($M_\odot$) & 0.01\\
Stellar mass $M_*$ ($M_\odot$) & 1 \\
Aspect ratio $H/R|_\mathrm{R=1\,au}$ & 0.05\\
Dead zone inner edge $R_\mathrm{dz,in}$ (au) & 0.1\\ 
Lever arm $\lambda$ & 3\\
Evolution of magnetic field $\omega$
& 0.5\\
$\dot{M}_\mathrm{PE,thick}$ ($M_\odot~\mathrm{yr^{-1}}$) & $10^{-10}$\\
$\dot{M}_\mathrm{PE,thin}$ ($M_\odot~\mathrm{yr^{-1}}$)& $10^{-9}$\\
\hline

\end{tabular}
\end{table}

\subsection{Stellar accretion rate}\label{sec:para_exp_acc}

\begin{figure*}
    \centering
    \includegraphics[width=\textwidth]{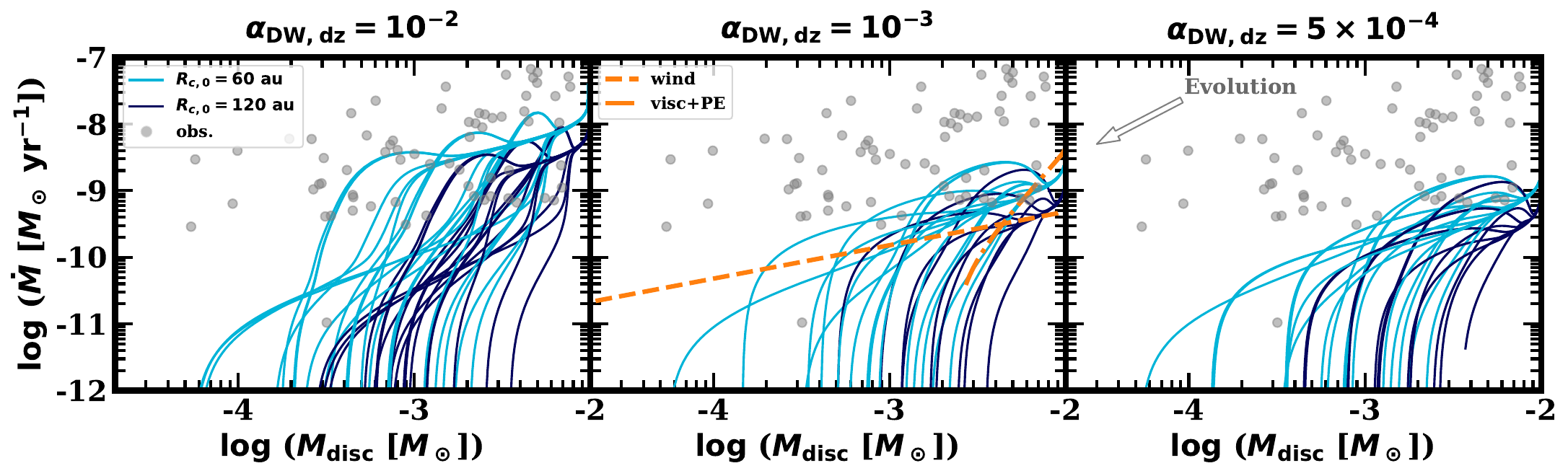}
    \caption{Evolution of all the 92 discs in the $\dot{M}_*$-$M_d$ plane. The three panels are for discs with $\alpha_\mathrm{DW,dz}=10^{-2}$ (left panel), $\alpha_\mathrm{DW,dz}=10^{-3}$ (middle panel) and $\alpha_\mathrm{DW,dz}=5\times10^{-4}$ (right panel), respectively. Each panel includes models with $R_\mathrm{c,0}=60$ (light blue lines) and $120~$au (dark blue lines), and observations from \citet{2022arXiv220309930M} for stars in the range of $0.3-1.2~M_\odot$ (grey dots). No upper limits for either stellar accretion rates or disc masses are included. Evolutionary tracks for two ``naive'' models -- pure viscosity+photoevaporation model ($\alpha_\mathrm{SS}=10^{-3}$, dash-dotted orange line) and pure wind model ($\alpha_\mathrm{DW}=10^{-3}$, dashed orange line) -- are overlaid in the middle panel. An arrow in the third panel indicates the evolutionary pathway is from the upper right to the lower left.}
    \label{fig:acc_rate}
\end{figure*}

The stellar accretion rate is one of observables for which we have a statistically large sample and that can be used to constrain the disc evolution model. We plot stellar accretion rates vs. disc gas masses of all the 92 models in Figure \ref{fig:acc_rate}, with comparison to observed stellar accretion rates and disc masses around stars with masses of $0.3-1.2~M_\odot$ \citep[from the compilation in][]{2022arXiv220309930M}. Discs with upper-limits (non-detections) on either stellar accretion rates or disc masses are excluded. Models are classified in three panels by their $\alpha_\mathrm{DW, dz}$, which determines initial stellar accretion rates together with the disc initial characteristic radius when $\alpha_\mathrm{SS,in}$, $\alpha_\mathrm{DW,in}$ and $\alpha_\mathrm{SS,dz}$ are fixed. Our models can explain intermediate mass discs ($3\times10^{-4}-10^{-2}~M_\odot$) with intermediate stellar accretion rates ($<2\times 10^{-8}~M_\odot~\mathrm{yr^{-1}}$) in the $\dot{M}_*-M_d$ plane. For a given initial disc mass, the upper limit of the stellar accretion rate can be elevated if a smaller $R_\mathrm{c,0}$ or a larger lever arm $\lambda$ is assumed.

\par The stellar accretion rates of ``hybrid'' models behave similarly to that of a purely viscous disc except the latter has a much longer evolutionary timescale (>12 Myr). On the contrary, the wind-only model follows a distinct evolutionary pathway. Its accretion rate can sustain a relatively high value when the disc mass is low, extending the evolutionary pathway to a region where no observational data has been obtained (the lower left corner in the $\dot{M}_*-M_d$ plane). However, if a larger lever arm is adopted for the pure wind model, it is able to explain low-mass discs ($\sim 10^{-4}~M_\odot$) observed with relatively high accretion rates ($\sim 10^{-9}~M_\odot~\mathrm{yr^{-1}}$).

\par As $M_d(t)$ should be a monotonically decreasing variable with time, small bumps exhibited in evolution tracks in the $\dot{M}_*-M_d$ plane indicate that $\dot{M}_*$ is not consistently declining with time for some models. This means some discs even after entering Class II still undergo small accretion \say{outbursts} due to the mass accumulation in the inner disc when the presence of \say{dead/wind zones} is taken into account. 

\subsection{Categorization of the surface density} \label{sec:para_exp_surf_cat}
As shown in Section \ref{sec:fiducial}, the relative change in $\alpha_\mathrm{DW}$ and $\alpha_\mathrm{SS}$ along the radius always leads to the creation of gas substructures. By visually inspecting substructures in the surface density profiles from group 1 simulations ($R_\mathrm{c,0}=60~$au and $R_\mathrm{dz,out}=30~$au), we can roughly classify them into three categories (Figure \ref{fig:ch1_sf_class}). 

\par When $\alpha_\mathrm{DW,dz}<10^{-2}$ (Category A), accretion driven by winds in the \say{dead/wind zone} is inefficient in transferring mass fed by the outer disc to the inner disc, and gas continually accumulates around the inner transition radius $R_\mathrm{dz,in}$, maintaining an overall surface density relatively higher than those of the other two categories. The fixed large $\alpha_\mathrm{SS,in}$ ($10^{-2}$) in the inner disc, set by default, efficiently fuels the central star, enabling quick consumption of the local gas. The contrast of the accretion rate on the two sides of the \say{dead/wind zone} inner edge forms a bump in the surface density. When $\alpha_\mathrm{DW,dz}= 10^{-2}$ (Category B and C), the accretion rates in the inner disc ($R\leq R_\mathrm{dz,in}$) and within the \say{dead/wind zone} ($R_\mathrm{dz,in}<R\leq R_\mathrm{dz,out}$) are comparable over the majority of the evolution and no substantial mass accumulates at the inner transition radii ($R_\mathrm{dz,in}$). The less significant change in the total $\alpha$ around $R_\mathrm{dz,in}$ in Category B and C renders a narrower spike in the surface density, instead of a wider bump. 

\par The morphology of the gas accumulation depends on the $\alpha$ assumed on the two sides of the ``dead/wind zone'' inner boundary, which, though not well constrained, are assigned reasonable values in our models. The width of the gas accumulation and the $\alpha$-transition itself both are several times wider than the local scale-height, making the excitation of Rossby wave instability less likely \citep{2009A&A...497..869L, 2012MNRAS.419.1701R}. But whether such a feature is stable or not should be studied in 2-D or 3-D simulations, which are beyond the scope of this study.

\par The morphology in the outer disc -- whether the gas is concentrated to a bump or not -- can further classify discs into Category B and C. When the outer disc is dominated by efficient expansion (large $\alpha_\mathrm{SS,out}$), mass is primarily moving further out and no significant mass is piled up (Category C). When the expansion is less efficient (small $\alpha_\mathrm{SS,out}$), the wind-driven accretion can \say{compensate} the spreading driven by viscosity to some extent, leading to more mass participating in the accretion and stocked up in the \say{dead/wind zone} outer edge (Category B). This process is also reflected in the smaller gas disc sizes in the middle panel of Figure \ref{fig:ch1_sf_class2} compared to those in the right panel. Regardless of the dominant mechanisms in the outer disc, a ``dip'' feature can be observed around the outer boundary of the ``dead/wind zone'' in all categories (see three panels of Figure \ref{fig:ch1_sf_class2}). This arises from the transition of $\alpha_\mathrm{DW}$ from larger to smaller values (see also Eq. \ref{eq:ch1_mdot_dw}).

\par For the 27 simulations in the first group, 18 cases belonging to Category A share the common feature that $\alpha_\mathrm{tot}=\alpha_\mathrm{DW}+\alpha_\mathrm{SS}$ in the \say{dead/wind zone} is not significantly larger or even smaller than $\alpha_\mathrm{tot}$ in the outer disc. Category B contains 6 simulations and they have $\alpha_\mathrm{tot}$ in the \say{dead/wind zone} considerably greater than that in the outer disc ($\alpha_\mathrm{tot,dz}/\alpha_\mathrm{tot,out}>10$). Three simulations are classified as Category C, where we require the \say{dead/wind zone} to be strongly influenced by the efficient wind $\alpha_\mathrm{DW,dz}=10^{-2}$ and the outer disc to be dominated by viscosity ($\alpha_\mathrm{SS,out}=10^{-2}$) initially. Similar classification is also applied to discs when their $R_\mathrm{c,0}$ and $R_\mathrm{dz,out}$ are extended to larger values for simulations in the second group.

\tikzstyle{startstop} = [rectangle, rounded corners, minimum width=3cm, minimum height=1cm,text centered, draw=black, fill=gray!20]
\tikzstyle{io} = [trapezium, trapezium left angle=70, trapezium right angle=110, minimum width=3cm, minimum height=1cm, text centered, draw=black, fill=blue!30]

\begin{figure*}
\centering
    \begin{subfigure}{\linewidth}
    \centering
    \begin{tikzpicture}[node distance=2cm]
    \node (start) [startstop] {$\alpha_\mathrm{DW,dz}\geq10^{-2}$ ?};
    \node (decision_a) [startstop, below of=start, xshift=-2cm] {Category A};
    \node (decision_b) [startstop, below of=start, xshift=2cm] {$\alpha_\mathrm{SS,out}\geq 10^{-2}$ ?};
    \node (decision_b1) [startstop, below of=decision_b, xshift=-2cm] {Category B};
    \node (decision_b2) [startstop, below of=decision_b, xshift=2cm] {Category C};
    \draw [arrow] (start) -- node[anchor=east] {No}(decision_a);
    \draw [arrow] (start) -- node[anchor=west] {Yes}(decision_b);
    \draw [arrow] (decision_b) -- node[anchor=east] {No}(decision_b1);
    \draw [arrow] (decision_b) -- node[anchor=west] {Yes}(decision_b2);
\end{tikzpicture}
    \caption{Classification flowchart.}
    \label{fig:ch1_sf_class1}
    \end{subfigure}

\bigskip
    \begin{subfigure}{\linewidth}
  \centering
  \includegraphics[width=\linewidth]{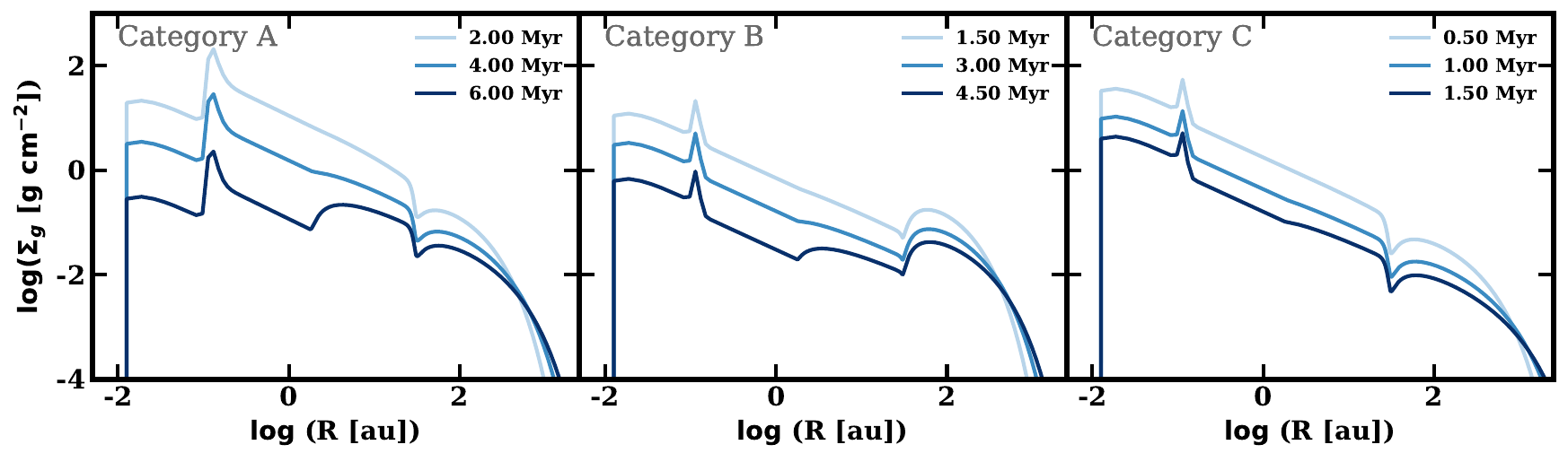}
  \caption{Three categories of the surface density profiles: profiles are selected from representative simulations of each category (Simulation 14 for Category A, Simulation 23 for Category B and Simulation 24 for Category C, see Table \ref{tb:parameters}) with equivalent time steps with respect to their own lifetimes to show the general behaviours until the effect of internal photoevaporation cannot be neglected.}\label{fig:ch1_sf_class2}
    \end{subfigure} 
\caption{Classification of surface density profiles and examples of each classification.}
\label{fig:ch1_sf_class}
\end{figure*}

\subsection{Disc spreading} \label{sec:para_exp_spread_fixed}
Three different radii are typically used to characterise the disc sizes: the characteristic radius $R_c$, beyond which the disc surface density drops exponentially; the outer radius $R_o$, a disc radius set by a certain surface density threshold; and the transition radius $R_t$ \citep{1998ApJ...495..385H, 2009ApJ...701..260I}, delimiting the accreting disc ($\dot{M}(R\leq R_t)\geq 0$) and the spreading disc ($\dot{M}(R>R_t)<0$). In this section, we explore the evolution of these radii in various combinations of $\alpha$ and discuss how they can be applied to understand observations.

\par The characteristic radius $R_c$ is commonly used to define the initial disc size. It keeps growing in the conventional viscous disc and remains unchanged in the magnetised wind disc \citep{2022MNRAS.512.2290T}. The outer radius $R_o$ increases in viscous discs and shrinks in wind-only discs (see the overlaid circles and triangles in the upper middle panel of Figure \ref{fig:sizes_allmodel}). The transition radius $R_t$ maintains its meaning only when the viscosity is considered as a purely wind-driven disc contracts at any radii at all times. Although the measurements of $R_o$ and $R_t$ are still straightforward when integrating the \say{dead/wind zone} into a \say{hybrid} disc, which entangles effects of winds and viscosity together, it can be challenging to trace the motion of $R_c$. The wind modifies the slope of the surface density profile and the presence of \say{dead/wind zones} creates substructures (see Section \ref{sec:fiducial}), jointly hindering the estimation of $R_c$ from simply fitting the surface density profile with a tapered power-law function. Therefore, we characterise $R_c$ for \say{hybrid} discs statistically. Detailed explanation of the method can be found in Appendix \ref{sec:appendix_r_c}.
 
 
\par For all models with $R_\mathrm{c,0}=60~$au and $R_\mathrm{dz,out}=30~$au, we measure these three radii every $0.5~$Myr. We deliberately choose a very small surface density threshold of $10^{-10}~\mathrm{g~cm^{-2}}$ for $R_o$ to accurately trace the outer disc motion. We defer the discussion of the selection of the surface density threshold to Section \ref{sec:disc_threshold}. We fit the variation of radii with time using a linear function, as suggested by visual inspection and analytical solution \citep{1998ApJ...495..385H}. The slopes of the fitting functions denoted as $\mathrm{d}R_c/\mathrm{d}t$, $\mathrm{d}R_o/\mathrm{d}t$ and $\mathrm{d}R_t/\mathrm{d}t$, are applied to characterise the expansion rates of $R_c$, $R_o$ and $R_t$, separately.

\par
\begin{figure*}
    \centering
    \includegraphics[width=1\textwidth]{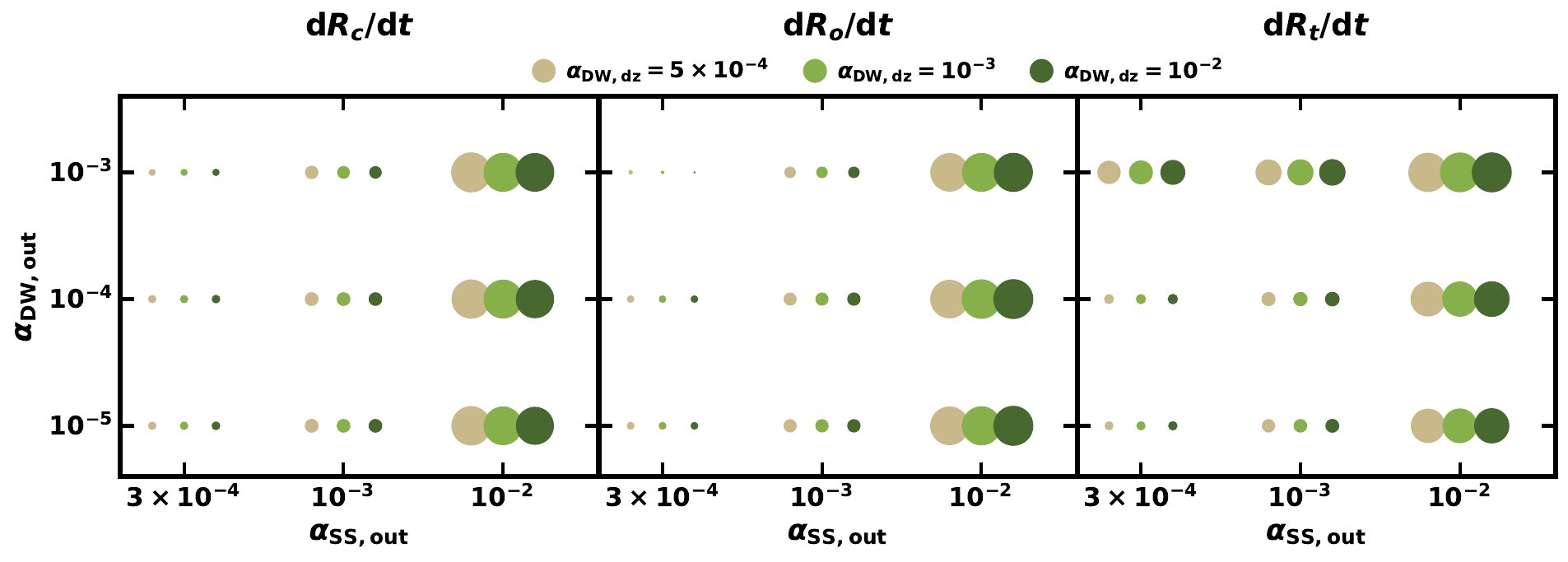}
    \caption{Expansion rates of three radii: $R_c$ (the radius beyond which the surface density drops exponentially), $R_o$ (the radius defined by a selected surface density threshold, here $\Sigma_\mathrm{thres}=10^{-10}~\mathrm{g~cm^{-2}}$) and $R_t$ (the radius delimiting accreting disc and expanding disc) for discs in group 1 simulations. Panels from the left to the right are for $\mathrm{d}R_c/\mathrm{d}t$, $\mathrm{d}R_o/\mathrm{d}t$ and $\mathrm{d}R_t/\mathrm{d}t$, respectively. Colours on each panel show the values of $\alpha_\mathrm{DW, dz}$ (light brown for $\alpha_\mathrm{DW, dz}=5\times10^{-4}$, light green for $\alpha_\mathrm{DW, dz}=10^{-3}$ and dark green for $\alpha_\mathrm{DW, dz}=10^{-2}$). The expansion rate is linearly encoded in the dot area, which is normalized to the maximum value in each panel. Irrespective of the radius used for characterisation, the expansion rate is dominated by $\alpha_\mathrm{SS,out}$ and moderately affected by $\alpha_\mathrm{DW,out}$ in some cases.}
    \label{fig:ch1_kc_ko_kt}
\end{figure*}

\begin{figure*}
    \centering
    \includegraphics[width=\textwidth]{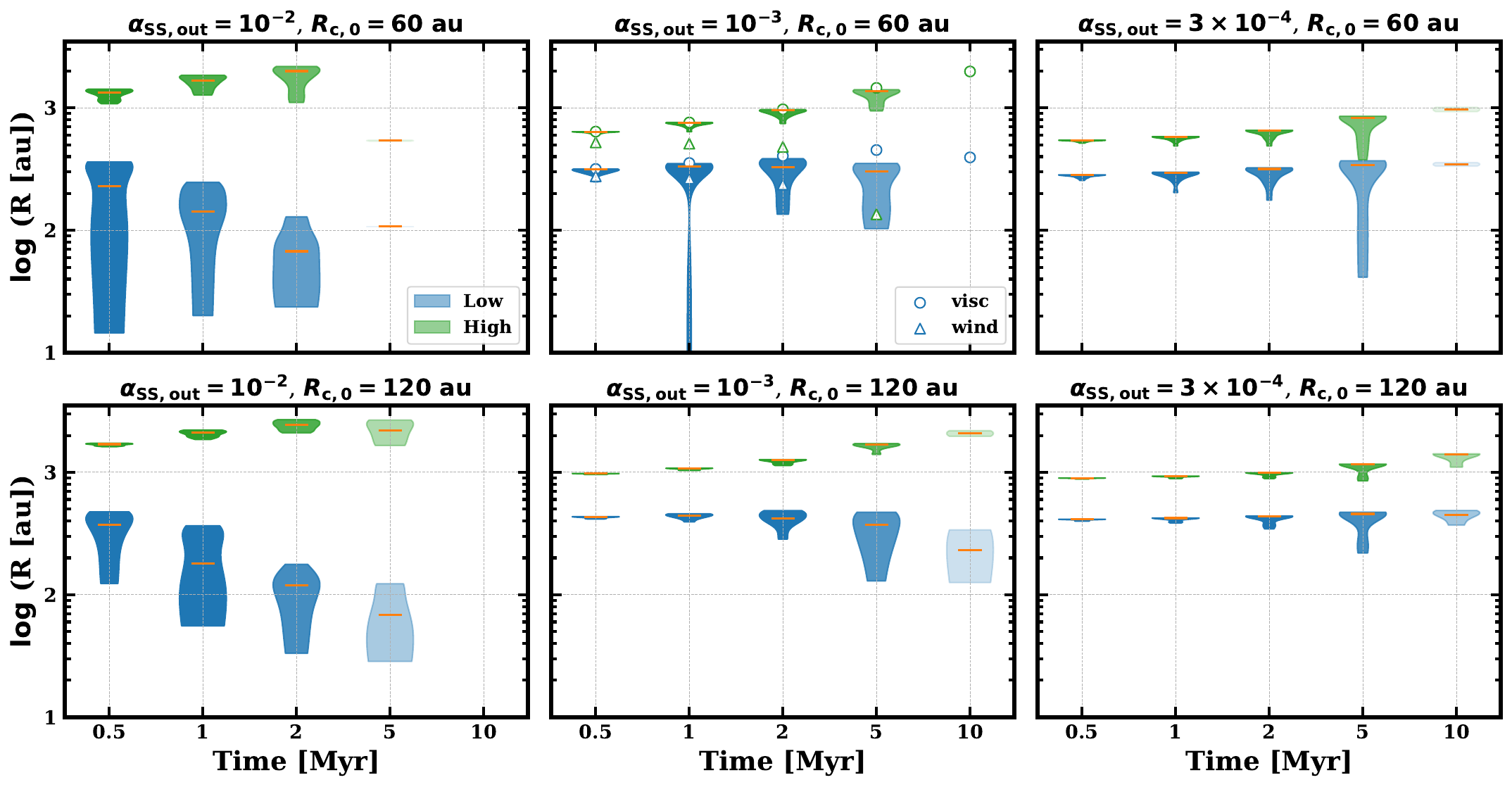}
    \caption{Disc sizes $R_o$ measured at 5 epochs (0.5, 1, 2, 5 and 10 Myr) via two surface density thresholds mimicking higher-sensitivity observations (green patches, $\Sigma_\mathrm{thres}=10^{-4}~\mathrm{g~cm^{-2}}$) limited by photodissociation of $\mathrm{{}^{12}CO}$, and lower-sensitivity ones (blue patches, $\Sigma_\mathrm{thres}=10^{-2}~\mathrm{g~cm^{-2}}$), for all models listed in Table \ref{tb:parameters}. The violin plot only shows the distribution of sizes. The number of discs remained in different stages, relative to the number of discs at $0.5~$Myr in each panel, is shown in the transparency. Higher transparency indicates more discs are dispersed at that time. The horizontal orange bar overplotted in each patch shows the median disc sizes. The long tail in the upper middle panel at $1~$Myr is due to one disc possessing a radius of $\lesssim 10$ au. The models are classified into 6 panels by their $\alpha_\mathrm{SS,out}$ and $R_\mathrm{c,0}$. The disc sizes of two ``naive'' models are plotted in the upper middle panel in open circles (viscous discs) and triangles (wind-only discs). Viscous discs (open circles) retain sizes similar to the median sizes inferred from ``hybrid'' models for both thresholds throughout the entire evolution.}
    \label{fig:sizes_allmodel}
\end{figure*}

 Figure \ref{fig:ch1_kc_ko_kt} shows clearly that the expansion rate increases with $\alpha_\mathrm{SS,out}$, and that $\alpha_\mathrm{DW, dz}$ has an almost negligible effect on the disc expansion rate regardless of which measurements we use. $\alpha_\mathrm{DW, out}$ also plays a minor role unless for $\mathrm{d}R_t/\mathrm{d}t$. When discs possess a large $\alpha_\mathrm{DW, out}$, it typically comes with a large $\mathrm{d}R_t/\mathrm{d}t$. This is because the efficient accretion driven by winds can partly offset the spreading caused by $\alpha_\mathrm{SS, out}$ in the outer disc and enlarge the region covered by an overall inflow ($\dot{M}_\mathrm{tot}(R)>0$), leaving the outermost part of the disc with less mass to spread more rapidly. 

\par The linear fitting function cannot always depict the evolution of disc sizes. When the gas disc size exhibits a trajectory with time analogous to a parabola, characterised by an initial increase followed by a subsequent decrease, the fitting still yields a positive expansion rate provided that the overall trend indicates growth. This is the case for discs with a wind-dominated outer part, displayed by the three dots on the top left of each panel, where $\alpha_\mathrm{DW,out}=10^{-3}$ and $\alpha_\mathrm{SS,out}=3\times10^{-4}$ (Simulations 7, 16 and 25 in Table \ref{tb:parameters}). $R_o$ of these discs does not start contracting until $\alpha_\mathrm{DW,out}/\alpha_\mathrm{SS,out}\gtrsim10$ due to the enhanced magnetic field induced by its own evolution. Contrary to $R_o$, both $R_c$ and $R_t$ keep increasing at all times (i.e., $\mathrm{d}R_c/\mathrm{d}t>0$, $\mathrm{d}R_t/\mathrm{d}t>0$).

\par Unlike $R_c$ and $R_t$, which are more meaningful from the theoretical perspective, $R_o$ is an observable quantity, which we can be traced via molecular line emission. $\mathrm{{}^{12}{CO}}$, the most abundant gas species after $\mathrm{H_2}$ in ISM, is accessible at millimetre wavelengths from the ground and is a suitable tracer for characterising the gas disc radius. The self-shielding from photodissociation by $\mathrm{{}^{12}{CO}}$ yields a nearly constant limit on the observable surface density of $\sim10^{-4}~\mathrm{g~cm^{-2}}$ \citep{1988ApJ...334..771V, 2019MNRAS.486.4829R, 2023MNRAS.518L..69T} when assuming an abundance of $10^{-4}$ relative to $\mathrm{H_2}$ \citep[e.g.,][]{1978ApJS...37..407D, 1982ApJ...262..590F, 1994ApJ...428L..69L}. We apply this threshold to mimic very high-sensitivity observations, which reach the fundamental sensitivity limit imposed by physical processes. In comparison, a higher threshold of $10^{-2}~\mathrm{g~cm^{-2}}$ is adopted to represent lower-sensitivity observations.

\par We measure the $\mathrm{{}^{12}{CO}}$ disc sizes for all models listed in Table \ref{tb:parameters} at 5 specific evolutionary stages (0.5, 1, 2, 5 and 10 Myr) by adopting two surface density thresholds discussed above, and show the results in Figure \ref{fig:sizes_allmodel}. We classify disc sizes by their values of $R_\mathrm{c,0}$ and $\alpha_\mathrm{SS,out}$. The domination over the disc expansion by the latter is illustrated in Figure \ref{fig:ch1_kc_ko_kt}. In Figure \ref{fig:sizes_allmodel}, discs characterised by a lower surface density threshold are more radially extended than those measured by a higher threshold when compared at the same age. Their sizes increase with time for given $\alpha_\mathrm{SS,out}$ and $R_\mathrm{c,0}$. Exceptions exist for discs with $\alpha_\mathrm{SS,out}=10^{-2}$, whose sizes drop from $2$ to $5~$Myr, tracing the switch-on of efficient photoevaporation at the end of evolution. Disc sizes traced with a higher threshold ($10^{-2}~\mathrm{g~cm^{-2}}$) decrease with time instead. This trend is particularly prominent for discs with large $\alpha_\mathrm{SS,out}$ ($10^{-2}$) and can be easily understood as they tend to be more radially extended and have a larger $R_t$ (a larger and positive $\mathrm{d}R_t/\mathrm{d}t$ in Figure \ref{fig:ch1_kc_ko_kt}). If the threshold surface density is higher than the surface density corresponding to $R_t$, it will trace a shrinking disc within $R_t$. This is mitigated for discs with smaller $\alpha_\mathrm{SS,out}$, whose $R_t$ at a given time corresponds to a higher surface density. They are more tolerant to the threshold we adopt for $R_o$. Interestingly, this tolerance may explain the smaller variations in disc sizes when $\alpha_\mathrm{SS,out}$ is smaller, and can also make discs with smaller $\alpha_\mathrm{SS,out}$ look larger than their counterparts with larger $\alpha_\mathrm{SS,out}$, bringing up confusion for disc size comparison when observations are not integrated for a sufficiently long time.

\par The disc size measurements taken here assume an ISM abundance of $\mathrm{{}^{12}CO}$. However, mounting evidences from observations show that $\mathrm{CO}$ is depleted in protoplanetary discs \citep[e.g.,][]{2013ApJ...776L..38F, 2016ApJ...823...91S, 2017A&A...599A.113M, 2017ApJ...844...99L}. Lockup of $\mathrm{CO}$ into ices or large solid bodies is required to explain this depletion in addition to freeze-out and photodissociation \cite[e.g.,][]{2017A&A...599A.113M}, inducing carbon depletion compared to the ISM value. The latter in the outer disc of Class II stars can also vary substantially among individuals \citep[e.g.,][]{2016A&A...592A..83K, 2022A&A...660A.126S}. These undoubtedly complex the disc gas size problem further.

\subsection{Disc lifetime} \label{sec:para_exp_life_fixed}

Various definitions of disc lifetimes exist in literature\footnote{\citet{2022MNRAS.512.2290T} defines the lifetime as the ratio of the disc mass to the stellar accretion rate; in observations, the lifetime for a cluster is estimated by extrapolation of the disc fraction against the disc/stellar age \citep[e.g.,][]{2001ApJ...553L.153H, 2005astro.ph.11083H, 2010A&A...510A..72F,
2015A&A...576A..52R, 2018MNRAS.477.5191R, 2021ApJ...921...72M}.}. The lifetime in this work is from the start of the simulation until either the disc is fully dispersed or the simulation is terminated due to reaching the time limit ($12~$Myr), which is shorter. We take $t=0$ in our models as the beginning of the Class II phase, so time used here is not directly comparable to observed ages for objects $\lesssim 0.5~$Myr.

\par
\begin{figure}
    \centering
    \includegraphics[width=0.46\textwidth]{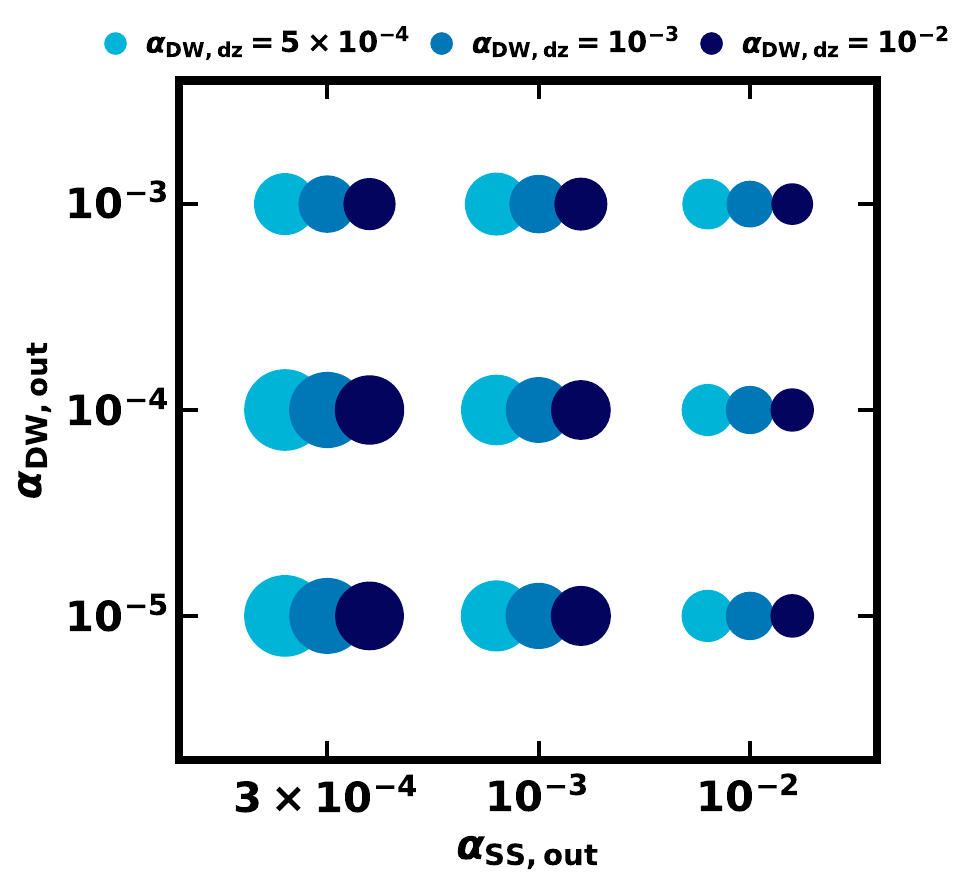}
    \caption{Disc lifetimes for discs in group 1 simulations (fixed $R_\mathrm{c,0}=60~$au and $R_\mathrm{dz,out}=30~$au). $\alpha_\mathrm{DW,dz}$ is coded in colours with darker blue indicating a larger $\alpha_\mathrm{DW,dz}$. The disc lifetime is linear to the dot area, with larger dots denoting longer lifetimes.}
    \label{fig:ch1_alpha_lft}
\end{figure}

 The lifetimes of 27 discs in the first group of simulations are shown in Figure \ref{fig:ch1_alpha_lft}, where we employ a similar illustration as Figure \ref{fig:ch1_kc_ko_kt}. We encode the lifetime in a way that is linearly proportional to the dot area and compress the dimension of $\alpha_\mathrm{DW, dz}$ into colours in the two-dimensional dot map.

\par In Figure \ref{fig:ch1_alpha_lft}, discs with larger $\alpha_\mathrm{SS,out}$ tend to have a shorter lifetime for a given combination of $\alpha_\mathrm{DW,dz}$ and $\alpha_\mathrm{DW,out}$. This is highlighted by much smaller dots in the third column than those with smaller $\alpha_\mathrm{SS, out}$ in the first two columns. This trend is underpinned when the $\alpha_\mathrm{DW, dz}$ is also large (darkest dots). This can be explained by the increasing radially average $\alpha$ when we increase the $\alpha$ in the ``dead/wind zone'' and in the outer disc. The minor responsibility of  $\alpha_\mathrm{DW, out}$ on the disc lifetime is partially due to its relatively smaller value than $\alpha_\mathrm{SS,out}$ assumed in this study.

\par However, regardless of the adopted combinations of $\alpha_\mathrm{DW}$ and $\alpha_\mathrm{SS}$, the disc lifetime is noticeably shortened after incorporating the magnetised wind (see Figure \ref{fig:cum_mass_loss}), implying the lifetimes of our \say{hybrid} models are generally akin to that of a purely wind-driven disc. This also means that equivalent angular momentum can be more efficiently transported away from discs by magnetised winds.

\par The lifetime increases for discs with larger $R_\mathrm{c,0}$ as both the stellar accretion rate and the wind extraction rate decrease due to the more radially-extended mass distribution. On the contrary, when the radially-averaged $\alpha_\mathrm{DW}$ increases with the enlarged ``dead/wind zone'', the lifetime does not decrease monotonically.  For several simulations, discs with other parameters the same except $R_\mathrm{dz,out}$ have their shortest lifetimes when the \say{dead/wind zone} size is intermediate ($75$ au, Simulations 28, 31, 33, 36, 58 and 61 in Table \ref{tb:parameters}). This is caused by the weak wind ($\alpha_\mathrm{DW,dz}<10^{-2}$) in the \say{dead/wind zone}. In this case, the locally accumulated gas can drive a minor accretion \say{outburst} -- a minor positive deviation from the original power-law accretion rate. If the surface density in the inner disc, after the outburst, abruptly becomes optically thin to the stellar radiation, an earlier turn-on of the rapid late-stage photoevaporation can reduce the disc lifetime. Discs with only intermediate-sized \say{dead/wind zones} fulfilling this condition therefore have the shortest lifetimes.

\subsection{Cumulative mass loss}\label{sec:para_exp_cum_fixed}

\begin{figure}
    \centering
    \includegraphics[width=0.44\textwidth]{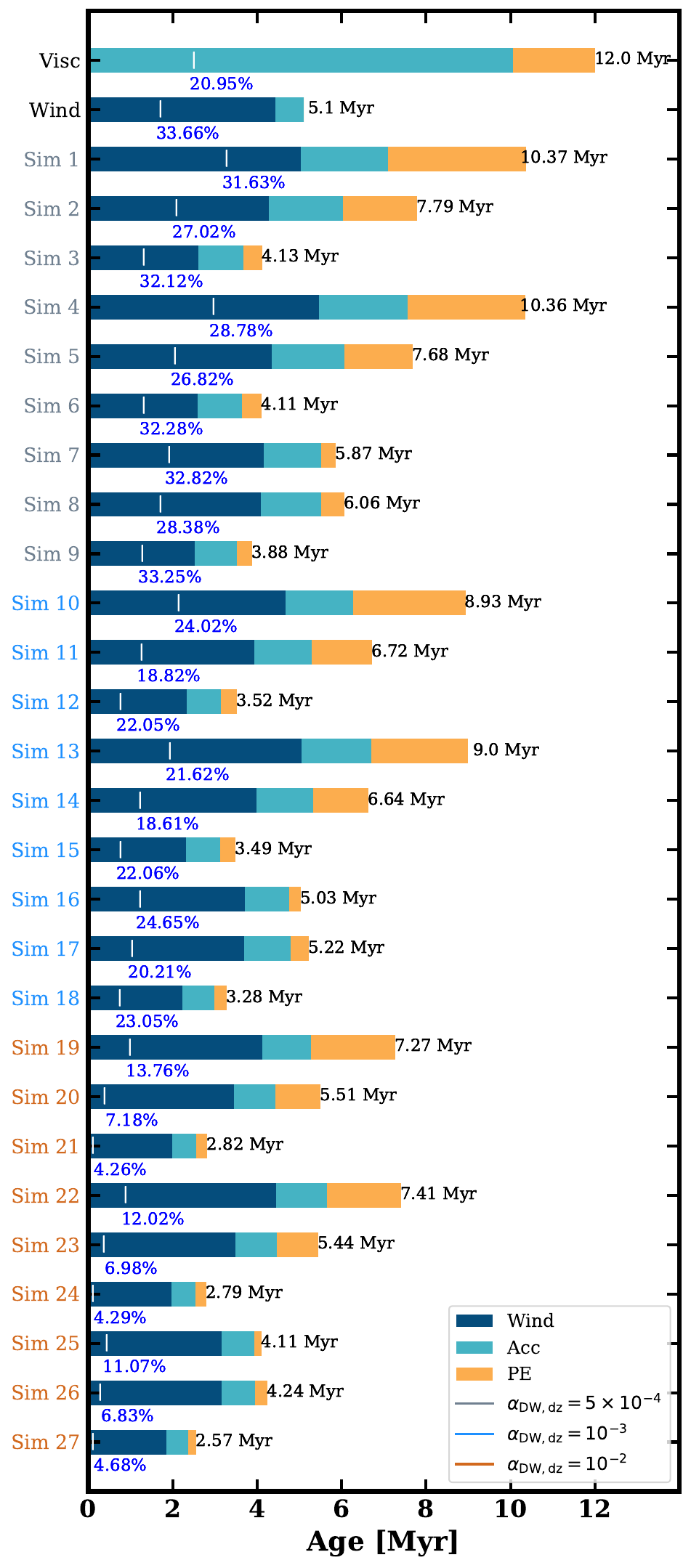}
    \caption{Disc lifetimes and cumulative mass loss fractions due to three mechanisms: wind extraction (dark blue), stellar accretion driven by viscosity and MHD winds (light blue), and internal photoevaporation (orange) for group 1 simulations (see specific values in Table \ref{tb:parameters}). Simulations are classified into three groups by their values of $\alpha_\mathrm{DW,dz}$. The first 9 simulations (in grey) below the two ``naive'' models have $\alpha_\mathrm{DW,dz}=5\times10^{-4}$, followed by 9 simulations (in blue) with $\alpha_\mathrm{DW,dz}=10^{-3}$. The 9 simulations shown in the bottom (in orange) have $\alpha_\mathrm{DW,dz}=10^{-2}$. The length of the bar serves as an indicator of the disc lifetime, whose value is also annotated at the end of each bar. The length of each segment for a specific model is proportional to the fraction of mass lost to the corresponding process. The short vertical white lines denote the time when the disc loses $60$ per cent of the total mass lost within $12$ Myr. The percentage of this duration relative to the disc lifetime is also noted in blue beneath the corresponding bar. When the disc lifetime exceeds the limit of $12$ Myr, we only consider gas that has been cleared from the disc.}
    \label{fig:cum_mass_loss}
\end{figure}

Three sinks of gas mass: stellar accretion (driven by viscosity and MHD winds), mass extraction by MHD winds, and mass loss by internal photoevaporation are considered in this work. Although the mass lost to each process is not traceable from observations, identifying them would help us to understand the dominant mass-loss mechanisms during evolution. The mass loss fraction by each component for each simulation is listed in Table \ref{tb:parameters} and visualised in Figure \ref{fig:cum_mass_loss} (for group 1 simulations) and Figure \ref{fig:table_visual} (for group 2 simulations). 

\par Most \say{hybrid} discs studied in this work lose a large proportion of gas to magnetised winds ($\gtrsim 55$ per cent) and to stellar accretion ($\sim 20$ per cent). They have  a time-scale and mass-loss budget analogous to those of the pure wind model. When the accretion and expansion in the outer disc are inefficient ($\alpha_\mathrm{DW, out}=10^{-5}$ or $10^{-4}$ and $\alpha_\mathrm{SS, out}=3\times10^{-4}$, Simulations 1, 4, 10, 13, 19 and 22 in Figure \ref{fig:cum_mass_loss} and Table \ref{tb:parameters}), the low viscosity and small wind torques do not transport the gas inwards efficiently, leaving more mass lost to photoevaporation at later stages. 

\par When we further separate the mass-loss process into two stages: the stage losing the first $60$ per cent of total mass (lost within $12~$Myr); and the stage losing the remaining 40 per cent. Except for the \say{naive} viscous model, the majority of gas in the first stage is extracted by winds and very little by photoevaporation, which has a low rate ($\simeq10^{-10}~M_\odot~\mathrm{yr^{-1}}$) in the early stage of evolution. The remaining $40$ per cent of gas is primarily removed either by wind extraction for shorter-lived discs, due to large $\alpha_\mathrm{SS}$ and $\alpha_\mathrm{DW}$ (see Section \ref{sec:para_exp_life_fixed}), or by photoevaporation for longer-lived discs. When we increase the \say{dead/wind zone} size, more mass is taken away by wind extraction due to its larger covering. This is partly the result of our chosen value of the lever arm $\lambda$. When the lever arm is adjusted to a higher value, more mass will be lost to stellar accretion instead of wind extraction (see Section \ref{sec:discussion_leverarm}).

\par We also notice from Figure \ref{fig:cum_mass_loss} that discs with small $\alpha_\mathrm{DW,dz}$ (Simulations 1-18) lose mass in a more steady approach than their counterparts with larger $\alpha_\mathrm{DW,dz}$ (Simulations 19-27). The former typically take $20-30$ per cent of their lifetimes to lose $60$ per cent of the total mass, while the latter require only $\lesssim 10$ per cent of their lifetimes to become comparably depleted. A similar pattern is also applicable when larger $R_\mathrm{c,0}$ and $R_\mathrm{dz,out}$ are adopted. This is determined by the higher extraction rate and accretion rate driven by strong winds in the intermediate disc ($\alpha_\mathrm{DW,dz}=10^{-2}$) when the initial surface density is higher.

\section{Discussion}\label{sec:discussion}
\subsection{Lever arm}\label{sec:discussion_leverarm}

\begin{figure*}
    \centering
    \includegraphics[width=\textwidth]{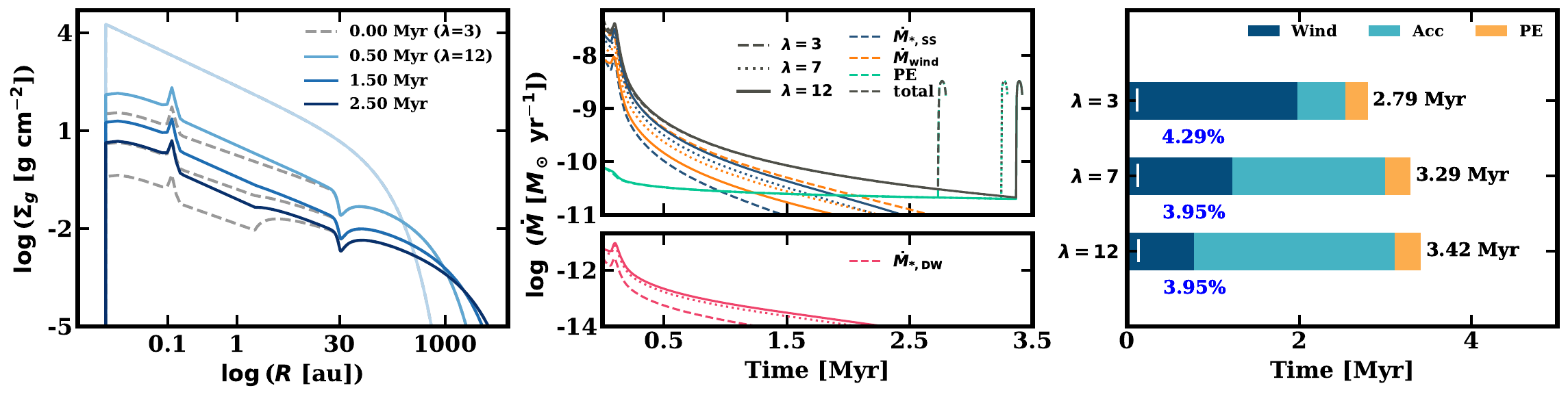}
    \caption{Left panel: Evolution of the surface density from the original fiducial model ($\lambda=3$, see Section \ref{sec:fiducial}, in dashed grey lines) in comparison to that from one modified fiducial model ($\lambda=12$, in solid blue lines) at the same time. Middle panel: Integrated mass-loss rates due to each component in the original ($\lambda=3$, dashed lines) and modified fiducial models (dotted lines for $\lambda=7$ and solid lines for $\lambda=12$). Contributions from wind-driven accretion (pink), and viscous accretion (dark blue), wind extraction (orange lines), photoevaporation (green) as well as the total of all of them (dark grey) are shown in the lower middle and upper middle panels, respectively, for visual convenience. Right panel: disc lifetimes and cumulative mass loss fractions by wind extraction (dark blue), stellar accretion (light blue) and photoevaporation (orange) for (modified) fiducial models with $\lambda=3$, $7$ and $12$, presented in a way similar to that introduced in Figure \ref{fig:cum_mass_loss}.}
    \label{fig:ch1_fid_lambda}
\end{figure*}

Recent observations and non-ideal MHD simulations consistently predict a small lever arm $\lambda$ and a small mass ejection-to-accretion ratio $f=\dot{M}_\mathrm{wind}/\dot{M}_\mathrm{acc}\sim0.1-1$ \citep{2014A&A...569A...5N, 2014ApJ...793....1Y, 2016ApJ...818..152B,  2020A&A...634L..12D, 2018ApJ...868...28F, 2020A&A...640A..82T}. In previous sections, our adopted lever arm ($\lambda=3$) gives rise to $f>1$\footnote{see Figure \ref{fig:cum_mass_loss} and Figure \ref{fig:table_visual}, where the dark blue bar is generally longer than the light blue bar, indicating that $\dot{M}_\mathrm{wind}$ is larger than $\dot{M}_\mathrm{*,SS}+\dot{M}_\mathrm{*,DW}$ averaged over time}. The analytical solution\footnote{$f=\dot{M}_\mathrm{wind}/\dot{M}_\mathrm{*,DW}=(R_c/R_\mathrm{in})^\xi-1$ from \citet{2020A&A...640A..82T, 2022MNRAS.512.2290T}. $\xi=1/[2(\lambda-1)]$ for the pure wind case, see also Eq. \ref{eq:wind_slope}.} based on a steady-state pure wind disc extending from $R_\mathrm{in}=0.01~$au to $R_c=60~$au predicts a lever arm of $\sim 7$ to achieve $f\sim1$. Therefore, we replace the lever arm in our fiducial model with $7$ and $12$. $R_\mathrm{in}$ here does not necessarily mean the disc inner edge but can be the inner radius of the wind-launching region instead. The fiducial model has a wind region originating from $0.1~$au (see Section \ref{sec:fiducial} and Figure \ref{fig:alpha_trans}). The comparison between the original and two modified fiducial models is shown in Figure \ref{fig:ch1_fid_lambda}.

\par The first panel of Figure \ref{fig:ch1_fid_lambda} illustrates when adopting a larger lever arm, less mass is taken away by winds from the intermediate region to drive a similar accretion rate (due to the fixed small $\alpha_\mathrm{DW,in}$, the middle panel), leaving the slope of the surface density closer to that of the initial profile (the left panel). Less mass loss in the \say{dead/wind zone} also means more mass will be accumulated in the inner disc, enhancing the viscous stellar accretion rate (the middle panel) and delaying the rapid clearing by internal photoevaporation. In contrast, the outer disc is governed by viscosity here, and the change in the lever arm does not affect the local mass distribution much. 

\par In the middle panel of Figure \ref{fig:ch1_fid_lambda}, the mass lost by wind-driven stellar accretion constitutes a negligible fraction of total mass loss, and this fraction is stable when varying the lever arm. This can be attributed to the imposed small $\alpha_\mathrm{DW,in}$ ($10^{-5}$), which suppresses the wind-driven accretion to the host star. But this also indicates that winds originating from radii larger than the disc inner edge drive local accretion instead of the stellar accretion, rendering the distribution of stellar accretion rates akin to that of viscous discs.

\par The radially integrated mass loss rate due to each component shown in the middle panel of Figure \ref{fig:ch1_fid_lambda} is similar to the Figure 3 presented in \citet{2023arXiv230413316K}, from which, we can infer whether a small or large lever arm is assumed by comparing the mass-loss rate by wind extraction with wind-driven stellar accretion rates. Differences between the middle panel of Figure \ref{fig:ch1_fid_lambda} and their Figure 3 arise from a more massive initial disc with a more compact mass distribution, and stronger photoevaporation over the majority of the disc lifetime adopted in \citet{2023arXiv230413316K}. 

\par We visualise the cumulative mass loss due to these three components: wind extraction, stellar accretion and photoevaporation in the right panel of Figure \ref{fig:ch1_fid_lambda}. Contrary to the fiducial model, where gas is mainly lost to wind extraction (see Section \ref{sec:para_exp_cum_fixed}), discs with larger $\lambda$ lose the majority of mass to stellar accretion due to the elevated viscous accretion rates and reduced wind extraction rates (the middle panel).

\par In summary, a change of the lever arm $\lambda$ can alter the slope of the gas surface density profile in the intermediate discs, modify the disc lifetime slightly, and change the ratio of mass lost by stellar accretion to that by wind extraction substantially. We caution readers here that the mass ejection-to-accretion ratio $f$ is sensitive to the extent of the wind-launching region, i.e. variations of either the inner or the outer wind-launching radius can alter $f$ by a factor of a few. Present observations constrain the inner launching radius of magnetised winds to $0.5$-$3~$au for Class II discs \citep[][and references therein]{2023ASPC..534..567P}. For a specific disc, the outer radius of the wind region is typically determined by $R_c$, beyond which the surface density drops sharply. Stricter constraints on the wind inner launching radius, which might vary from disc to disc, are necessary to understand the relative importance of mass loss due to wind extraction and stellar accretion.

\subsection{Surface density-adaptive ``dead/wind zone''}\label{sec:disc_DZsize}

The \say{dead/wind zone} size is fixed for all \say{hybrid} models during the entire evolution. However, a more realistic treatment should be one evolving with the surface density. A decreasing surface density due to evolution alleviates the difficulty of ionizing electrons in the disc midplane, yielding a progressively smaller MRI-quenched region. To test this, we follow \citet{2016A&A...596A..81P} and define the ``dead/wind zone'' outer edge by radii corresponding to $\Sigma_g=0.5~\mathrm{g~cm^{-2}}$. We implement this by tracing the radius \say{on the fly} in simulations. The varying $R_\mathrm{dz,out}$ changes the width of the outer boundary transition ($w_\mathrm{out}$, see Section \ref{sec:model_dz}) slightly but does not alter the overall profile. A lower limit of $10~$au is imposed to the \say{dead/wind zone} outer edge to sustain low turbulence around tens of au as estimated from observations \citep[][and references therein]{2023NewAR..9601674R}.

\par
\begin{figure}
    \centering
    \includegraphics[width=0.45\textwidth]{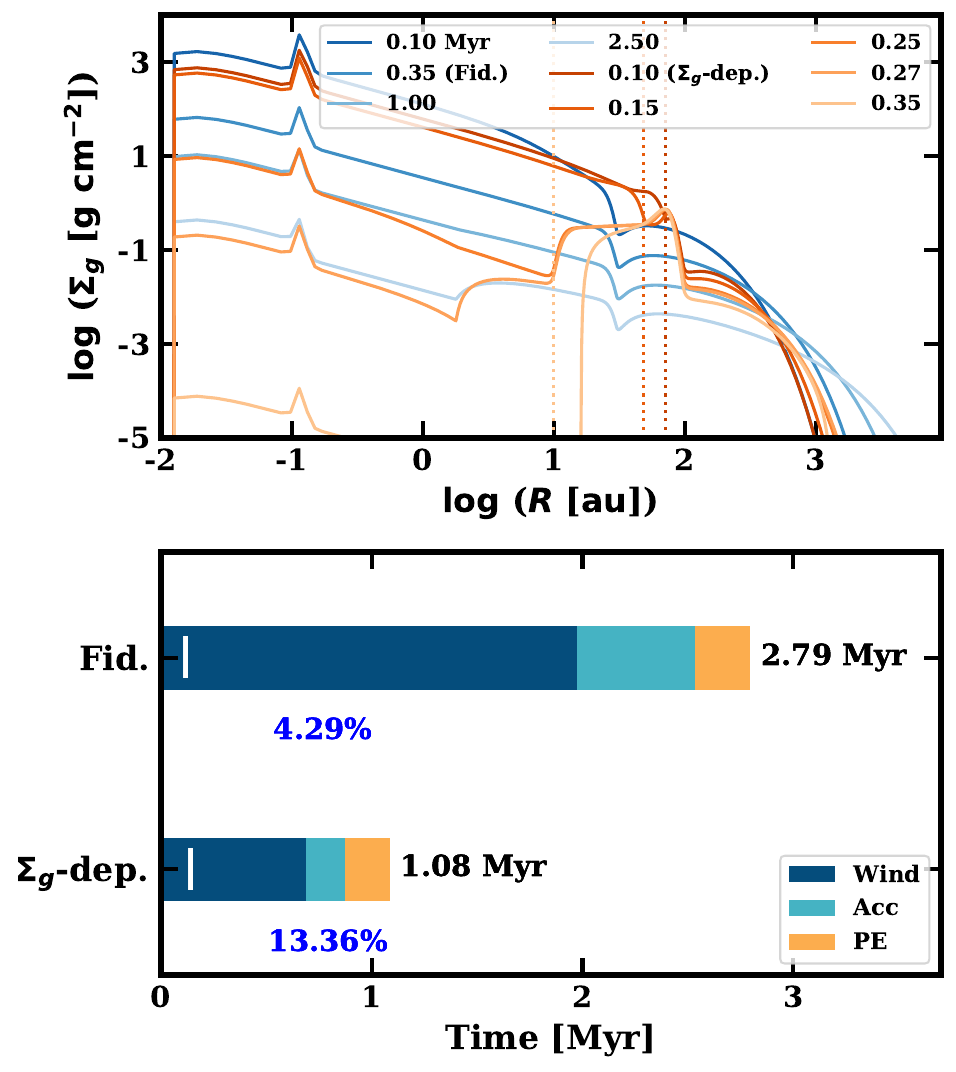}
    \caption{Upper panel: Surface densities for the fiducial model (blue lines) and the $\Sigma_g$-dependent model (orange lines), plotted at different times due to the large disparity in their disc lifetimes. The time-varying ``dead/wind zone'' outer edges are indicated with dotted vertical lines, coded in the same colour as those of their corresponding surface densities, until reaching the lower limit of $10$ au. Lower panel: the disc lifetime and cumulative mass loss are presented in the same way as described in Figure \ref{fig:cum_mass_loss}.} 
    \label{fig:adp_dz}
\end{figure}

 The upper panel of Figure \ref{fig:adp_dz} shows surface density profiles compared between the fiducial model and the $\Sigma_g$-dependent model. More complex time-varying substructures in gas are formed caused by the inwardly moving \say{dead/wind zone} outer edge. The disc lifetime is also significantly shortened for the $\Sigma_g$-dependent model due to the initially larger wind-dominated \say{dead/wind zone}. The outer edge rapidly drifts from the initial $\sim 94$ au to the manually imposed lower limit of $10~$au within $0.2$ Myr. This takes $20$ per cent of the total disc lifetime ($\sim 1$ Myr), indicating that the evolution is slowed down by the shrinking \say{dead/wind zone}. Further comparison with a disc that has a fixed \say{dead/wind zone} outer edge at $94~$au, but a much shorter lifetime, also validates this statement. 

\par Though the lifetime is more than halved after adoption of the $\Sigma_g$-dependent \say{dead/wind zone}, the cumulative mass loss fraction by winds for it ($\sim 64$ per cent) is marginally lower than for the fiducial model ($\sim 71$ per cent) as the former has a smaller \say{dead/wind zone} averaged over time. Nevertheless, this does not alter our conclusion in Section \ref{sec:para_exp_cum_fixed} that discs in our \say{hybrid} models primarily lose mass in a way akin to a pure wind model. As the inclusion of a $\Sigma_g$-dependent \say{dead/wind zone} changes the disc lifetime substantially, a better constraint on the \say{dead/wind zone} sizes can improve our understanding of the window left for planet formation in protoplanetary discs.

\subsection{Sensitivity of the disc size \texorpdfstring{$\boldsymbol{R_o}$}{Ro} to the threshold surface density} \label{sec:disc_threshold}

\par The outer radius $R_o$ is determined by the imposed surface density threshold. Incorrect selection of the threshold can lead to mis-interpreting how the disc size changes over time (see Section \ref{sec:para_exp_spread_fixed}). Hence, it is necessary to examine the sensitivity of $R_o$ to the surface density threshold.

\par We select 6 thresholds $\Sigma_\mathrm{thres}$, ranging from $10^{-12}$ to $10^{-2}~\mathrm{g~cm^{-2}}$ in steps of 2 dex, to trace $R_o$ for all simulations in this work every $0.1$ Myr except those with lifetimes shorter than $\sim 1$ Myr.

\par
\begin{figure*}
    \centering
    \includegraphics[width=\textwidth]{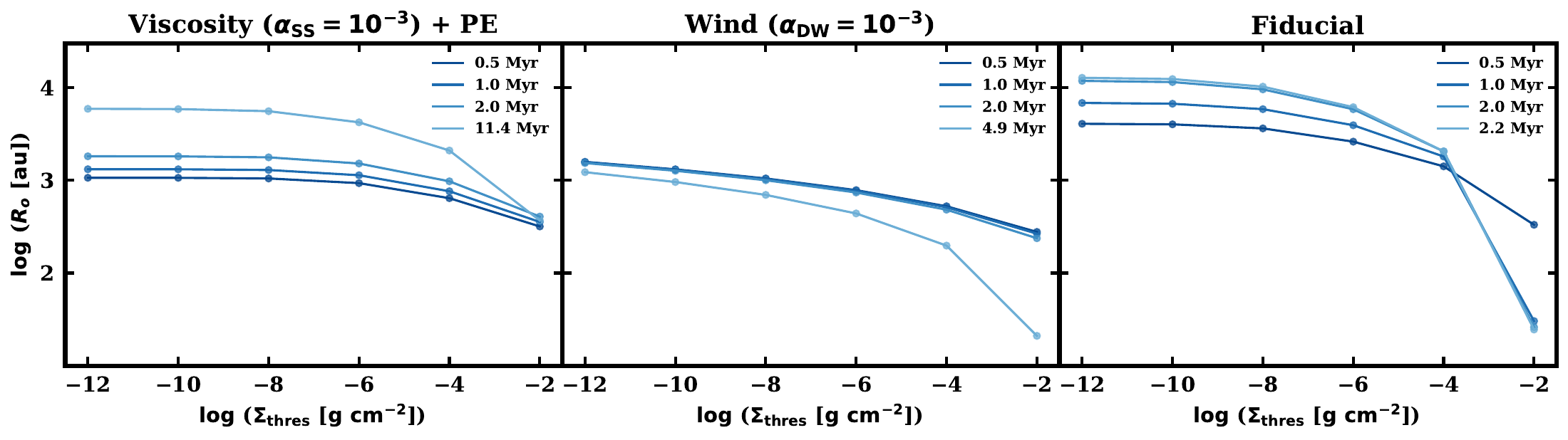}
    \caption{The disc outer radius $R_o$ measured with six thresholds ranging from $10^{-12}$ to $10^{-2}~\mathrm{g~cm^{-2}}$ for the two ``naive'' models: viscous accretion with internal photoevaporation ($\alpha_\mathrm{SS}=10^{-3}$, left panel) and wind-driven accretion ($\alpha_\mathrm{DW}=10^{-3}$, middle panel); and our fiducial model (right panel, see Section \ref{sec:fiducial}). Radii are measured at four epochs for each disc. The first three epochs are at $0.5~$Myr, $1.0~$Myr and $2.0~$Myr and the last epoch is customised for each disc to capture the disc size $\sim 0.5~$Myr before its clearance or it reaches the time limit ($12~$Myr).}
    \label{fig:r_o_threshold}
\end{figure*}

 Figure \ref{fig:r_o_threshold} shows $R_o$ traced by different thresholds for ``hybrid'' discs (represented by the fiducial model) and two ``naive'' models. A threshold of $10^{-2}~\mathrm{g~cm^{-2}}$ can effectively trace the disc expansion or contraction for \say{naive} models, but will misleadingly trace a shrinking disc for the fiducial model when the outer disc is in fact spreading. A slightly smaller threshold of $10^{-3}~\mathrm{g~cm^{-2}}$ still fails to trace the motion of more than half of discs that are wrongly traced by $\Sigma_\mathrm{thres}=10^{-2}~\mathrm{g~cm^{-2}}$ in Table \ref{tb:parameters}. The outer disc behaviour is only captured accurately when a threshold of $\lesssim 10^{-4}~\mathrm{g~cm^{-2}}$ is adopted. This value is quite close to the maximum sensitivity limited by photodissociation of $\mathrm{{}^{12}CO}$. 

\par A lower threshold could be achieved by observing neutral atomic carbon, found in a thin layer sandwiched between the carbon ionization front and the $\mathrm{{}^{12}CO}$ region \citep{1985ApJ...291..722T}. Recent observations suggest it originates from a more elevated layer than $\mathrm{{}^{12}CO}$ and its isotopologues \citep{2023arXiv231116233L}. However, the low signal-to-noise ratio in the outer disc in real observations \citep{2023arXiv231116233L} may limit its capability to accurately trace the disc at even larger radii than $\mathrm{{}^{12}CO}$ can. 

\par The almost constant $R_o$ with decreasing thresholds when $\Sigma_\mathrm{thres}<10^{-8}~\mathrm{g~cm^{-2}}$ shown in the left and right panels of Figure \ref{fig:r_o_threshold} arises from the simplified photoevaporation prescription adopted in our model, which efficiently removes gas with $\Sigma_g<10^{-8}~\mathrm{g~cm^{-2}}$, resulting in a very sharp outer edge at these low surface densities.

\par We further investigate the robustness of the threshold of $10^{-4}~\mathrm{g~cm^{-2}}$ to discs with a few combinations of $\alpha$ discussed before, but with smaller $R_{c,0}$ and $R_\mathrm{dz,out}$ ($R_{c,0}=10/30~$au with $R_\mathrm{dz,out}$ being 0.5/1.5 $R_{c,0}$), and to the initially more compact disc ($d\log\Sigma_g/d\log R=-3/2$). All of these results validate the threshold of $10^{-4}~\mathrm{g~cm^{-2}}$ for accurately tracing the pattern of $R_o$. We therefore caution that observations with lower sensitivity may not accurately capture the evolution of the outer edges of real discs \citep[see also][]{2022ApJ...926...61T}.

\section{Population synthesis}\label{sec:disc_pop}

In the previous sections, we discussed discs of a single initial mass ($0.01~M_\odot$), with two different initial characteristic radii $R_\mathrm{c,0}$ ($60~$au and $120~$au) and three different \say{dead/wind zone} outer edges $R_\mathrm{dz,out}$ ($30~$au, $75~$au and $135~$au). However, star-disc systems which form and evolve in distinct environments tend to have different initial conditions. The local radiation and magnetic fields also possibly influence the values of $\alpha_\mathrm{SS}$ and $\alpha_\mathrm{DW}$, and the \say{dead/wind zone} sizes. We refer to the variations in initial properties among individual discs as \say{personalities} of discs. Although we do not have much knowledge of $\alpha_\mathrm{DW}$, measurements of $\alpha_\mathrm{SS}$ inferred from observations suggest a relatively large range of values \citep{2023NewAR..9601674R}. Furthermore, we lack observational constraints on the \say{dead/wind zone} outer edges. All the undetermined factors above affect the disc properties discussed in Section \ref{sec:explor}, and hence disc demographics. 

\par In the following section, we implement two small-scale population syntheses based on our \say{hybrid} disc models to address whether groups of discs possessing different \say{personalities} still exhibit the observable disc expansion or contraction predicted by the \say{naive} models \citep[e.g.,][]{1974MNRAS.168..603L, 2022MNRAS.512.2290T}.

\subsection{Methods}
We assume discs in the first population have various disc masses, characteristic radii and \say{dead/wind zone} fractions, but the same transition profile, i.e. identical combinations of $\alpha_\mathrm{DW}$ and $\alpha_\mathrm{SS}$. We assume a combination of moderate viscous and wind torques from above discussion and adopt $\alpha_\mathrm{DW,dz}=10^{-3}$, $\alpha_\mathrm{DW,out}=10^{-4}$ and $\alpha_\mathrm{SS,out}=10^{-3}$.

\par We draw 1000 radii from exponentially distributed characteristic radii ranging from 20 to 200 au with steps of 5 au, which cover the majority of gas disc sizes measured by $\mathrm{{}^{12}CO}$ (see Appendix \ref{sec:appendix_size} for collection of $98$ $\mathrm{{}^{12}CO}$ disc sizes.) The exponential distribution is described by $p(r_i)=\exp{(-3\log(r_i))}/\sum_{i}\exp{(-3\log(r_i))}$, where $p(r_i)$ is the probability of the characteristic radius $r_i$. The exponential distribution is also an assumption based on Figure \ref{fig:disc_size}. Although high-resolution studies of discs from ALMA Large Programs suggest both Class 0/I and Class II discs traced by ${}^{12}\mathrm{CO}$ can spread to hundreds of au \citep[e.g.,][]{2021ApJS..257....3L, 2023ApJ...951...10V, 2023ApJ...951...11Y}, these samples were selected in various ways, and are generally biased towards more extended discs. Roughly 50 per cent of ${}^{12}\mathrm{CO}$ discs from an incomplete collection in Figure \ref{fig:disc_size} having sizes larger than $\sim150~$au can partially justify the bias. Considering small and faint discs are less likely to be detected in gas, discs with smaller sizes are likely to take an even larger fraction.

\par We assume a uniform distribution of the ratio of the \say{dead/wind zone} size to the disc characteristic radius, from $10$ to $120$ per cent with steps of $10$ per cent, for the poorly constrained ``dead/wind zone'' sizes. For example, discs with $R_\mathrm{c,0}=60~$au have a ``dead/wind zone'' from $0.1$ to $12$ au if it takes 20 per cent of the characteristic radius.

\par We simply assume a binary uniform distribution for the disc mass ($0.01~M_\odot$ and $0.05~M_\odot$), as discs with other parameters the same but only the disc mass different exhibit a scaling relation regarding disc sizes. 

\par In the second population, we extend the dimensions of disc \say{personalities} by additionally varying $\alpha$. We uniformly draw 1000 samples for $\alpha_\mathrm{DW,dz}$, $\alpha_\mathrm{DW,out}$ and $\alpha_\mathrm{SS,out}$ from values adopted in Section \ref{sec:explor} (see also Table \ref{tb:ch1_para}), respectively, and combine them as 1000 sets of $\alpha$ for discs. We then integrate samples of $\alpha$-combinations into the initial properties of the first population to constitute the second one. The distributions of parameters sampled for the first (only the upper panels) and the second populations (all the panels) can be found in Figure \ref{fig:pop_syn_distri}. 

\par We characterise disc sizes of two populations by $R_o$ and adopt thresholds of $\Sigma_g=10^{-2}~\mathrm{g~cm^{-2}}$ and $10^{-4}~\mathrm{g~cm^{-2}}$ to mimic observations taken with low and high sensitivity as in Section \ref{sec:para_exp_spread_fixed}. We randomly sample 100 disc sizes from each population at ages between $0.1$ and $10$ Myr, and plot them against the disc age. We include samples having a disc size of $0~$au, and samples that are coincidentally selected multiple times from the same model. The former represent discs that have dispersed by the time of observations (the disc lifetime is shorter than the specified time) and the latter represent discs with the same \say{personalities}. 

\subsection{Results}
\par
\begin{figure*}
\centering
    \begin{subfigure}{\linewidth}
    \centering
    \includegraphics[width=0.75\linewidth]{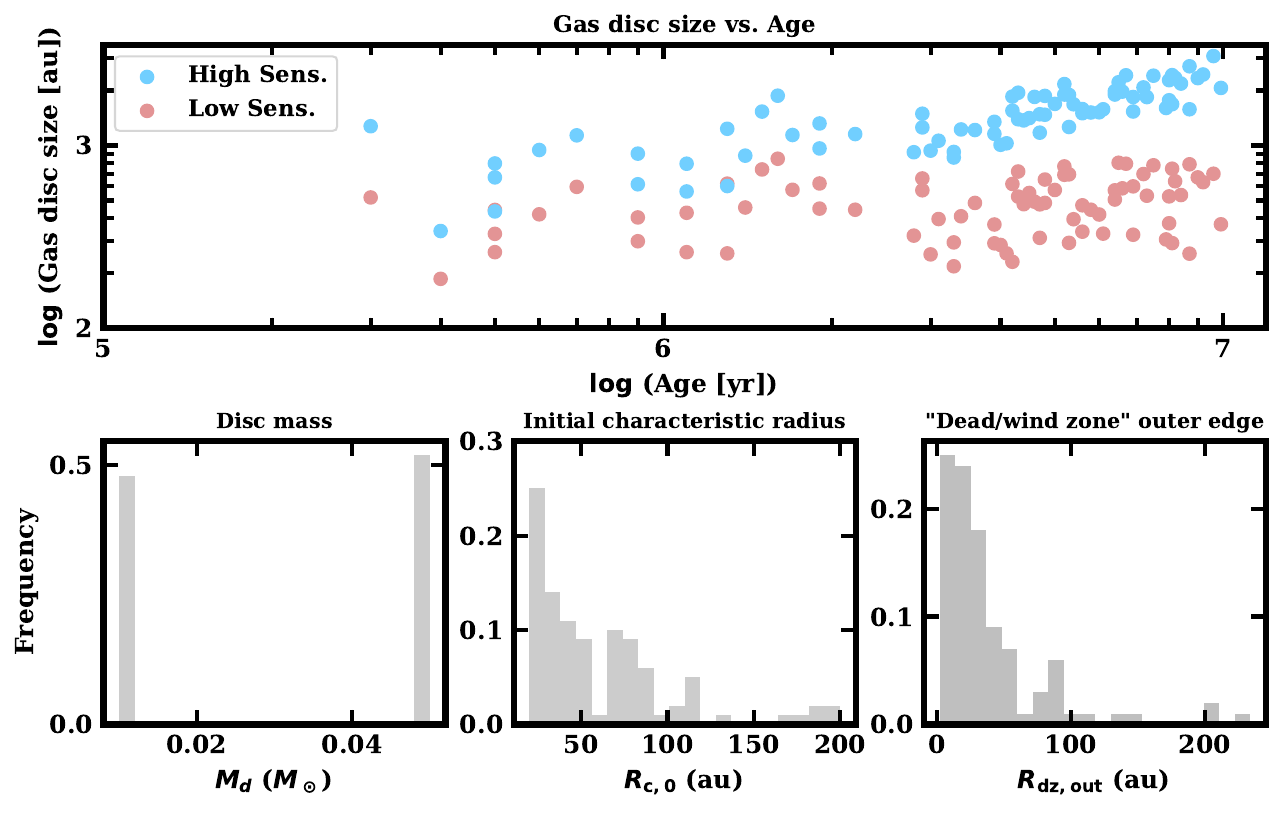}
  \caption{The disc size vs. age diagram for 100 samples randomly drawn from the first population (fixed $\alpha$). Distributions of initial disc masses (lower left), initial characteristic radii (lower middle) and ``dead/wind zone'' outer edges (lower left) for 100 samples are shown in the bottom row.}\label{fig:size_vs_age_fa}
    \end{subfigure}

\bigskip
    \begin{subfigure}{\linewidth}
  \centering
  \includegraphics[width=\linewidth]{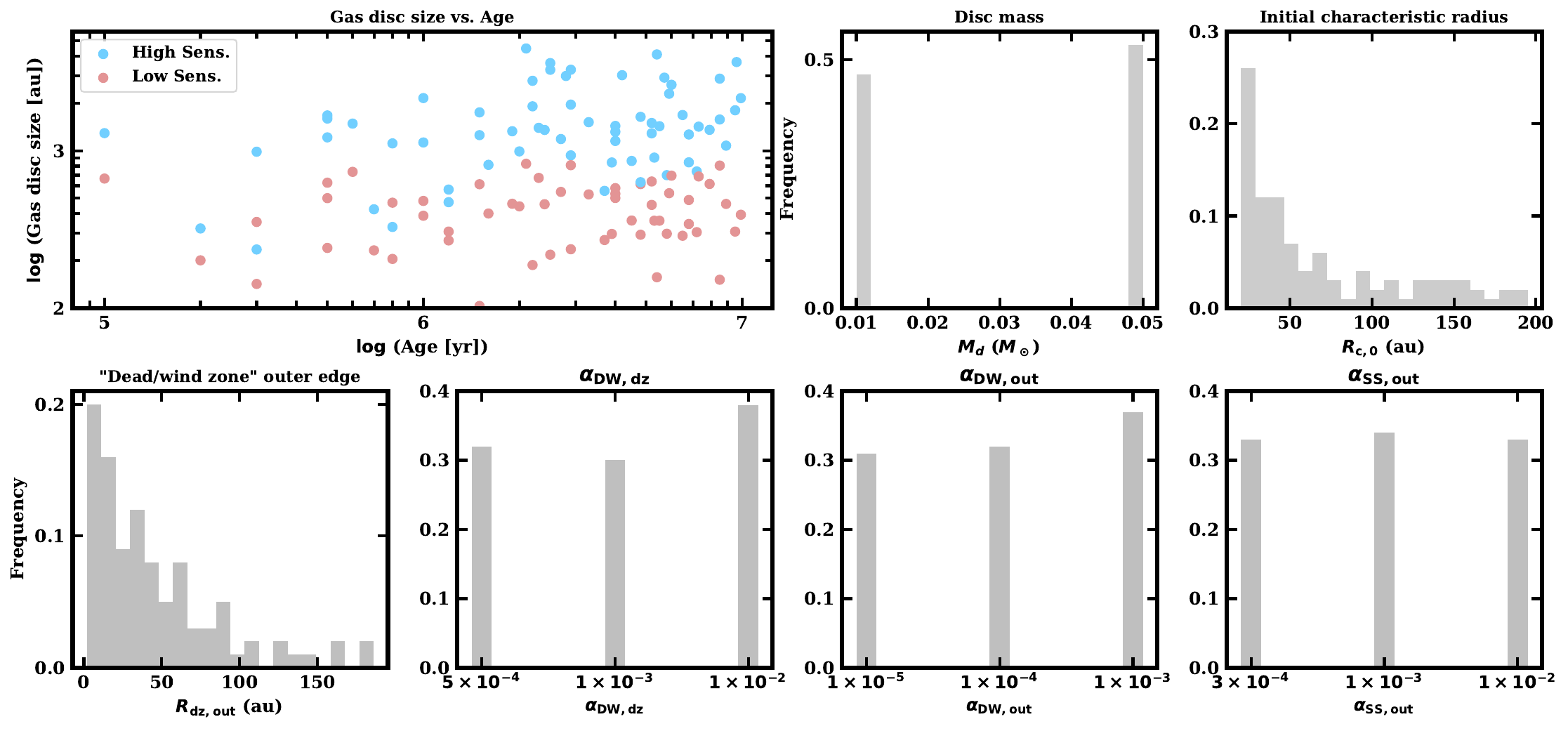}
  \caption{The disc size vs. age diagram for 100 samples randomly drawn from the second population (varying $\alpha$). Distributions of initial disc masses (upper middle), initial characteristic radii (upper right),``dead/wind zone'' outer edges (lower left) and $\alpha$ (three panels on the right side of the bottom row) for 100 samples are shown around.}\label{fig:size_vs_age_full}
    \end{subfigure} 
\caption{Disc size vs. age diagrams for the first population (Figure \ref{fig:size_vs_age_fa}), where we assume $\alpha_\mathrm{DW,dz}=10^{-3}$, $\alpha_\mathrm{DW,out}=10^{-4}$ and $\alpha_\mathrm{SS,out}=10^{-3}$ for all the discs, and the second population (Figure \ref{fig:size_vs_age_full}), where we also vary $\alpha_\mathrm{DW}$ and $\alpha_\mathrm{SS}$ among individual discs. 100 samples are drawn from each population and their sizes are measured by $R_o$ with thresholds of $10^{-2}~\mathrm{g~cm^{-2}}$(pink dots) and $10^{-4}~\mathrm{g~cm^{-2}}$ (blue dots), to mimic observations with lower and higher sensitivity. Distributions of initial disc masses, initial characteristic radii, "dead/wind zone`` outer edges (and $\alpha$) for samples shown in the main diagram are also plotted.}
\label{fig:size_vs_age}
\end{figure*}

 Figure \ref{fig:size_vs_age} shows gas sizes vs. disc ages for a single draw from two populations, accompanied by the distributions of disc properties for each draw. We see for both populations that discs measured by higher-sensitivity observations generally have larger sizes. This is consistent with our conclusion drawn in Section \ref{sec:para_exp_spread_fixed}. In Figure \ref{fig:size_vs_age_fa}, where discs have the same combinations of $\alpha_\mathrm{SS}$ and $\alpha_\mathrm{DW}$, gas sizes measured by high-sensitivity observations (blue dots) increase slightly over time, aligning with expectation from a viscosity-dominated outer disc, which is the case assumed in our models. The increasing gas sizes can also be partly attributed to the tendency for larger discs to survive for a longer time (Section \ref{sec:para_exp_life_fixed}). This increasing trend nearly vanishes when discs are observed with lower sensitivity (pink dots) due to the incapability of a higher threshold to accurately trace the outer disc motion (Section \ref{sec:disc_threshold} and also the right panel of Figure \ref{fig:r_o_threshold}). When we also consider varying combinations of $\alpha$ (Figure \ref{fig:size_vs_age_full}), discs with similar ages have more diverse sizes, represented by more scattered dots in the upper left panel of Figure \ref{fig:size_vs_age_full} than in Figure \ref{fig:size_vs_age_fa}. The large scatter in Class II disc sizes has been also been observed in \citet{2018ApJ...864..168N} and \citet{2022ApJ...931....6L}. This scattering due to disc “personalities” makes the increasing radii over time shown in higher-sensitivity observations in Figure \ref{fig:size_vs_age_full} even weaker. Therefore, capturing how disc sizes change with time can be challenging even when we ignore the uncertainties existed in age estimation and radius measurement, as it requires high-sensitivity observations, which approach the limitation imposed by photodissociation of ${}^{12}\mathrm{CO}$ (Section \ref{sec:para_exp_spread_fixed}), for both populations studied here.

\par We repeatedly draw 100 gas disc sizes from the synthesised population for 100 times. We see weak variations in the overall pattern in the disc size--age diagram depending on the randomly-selected samples. This may make it difficult to conclude which mechanisms drive the motion of outer discs given the selected samples.

\par It is worth noting that the two populations discussed here are based on different assumptions regarding $\alpha$. The first assumes universal combinations of $\alpha$, including $\alpha_\mathrm{SS,out}$, which dominates the disc expansion (see Section \ref{sec:para_exp_spread_fixed}), while the second assumes varying $\alpha$ among individuals. It is likely that a more realistic case is in between, but measurements of $\alpha_\mathrm{SS}$ in the outer disc by ALMA observations vary by orders of magnitude \citep{2015ApJ...813...99F, 2017ApJ...843..150F, 2018ApJ...856..117F, 2020ApJ...895..109F, 2018ApJ...864..133T}, and are limited in constraining the distribution of $\alpha_\mathrm{SS}$ \citep{2022MNRAS.515.2548A}.

\par A more detailed and larger-scale population synthesis (such as \citet{2023A&A...673A..78E}), which is out of the scope of our toy population study, has the potential to constrain the preferred disc properties, such as $\alpha_\mathrm{SS}$, $\alpha_\mathrm{DW}$, the lever arm $\lambda$ and even the dead zone size, based on present theories by comparing with observations statistically. However, our limited knowledge on disc fundamental properties, such as distributions of disc masses and sizes, which are inputs to population modelling, and biased observations as references, may limit the usefulness of such comparison.

\section{Implications and limitations}\label{sec:impli_and_limit}
\subsection{Observational implications}\label{sec:disc_obs}

Previous research seeking mechanisms responsible for the angular momentum transport associates the mechanisms with either gas disc sizes \citep[e.g.,][]{2018ApJ...864..168N, 2022arXiv220309930M, 2022ApJ...926...61T, 2022MNRAS.514.1088Z} or the stellar accretion rates \citep{2023MNRAS.524.3948A}. When gas discs spread over time, transport of angular momentum is attributed to viscosity; otherwise, magnetised winds are considered instead. The distribution of stellar accretion rates also serves as a proxy for the two mechanisms. However, these two observational diagnostics may only trace local disc physics (in the outer and inner disc, respectively) when a more realistic disc model accounting for the \say{dead/wind zone} is employed.

\par The incorporation of MHD winds in the disc alters the disc lifetime and the dominant process of mass removal (Section \ref{sec:para_exp_life_fixed} and \ref{sec:para_exp_cum_fixed}) from the traditional viscous disc in this study, indicating that winds can remove angular momentum with higher efficiency\footnote{This is likely a consequence of the well known fact that the lifetime of a wind-driven disc is significantly shorter than that of a viscous disc for the same $\alpha$.}. But in some \say{hybrid} models, the inner and outer discs still behave like a viscous disc, i.e., accreting for the former (Section \ref{sec:para_exp_acc}), and expanding (Section \ref{sec:para_exp_spread_fixed}) for the latter. That is to say, for an individual disc evolving similarly to some \say{hybrid} models, even if we can observe its gas size growing, or its accretion rate behaving like that of a viscous disc over an unrealistically long time (a few million years), we can only conclude that viscosity dominates the expansion in the outer disc or the stellar accretion in the inner disc. The problem becomes more complicated when we are limited to observing demographic ``snapshots'' of evolving populations. In such a case, we cannot ignore the pitfall presented by disc \say{personalities}, which make statistically identifying how gas disc sizes vary with time challenging (Section \ref{sec:disc_pop}).

\par Previous studies investigating the dominant mechanisms over disc evolution simply assume a homogeneous $\alpha_\mathrm{DW}$ or $\alpha_\mathrm{SS}$ for the entire disc and remain ambiguous in the use of \say{disc evolution}. When a \say{hybrid} disc with \say{dead/wind zones} is considered, angular momentum can be transported by different mechanisms in different regions of one disc. \say{Disc evolution} can point to the evolution of stellar accretion rates, disc sizes and also mass loss fractions by different processes, which can be distinct from angular momentum transport. Nevertheless, characterising disc sizes and stellar accretion rates, and studying them in demographics still remain crucial. Although they have limited capability in signifying the major contributor to the \textit{global} angular momentum transport, i.e., how the angular momentum is transported at any radii of a specific disc, they do inform the dominant mechanisms of \textit{local} angular momentum transport, i.e., how the angular momentum is transported in the very inner disc, in the intermediate disc, and in the outer disc.

\subsection{Limitations}\label{sec:disc_lim}
The models presented in this work are relatively simple and are not able to precisely reproduce complete \say{personalities} of protoplanetary discs. One of the major uncertainties is from our lack of constraints on the strengths and configurations of magnetic fields, and their evolution. The lever arm $\lambda$ is assumed to be a time-independent parameter, and the evolution of magnetic fields ($\alpha_\mathrm{DW}\propto \Sigma_g^{-\omega}$) is treated in an oversimplified way in our study. External radiation and disc-disc interaction in dense environments are efficient in modifying the disc sizes \citep[e.g.,][]{2016ApJ...828...48V, 2018MNRAS.478.2700W, 2022MNRAS.514.2315C}, but are also not considered here. The outer disc expansion due to magnetic fields beyond the radius truncated by external photoevaporation \citep{2021ApJ...922..201Y} is not included in our wind analytical solution. Additionally, we only consider \say{hybrid} discs as discs simultaneously driven by viscosity and winds, and the dominant mechanism only varies with locations. A more realistic case might be that the dominant mechanism also varies with time \citep{2022ApJ...931....6L}, i.e. the majority of angular momentum is probably transported by different mechanisms at different times.

\par We have also not explored the interaction between gas and dust in the disc evolution. While small dust is well-coupled to the gas, larger dust, which suffers radial drift \citep[e.g.,][]{2010A&A...513A..79B}, behaves differently from the gas. The dust/gas dynamics can be further complicated by the coagulation and fragmentation of particles, which can change the size distribution of dust \citep[e.g.,][]{2010A&A...513A..79B}, and by the dust back-reaction on the gas when the dust-to-gas ratio is non-negligible \citep[e.g.,][]{2018MNRAS.479.4187D}. All of these result in significant differences between the radial distributions of dust and gas. While ALMA now allows gas observations with higher resolution and sensitivity, the vast majority of observations still only trace the dust. Inclusion of dust components in future studies, with the aid of radiative transfer techniques, would potentially allow us to study how dust evolves in ``hybrid'' discs with \say{dead/wind zone} models. Meanwhile, a full scale population synthesis factoring in disc \say{personalities} can provide insight into correlations we inferred from observations \citep[e.g.,][]{2022A&A...661A..66Z, 2024arXiv240514501D}{}{}.

\section{Conclusion}\label{sec:conclusion}
In this paper we have run a suite of 1-D gas simulations (a total of 92 individual models) to study the evolution of ``hybrid'' protoplanetary discs regulated by radially varying $\alpha$-parametrized viscosity ($\alpha_\mathrm{SS}(r)$) and magneto-hydrodynamic winds ($\alpha_\mathrm{DW}(r)$), as well as internal photoevaporation. Our models are broadly consistent with current understanding of protoplanetary discs in terms of several properties, such as stellar accretion rates, gas disc sizes and lifetimes. We vary $\alpha_\mathrm{SS}$, $\alpha_\mathrm{DW}$, the disc initial characteristic radius $R_\mathrm{c,0}$ and the \say{dead/wind zone} outer edge $R_\mathrm{dz,out}$ in the ``hybrid'' models, and compare the evolution of their properties with those of ``naive'' models (purely viscous and wind-only discs). This understanding of ``hybrid'' discs is further applied to the population level to examine the effectiveness of gas disc sizes in differentiating the dominant mechanisms transporting angular momentum. We summarise the main results as follows:

\begin{itemize}
    \item The radially varying $\alpha$ invariably creates gas substructures around the inner ($R_\mathrm{dz,in}$) and outer ($R_\mathrm{dz,out}$) edges of the \say{dead/wind zone}. The disc surface density profiles from models in this study can be classified into three categories by their morphologies. However, we caution that the stability of these substructures requires investigation with 2-D and 3-D hydrodynamic simulations.

    \item Comparison with \say{naive} models shows that \say{hybrid} discs behave mainly like viscous discs in terms of stellar accretion rates and disc expansion, but behave like wind-driven discs in terms of cumulative mass loss and lifetimes.
    
    \item We measure disc sizes in three ways: the characteristic radius $R_c$, beyond which the surface density drops sharply; the transition radius $R_t$, delimiting the accreting ($\dot{M}(R)>0$) inner disc from the spreading ($\dot{M}(R)<0$) outer disc; and the outer radius $R_o$, defined by a threshold surface density. The first two consistently increase for all the ``hybrid'' models explored here, while the third contracts when magnetised winds dominate the outer disc (when the parameterization of the magnetic field evolution leads to $\alpha_\mathrm{DW,out}$/$\alpha_\mathrm{SS,out}>10$ at late times.)
    
    \item Winds originating from a radius larger than the disc inner edge may only be able to drive local accretion where winds dominate. The fact that viscosity still drives the observed stellar accretion rate for “hybrid” discs places obstacles in differentiating two mechanisms by the distribution of stellar accretion rates.
    
    \item We conducted two small-scale population syntheses, with the first fixing $\alpha$ but varying initial disc masses, initial characteristic radii and \say{dead/wind zone} outer edges, and the second additionally varying $\alpha$. The gas disc expansion over time vanishes unless discs are observed at very high-sensitivity ($\Sigma_\mathrm{thres}=10^{-4}~\mathrm{g~cm^{-2}}$), which approaches the limitation set by photodissociation of ${}^{12}\mathrm{CO}$. This reveals that identifying the dominant mechanism of angular momentum transport in the outer disc from measuring disc sizes in ``snapshot'' demographics can be more challenging than previously thought.

    \item Our \say{hybrid} models show that the inclusion of magnetised winds substantially changes the disc evolution time-scale, and the cumulative mass loss fractions by different physical processes. This implies that winds may transport angular momentum more efficiently than viscosity does. However, the physical processes dominating angular momentum transport can differ from those governing stellar accretion and disc expansion. As a result, stellar accretion rates and gas disc sizes may be less valid proxies of \textit{global} angular momentum transport but good indicators for the \textit{local} angular momentum transport. Other observable diagnostics should be considered jointly in order to determine the dominant mechanism in transporting the majority of angular momentum in the disc.
\end{itemize}

\section*{Acknowledgements}
ST acknowledges the University of Leicester for a Future 100 Studentship. RA acknowledges funding from the Science \& Technology Facilities Council (STFC) through Consolidated Grant ST/W000857/1. This project has received funding from the Fondazione Cariplo, grant no. 2022-1217, and the European Research Council (ERC) under the European Union’s Horizon Europe Research \& Innovation Programme under grant agreement no. 101039651 (DiscEvol). Views and opinions expressed are however those of the author(s) only, and do not necessarily reflect those of the European Union or the European Research Council Executive Agency. Neither the European Union nor the granting authority can be held responsible for them. This work is strongly benefitted from the Core2disk-III residential program of Institut Pascal at Universit\'e Paris-Saclay, with the support of the program ``Investissements d’avenir'' ANR-11-IDEX-0003-01. This research used the ALICE High Performance Computing Facility at the University of Leicester.

\section*{Data Availability}
The observational data used in this paper are from the compilation
of \citet{2022arXiv220309930M}, and are publicly available at
\url{http://ppvii.org/chapter/15/}. Data generated in simulations and codes reproducing figures in this work are available on reasonable request to the corresponding author. This work made use of \textsc{Jupyter} \citep{jupyter}, \textsc{Matplotlib} \citep{matplotlib}, \textsc{Numpy} \citep{numpy, numpy2}, \textsc{Scipy} \citep{scipy2020}, \textsc{Astropy} \citep{astropy} and \textsc{Pandas} \citep{pandas2010, pandas2020}.



\bibliographystyle{mnras}
\bibliography{reference} 

\begin{thebibliography}{}
\makeatletter
\relax
\def\mn@urlcharsother{\let\do\@makeother \do\$\do\&\do\#\do\^\do\_\do\%\do\~}
\def\mn@doi{\begingroup\mn@urlcharsother \@ifnextchar [ {\mn@doi@} {\mn@doi@[]}}
\def\mn@doi@[#1]#2{\def\@tempa{#1}\ifx\@tempa\@empty \href {http://dx.doi.org/#2} {doi:#2}\else \href {http://dx.doi.org/#2} {#1}\fi \endgroup}
\def\mn@eprint#1#2{\mn@eprint@#1:#2::\@nil}
\def\mn@eprint@arXiv#1{\href {http://arxiv.org/abs/#1} {{\tt arXiv:#1}}}
\def\mn@eprint@dblp#1{\href {http://dblp.uni-trier.de/rec/bibtex/#1.xml} {dblp:#1}}
\def\mn@eprint@#1:#2:#3:#4\@nil{\def\@tempa {#1}\def\@tempb {#2}\def\@tempc {#3}\ifx \@tempc \@empty \let \@tempc \@tempb \let \@tempb \@tempa \fi \ifx \@tempb \@empty \def\@tempb {arXiv}\fi \@ifundefined {mn@eprint@\@tempb}{\@tempb:\@tempc}{\expandafter \expandafter \csname mn@eprint@\@tempb\endcsname \expandafter{\@tempc}}}

\bibitem[\protect\citeauthoryear{{ALMA Partnership} et~al.,}{{ALMA Partnership} et~al.}{2015}]{2015ApJ...808L...3A}
{ALMA Partnership} et~al., 2015, \mn@doi [\apjl] {10.1088/2041-8205/808/1/L3}, 808, L3

\bibitem[\protect\citeauthoryear{Alcal{\'{a}} et~al.,}{Alcal{\'{a}} et~al.}{2014}]{2014A&A...561A...2A}
Alcal{\'{a}} J.,  et~al., 2014, \mn@doi [\aap] {10.1051/0004-6361/201322254}, 561, A2

\bibitem[\protect\citeauthoryear{Alcal{\'{a}} et~al.,}{Alcal{\'{a}} et~al.}{2017}]{2017A&A...600A..20A}
Alcal{\'{a}} J.,  et~al., 2017, \mn@doi [\aap] {10.1051/0004-6361/201629929}, 600, A20

\bibitem[\protect\citeauthoryear{Alessi \& Pudritz}{Alessi \& Pudritz}{2022}]{2022MNRAS.515.2548A}
Alessi M.,  Pudritz R.~E.,  2022, \mn@doi [\mnras] {10.1093/mnras/stac1782}, 515, 2548

\bibitem[\protect\citeauthoryear{{Alexander}}{{Alexander}}{2012}]{2012ApJ...757L..29A}
{Alexander} R.,  2012, \mn@doi [\apjl] {10.1088/2041-8205/757/2/L29}, \href {https://ui.adsabs.harvard.edu/abs/2012ApJ...757L..29A} {757, L29}

\bibitem[\protect\citeauthoryear{{Alexander}, {Rosotti}, {Armitage}, {Herczeg}, {Manara}  \& {Tabone}}{{Alexander} et~al.}{2023}]{2023MNRAS.524.3948A}
{Alexander} R.,  {Rosotti} G.,  {Armitage} P.~J.,  {Herczeg} G.~J.,  {Manara} C.~F.,   {Tabone} B.,  2023, \mn@doi [\mnras] {10.1093/mnras/stad1983}, \href {https://ui.adsabs.harvard.edu/abs/2023MNRAS.524.3948A} {524, 3948}

\bibitem[\protect\citeauthoryear{{Andrews} et~al.,}{{Andrews} et~al.}{2018}]{2018ApJ...869L..41A}
{Andrews} S.~M.,  et~al., 2018, \mn@doi [\apjl] {10.3847/2041-8213/aaf741}, \href {https://ui.adsabs.harvard.edu/abs/2018ApJ...869L..41A} {869, L41}

\bibitem[\protect\citeauthoryear{{Ansdell} et~al.,}{{Ansdell} et~al.}{2018}]{2018ApJ...859...21A}
{Ansdell} M.,  et~al., 2018, \mn@doi [\apj] {10.3847/1538-4357/aab890}, \href {https://ui.adsabs.harvard.edu/abs/2018ApJ...859...21A} {859, 21}

\bibitem[\protect\citeauthoryear{Antilen, Casassus, Cieza  \& Gonz{\'{a}}lez-Ruilova}{Antilen et~al.}{2023}]{2023MNRAS.522.2611A}
Antilen J.,  Casassus S.,  Cieza L.~A.,   Gonz{\'{a}}lez-Ruilova C.,  2023, \mn@doi [\mnras] {10.1093/mnras/stad975}, 522, 2611

\bibitem[\protect\citeauthoryear{Bai}{Bai}{2013}]{2013ApJ...772...96B}
Bai X.-N.,  2013, \mn@doi [\apj] {10.1088/0004-637X/772/2/96}, 772, 96

\bibitem[\protect\citeauthoryear{{Bai}}{{Bai}}{2015}]{2015ApJ...798...84B}
{Bai} X.-N.,  2015, \mn@doi [\apj] {10.1088/0004-637X/798/2/84}, \href {https://ui.adsabs.harvard.edu/abs/2015ApJ...798...84B} {798, 84}

\bibitem[\protect\citeauthoryear{{Bai} \& {Stone}}{{Bai} \& {Stone}}{2013}]{2013ApJ...769...76B}
{Bai} X.-N.,  {Stone} J.~M.,  2013, \mn@doi [\apj] {10.1088/0004-637X/769/1/76}, \href {https://ui.adsabs.harvard.edu/abs/2013ApJ...769...76B} {769, 76}

\bibitem[\protect\citeauthoryear{{Bai}, {Ye}, {Goodman}  \& {Yuan}}{{Bai} et~al.}{2016}]{2016ApJ...818..152B}
{Bai} X.-N.,  {Ye} J.,  {Goodman} J.,   {Yuan} F.,  2016, \mn@doi [\apj] {10.3847/0004-637X/818/2/152}, \href {https://ui.adsabs.harvard.edu/abs/2016ApJ...818..152B} {818, 152}

\bibitem[\protect\citeauthoryear{Balbus \& Hawley}{Balbus \& Hawley}{1991}]{1991ApJ...376..214B}
Balbus S.~A.,  Hawley J.~F.,  1991, \mn@doi [\apj] {10.1086/170270}, 376, 214

\bibitem[\protect\citeauthoryear{{Barenfeld}, {Carpenter}, {Sargent}, {Isella}  \& {Ricci}}{{Barenfeld} et~al.}{2017}]{2017ApJ...851...85B}
{Barenfeld} S.~A.,  {Carpenter} J.~M.,  {Sargent} A.~I.,  {Isella} A.,   {Ricci} L.,  2017, \mn@doi [\apj] {10.3847/1538-4357/aa989d}, \href {https://ui.adsabs.harvard.edu/abs/2017ApJ...851...85B} {851, 85}

\bibitem[\protect\citeauthoryear{{Bath} \& {Pringle}}{{Bath} \& {Pringle}}{1981}]{1981MNRAS.194..967B}
{Bath} G.~T.,  {Pringle} J.~E.,  1981, \mn@doi [\mnras] {10.1093/mnras/194.4.967}, \href {https://ui.adsabs.harvard.edu/abs/1981MNRAS.194..967B} {194, 967}

\bibitem[\protect\citeauthoryear{Bi et~al.,}{Bi et~al.}{2020}]{2020ApJ...895L..18B}
Bi J.,  et~al., 2020, \mn@doi [\apjl] {10.3847/2041-8213/ab8eb4}, 895, L18

\bibitem[\protect\citeauthoryear{Birnstiel, Dullemond  \& Brauer}{Birnstiel et~al.}{2010}]{2010A&A...513A..79B}
Birnstiel T.,  Dullemond C.,   Brauer F.,  2010, \mn@doi [\aap] {10.1051/0004-6361/200913731}, 513, A79

\bibitem[\protect\citeauthoryear{{Blandford} \& {Payne}}{{Blandford} \& {Payne}}{1982}]{1982MNRAS.199..883B}
{Blandford} R.~D.,  {Payne} D.~G.,  1982, \mn@doi [\mnras] {10.1093/mnras/199.4.883}, \href {https://ui.adsabs.harvard.edu/abs/1982MNRAS.199..883B} {199, 883}

\bibitem[\protect\citeauthoryear{Booth et~al.,}{Booth et~al.}{2021}]{2021ApJS..257...16B}
Booth A.~S.,  et~al., 2021, \mn@doi [\apjs] {10.3847/1538-4365/ac1ad4}, 257, 16

\bibitem[\protect\citeauthoryear{Bruderer}{Bruderer}{2013}]{2013A&A...559A..46B}
Bruderer S.,  2013, \mn@doi [\aap] {10.1051/0004-6361/201321171}, 559, A46

\bibitem[\protect\citeauthoryear{Bruderer, van Dishoeck, Doty  \& Herczeg}{Bruderer et~al.}{2012}]{2012A&A...541A..91B}
Bruderer S.,  van Dishoeck E.,  Doty S.,   Herczeg G.,  2012, \mn@doi [\aap] {10.1051/0004-6361/201118218}, 541, A91

\bibitem[\protect\citeauthoryear{Casassus et~al.,}{Casassus et~al.}{2021}]{2021MNRAS.507.3789C}
Casassus S.,  et~al., 2021, \mn@doi [\mnras] {10.1093/mnras/stab2359}, 507, 3789

\bibitem[\protect\citeauthoryear{{Chiang} \& {Goldreich}}{{Chiang} \& {Goldreich}}{1997}]{1997ApJ...490..368C}
{Chiang} E.~I.,  {Goldreich} P.,  1997, \mn@doi [\apj] {10.1086/304869}, \href {https://ui.adsabs.harvard.edu/abs/1997ApJ...490..368C} {490, 368}

\bibitem[\protect\citeauthoryear{Cieza et~al.,}{Cieza et~al.}{2019}]{2019MNRAS.482..698C}
Cieza L.~A.,  et~al., 2019, \mn@doi [\mnras] {10.1093/mnras/sty2653}, 482, 698

\bibitem[\protect\citeauthoryear{Clarke, Gendrin  \& Sotomayor}{Clarke et~al.}{2001}]{2001MNRAS.328..485C}
Clarke C.,  Gendrin A.,   Sotomayor M.,  2001, \mn@doi [\mnras] {10.1046/j.1365-8711.2001.04891.x}, 328, 485

\bibitem[\protect\citeauthoryear{Cleeves, Adams  \& Bergin}{Cleeves et~al.}{2013}]{2013ApJ...772....5C}
Cleeves L.~I.,  Adams F.~C.,   Bergin E.~A.,  2013, \mn@doi [\apj] {10.1088/0004-637X/772/1/5}, 772, 5

\bibitem[\protect\citeauthoryear{Coleman \& Haworth}{Coleman \& Haworth}{2022}]{2022MNRAS.514.2315C}
Coleman G.~A.,  Haworth T.~J.,  2022, \mn@doi [\mnras] {10.1093/mnras/stac1513}, 514, 2315

\bibitem[\protect\citeauthoryear{{Delussu}, {Birnstiel}, {Miotello}, {Pinilla}, {Rosotti}  \& {Andrews}}{{Delussu} et~al.}{2024}]{2024arXiv240514501D}
{Delussu} L.,  {Birnstiel} T.,  {Miotello} A.,  {Pinilla} P.,  {Rosotti} G.,   {Andrews} S.~M.,  2024, \mn@doi [arXiv e-prints] {10.48550/arXiv.2405.14501}, \href {https://ui.adsabs.harvard.edu/abs/2024arXiv240514501D} {p. arXiv:2405.14501}

\bibitem[\protect\citeauthoryear{{Dickman}}{{Dickman}}{1978}]{1978ApJS...37..407D}
{Dickman} R.~L.,  1978, \mn@doi [\apjs] {10.1086/190535}, \href {https://ui.adsabs.harvard.edu/abs/1978ApJS...37..407D} {37, 407}

\bibitem[\protect\citeauthoryear{Dipierro et~al.,}{Dipierro et~al.}{2018a}]{2018MNRAS.475.5296D}
Dipierro G.,  et~al., 2018a, \mn@doi [\mnras] {10.1093/mnras/sty181}, 475, 5296

\bibitem[\protect\citeauthoryear{Dipierro, Laibe, Alexander  \& Hutchison}{Dipierro et~al.}{2018b}]{2018MNRAS.479.4187D}
Dipierro G.,  Laibe G.,  Alexander R.,   Hutchison M.,  2018b, \mn@doi [\mnras] {10.1093/mnras/sty1701}, 479, 4187

\bibitem[\protect\citeauthoryear{Eisner et~al.,}{Eisner et~al.}{2018}]{2018ApJ...860...77E}
Eisner J.,  et~al., 2018, \mn@doi [\apj] {10.3847/1538-4357/aac3e2}, 860, 77

\bibitem[\protect\citeauthoryear{{Emsenhuber}, {Burn}, {Weder}, {Monsch}, {Picogna}, {Ercolano}  \& {Preibisch}}{{Emsenhuber} et~al.}{2023}]{2023A&A...673A..78E}
{Emsenhuber} A.,  {Burn} R.,  {Weder} J.,  {Monsch} K.,  {Picogna} G.,  {Ercolano} B.,   {Preibisch} T.,  2023, \mn@doi [\aap] {10.1051/0004-6361/202244767}, \href {https://ui.adsabs.harvard.edu/abs/2023A&A...673A..78E} {673, A78}

\bibitem[\protect\citeauthoryear{Ercolano, Picogna, Monsch, Drake  \& Preibisch}{Ercolano et~al.}{2021}]{2021MNRAS.508.1675E}
Ercolano B.,  Picogna G.,  Monsch K.,  Drake J.~J.,   Preibisch T.,  2021, \mn@doi [\mnras] {10.1093/mnras/stab2590}, 508, 1675

\bibitem[\protect\citeauthoryear{Facchini et~al.,}{Facchini et~al.}{2019}]{2019A&A...626L...2F}
Facchini S.,  et~al., 2019, \mn@doi [\aap] {10.1051/0004-6361/201935496}, 626, L2

\bibitem[\protect\citeauthoryear{{Fang} et~al.,}{{Fang} et~al.}{2018}]{2018ApJ...868...28F}
{Fang} M.,  et~al., 2018, \mn@doi [\apj] {10.3847/1538-4357/aae780}, \href {https://ui.adsabs.harvard.edu/abs/2018ApJ...868...28F} {868, 28}

\bibitem[\protect\citeauthoryear{{Favre}, {Cleeves}, {Bergin}, {Qi}  \& {Blake}}{{Favre} et~al.}{2013}]{2013ApJ...776L..38F}
{Favre} C.,  {Cleeves} L.~I.,  {Bergin} E.~A.,  {Qi} C.,   {Blake} G.~A.,  2013, \mn@doi [\apjl] {10.1088/2041-8205/776/2/L38}, \href {https://ui.adsabs.harvard.edu/abs/2013ApJ...776L..38F} {776, L38}

\bibitem[\protect\citeauthoryear{Fedele, van~den Ancker, Henning, Jayawardhana  \& Oliveira}{Fedele et~al.}{2010}]{2010A&A...510A..72F}
Fedele D.,  van~den Ancker M.,  Henning T.,  Jayawardhana R.,   Oliveira J.,  2010, \mn@doi [\aap] {10.1051/0004-6361/200912810}, 510, A72

\bibitem[\protect\citeauthoryear{Ferreira}{Ferreira}{1997}]{1997A&A...319..340F}
Ferreira J.,  1997, \mn@doi [\aap] {10.48550/arXiv.astro-ph/9607057}, 319, 340

\bibitem[\protect\citeauthoryear{Flaherty, Hughes, Rosenfeld, Andrews, Chiang, Simon, Kerzner  \& Wilner}{Flaherty et~al.}{2015}]{2015ApJ...813...99F}
Flaherty K.~M.,  Hughes A.~M.,  Rosenfeld K.~A.,  Andrews S.~M.,  Chiang E.,  Simon J.~B.,  Kerzner S.,   Wilner D.~J.,  2015, \mn@doi [\apj] {10.1088/0004-637X/813/2/99}, 813, 99

\bibitem[\protect\citeauthoryear{Flaherty et~al.,}{Flaherty et~al.}{2017}]{2017ApJ...843..150F}
Flaherty K.~M.,  et~al., 2017, \mn@doi [\apj] {10.3847/1538-4357/aa79f9}, 843, 150

\bibitem[\protect\citeauthoryear{Flaherty, Hughes, Teague, Simon, Andrews  \& Wilner}{Flaherty et~al.}{2018}]{2018ApJ...856..117F}
Flaherty K.~M.,  Hughes A.~M.,  Teague R.,  Simon J.~B.,  Andrews S.~M.,   Wilner D.~J.,  2018, \mn@doi [\apj] {10.3847/1538-4357/aab615}, 856, 117

\bibitem[\protect\citeauthoryear{Flaherty et~al.,}{Flaherty et~al.}{2020}]{2020ApJ...895..109F}
Flaherty K.,  et~al., 2020, \mn@doi [\apj] {10.3847/1538-4357/ab8cc5}, 895, 109

\bibitem[\protect\citeauthoryear{Flock, Fromang, Turner  \& Benisty}{Flock et~al.}{2016}]{2016ApJ...827..144F}
Flock M.,  Fromang S.,  Turner N.,   Benisty M.,  2016, \mn@doi [\apj] {10.3847/0004-637X/827/2/144}, 827, 144

\bibitem[\protect\citeauthoryear{Font, McCarthy, Johnstone  \& Ballantyne}{Font et~al.}{2004}]{2004ApJ...607..890F}
Font A.~S.,  McCarthy I.~G.,  Johnstone D.,   Ballantyne D.~R.,  2004, \mn@doi [\apj] {10.1086/383518}, 607, 890

\bibitem[\protect\citeauthoryear{{Frerking}, {Langer}  \& {Wilson}}{{Frerking} et~al.}{1982}]{1982ApJ...262..590F}
{Frerking} M.~A.,  {Langer} W.~D.,   {Wilson} R.~W.,  1982, \mn@doi [\apj] {10.1086/160451}, \href {https://ui.adsabs.harvard.edu/abs/1982ApJ...262..590F} {262, 590}

\bibitem[\protect\citeauthoryear{{Gammie}}{{Gammie}}{1996}]{1996ApJ...457..355G}
{Gammie} C.~F.,  1996, \mn@doi [\apj] {10.1086/176735}, \href {https://ui.adsabs.harvard.edu/abs/1996ApJ...457..355G} {457, 355}

\bibitem[\protect\citeauthoryear{{G{\'a}rate}, {Birnstiel}, {Stammler}  \& {G{\"u}nther}}{{G{\'a}rate} et~al.}{2019}]{2019ApJ...871...53G}
{G{\'a}rate} M.,  {Birnstiel} T.,  {Stammler} S.~M.,   {G{\"u}nther} H.~M.,  2019, \mn@doi [\apj] {10.3847/1538-4357/aaf4fc}, \href {https://ui.adsabs.harvard.edu/abs/2019ApJ...871...53G} {871, 53}

\bibitem[\protect\citeauthoryear{{G{\'a}rate} et~al.,}{{G{\'a}rate} et~al.}{2021}]{2021A&A...655A..18G}
{G{\'a}rate} M.,  et~al., 2021, \mn@doi [\aap] {10.1051/0004-6361/202141444}, \href {https://ui.adsabs.harvard.edu/abs/2021A&A...655A..18G} {655, A18}

\bibitem[\protect\citeauthoryear{{Haisch}, Lada  \& Lada}{{Haisch} et~al.}{2001}]{2001ApJ...553L.153H}
{Haisch} Lada E.~A.,   Lada C.~J.,  2001, \mn@doi [\apjl] {10.1086/320685}, 553, L153

\bibitem[\protect\citeauthoryear{{Harris} et~al.,}{{Harris} et~al.}{2020}]{numpy2}
{Harris} C.~R.,  et~al., 2020, \mn@doi [\nat] {10.1038/s41586-020-2649-2}, \href {https://ui.adsabs.harvard.edu/abs/2020Natur.585..357H} {585, 357}

\bibitem[\protect\citeauthoryear{{Hartmann}, {Calvet}, {Gullbring}  \& {D'Alessio}}{{Hartmann} et~al.}{1998}]{1998ApJ...495..385H}
{Hartmann} L.,  {Calvet} N.,  {Gullbring} E.,   {D'Alessio} P.,  1998, \mn@doi [\apj] {10.1086/305277}, \href {https://ui.adsabs.harvard.edu/abs/1998ApJ...495..385H} {495, 385}

\bibitem[\protect\citeauthoryear{Hawley, Gammie  \& Balbus}{Hawley et~al.}{1995}]{1995ApJ...440..742H}
Hawley J.~F.,  Gammie C.~F.,   Balbus S.~A.,  1995, \mn@doi [\apj] {10.1086/175311}, 440, 742

\bibitem[\protect\citeauthoryear{Hillenbrand}{Hillenbrand}{2005}]{2005astro.ph.11083H}
Hillenbrand L.~A.,  2005, \mn@doi [arXiv e-prints] {10.48550/arXiv.astro-ph/0511083}, pp astro--ph/0511083

\bibitem[\protect\citeauthoryear{Huang, {\"{O}}berg  \& Andrews}{Huang et~al.}{2016}]{2016ApJ...823L..18H}
Huang J.,  {\"{O}}berg K.~I.,   Andrews S.~M.,  2016, \mn@doi [\apjl] {10.3847/2041-8205/823/1/L18}, 823, L18

\bibitem[\protect\citeauthoryear{Huang et~al.,}{Huang et~al.}{2018}]{2018ApJ...869L..42H}
Huang J.,  et~al., 2018, \mn@doi [\apjl] {10.3847/2041-8213/aaf740}, 869, L42

\bibitem[\protect\citeauthoryear{Hunter}{Hunter}{2007}]{matplotlib}
Hunter J.~D.,  2007, \mn@doi [Computing in Science & Engineering] {10.1109/MCSE.2007.55}, 9, 90

\bibitem[\protect\citeauthoryear{{Isella}, {Carpenter}  \& {Sargent}}{{Isella} et~al.}{2009}]{2009ApJ...701..260I}
{Isella} A.,  {Carpenter} J.~M.,   {Sargent} A.~I.,  2009, \mn@doi [\apj] {10.1088/0004-637X/701/1/260}, \href {https://ui.adsabs.harvard.edu/abs/2009ApJ...701..260I} {701, 260}

\bibitem[\protect\citeauthoryear{{Kama} et~al.,}{{Kama} et~al.}{2016}]{2016A&A...592A..83K}
{Kama} M.,  et~al., 2016, \mn@doi [\aap] {10.1051/0004-6361/201526991}, \href {https://ui.adsabs.harvard.edu/abs/2016A&A...592A..83K} {592, A83}

\bibitem[\protect\citeauthoryear{{Kenyon} \& {Hartmann}}{{Kenyon} \& {Hartmann}}{1987}]{1987ApJ...323..714K}
{Kenyon} S.~J.,  {Hartmann} L.,  1987, \mn@doi [\apj] {10.1086/165866}, \href {https://ui.adsabs.harvard.edu/abs/1987ApJ...323..714K} {323, 714}

\bibitem[\protect\citeauthoryear{Kluyver et~al.,}{Kluyver et~al.}{2016}]{jupyter}
Kluyver T.,  et~al., 2016, in Loizides F.,  Scmidt B.,  eds, Positioning and Power in Academic Publishing: Players, Agents and Agendas. IOS Press, pp 87--90, \url {https://eprints.soton.ac.uk/403913/}

\bibitem[\protect\citeauthoryear{{Komaki}, {Fukuhara}, {Suzuki}  \& {Yoshida}}{{Komaki} et~al.}{2023}]{2023arXiv230413316K}
{Komaki} A.,  {Fukuhara} S.,  {Suzuki} T.~K.,   {Yoshida} N.,  2023, \mn@doi [arXiv e-prints] {10.48550/arXiv.2304.13316}, \href {https://ui.adsabs.harvard.edu/abs/2023arXiv230413316K} {p. arXiv:2304.13316}

\bibitem[\protect\citeauthoryear{Kratter \& Lodato}{Kratter \& Lodato}{2016}]{2016ARA&A..54..271K}
Kratter K.,  Lodato G.,  2016, \mn@doi [\araa] {10.1146/annurev-astro-081915-023307}, 54, 271

\bibitem[\protect\citeauthoryear{Kurtovic et~al.,}{Kurtovic et~al.}{2021}]{2021A&A...645A.139K}
Kurtovic N.,  et~al., 2021, \mn@doi [\aap] {10.1051/0004-6361/202038983}, 645, A139

\bibitem[\protect\citeauthoryear{{Lacy}, {Knacke}, {Geballe}  \& {Tokunaga}}{{Lacy} et~al.}{1994}]{1994ApJ...428L..69L}
{Lacy} J.~H.,  {Knacke} R.,  {Geballe} T.~R.,   {Tokunaga} A.~T.,  1994, \mn@doi [\apjl] {10.1086/187395}, \href {https://ui.adsabs.harvard.edu/abs/1994ApJ...428L..69L} {428, L69}

\bibitem[\protect\citeauthoryear{Law et~al.,}{Law et~al.}{2021}]{2021ApJS..257....3L}
Law C.~J.,  et~al., 2021, \mn@doi [\apjs] {10.3847/1538-4365/ac1434}, 257, 3

\bibitem[\protect\citeauthoryear{Law et~al.,}{Law et~al.}{2022}]{2022ApJ...932..114L}
Law C.~J.,  et~al., 2022, \mn@doi [\apj] {10.3847/1538-4357/ac6c02}, 932, 114

\bibitem[\protect\citeauthoryear{Law, Alarc{\'{o}}n, Cleeves, {\"{O}}berg  \& Paneque-Carre{\~{n}}o}{Law et~al.}{2023a}]{2023arXiv231116233L}
Law C.~J.,  Alarc{\'{o}}n F.,  Cleeves L.~I.,  {\"{O}}berg K.~I.,   Paneque-Carre{\~{n}}o T.,  2023a, arXiv e-prints, p. arXiv:2311.16233

\bibitem[\protect\citeauthoryear{Law et~al.,}{Law et~al.}{2023b}]{2023ApJ...948...60L}
Law C.~J.,  et~al., 2023b, \mn@doi [\apj] {10.3847/1538-4357/acb3c4}, 948, 60

\bibitem[\protect\citeauthoryear{Lin \& Pringle}{Lin \& Pringle}{1987}]{1987MNRAS.225..607L}
Lin D.,  Pringle J.,  1987, \mn@doi [\mnras] {10.1093/mnras/225.3.607}, 225, 607

\bibitem[\protect\citeauthoryear{{Long} et~al.,}{{Long} et~al.}{2017}]{2017ApJ...844...99L}
{Long} F.,  et~al., 2017, \mn@doi [\apj] {10.3847/1538-4357/aa78fc}, \href {https://ui.adsabs.harvard.edu/abs/2017ApJ...844...99L} {844, 99}

\bibitem[\protect\citeauthoryear{{Long} et~al.,}{{Long} et~al.}{2022}]{2022ApJ...931....6L}
{Long} F.,  et~al., 2022, \mn@doi [\apj] {10.3847/1538-4357/ac634e}, \href {https://ui.adsabs.harvard.edu/abs/2022ApJ...931....6L} {931, 6}

\bibitem[\protect\citeauthoryear{Louvet, Dougados, Cabrit, Mardones, M{\'{e}}nard, Tabone, Pinte  \& Dent}{Louvet et~al.}{2018}]{2018A&A...618A.120L}
Louvet F.,  Dougados C.,  Cabrit S.,  Mardones D.,  M{\'{e}}nard F.,  Tabone B.,  Pinte C.,   Dent W.,  2018, \mn@doi [\aap] {10.1051/0004-6361/201731733}, 618, A120

\bibitem[\protect\citeauthoryear{{Lynden-Bell} \& {Pringle}}{{Lynden-Bell} \& {Pringle}}{1974}]{1974MNRAS.168..603L}
{Lynden-Bell} D.,  {Pringle} J.~E.,  1974, \mn@doi [\mnras] {10.1093/mnras/168.3.603}, \href {https://ui.adsabs.harvard.edu/abs/1974MNRAS.168..603L} {168, 603}

\bibitem[\protect\citeauthoryear{{Lyra}, {Johansen}, {Zsom}, {Klahr}  \& {Piskunov}}{{Lyra} et~al.}{2009}]{2009A&A...497..869L}
{Lyra} W.,  {Johansen} A.,  {Zsom} A.,  {Klahr} H.,   {Piskunov} N.,  2009, \mn@doi [\aap] {10.1051/0004-6361/200811265}, \href {https://ui.adsabs.harvard.edu/abs/2009A&A...497..869L} {497, 869}

\bibitem[\protect\citeauthoryear{Manara et~al.,}{Manara et~al.}{2016}]{2016A&A...591L...3M}
Manara C.,  et~al., 2016, \mn@doi [\aap] {10.1051/0004-6361/201628549}, 591, L3

\bibitem[\protect\citeauthoryear{Manara et~al.,}{Manara et~al.}{2017}]{2017A&A...604A.127M}
Manara C.,  et~al., 2017, \mn@doi [\aap] {10.1051/0004-6361/201630147}, 604, A127

\bibitem[\protect\citeauthoryear{Manara et~al.,}{Manara et~al.}{2020}]{2020A&A...639A..58M}
Manara C.,  et~al., 2020, \mn@doi [\aap] {10.1051/0004-6361/202037949}, 639, A58

\bibitem[\protect\citeauthoryear{Manara, Ansdell, Rosotti, Hughes, Armitage, Lodato  \& Williams}{Manara et~al.}{2022}]{2022arXiv220309930M}
Manara C.,  Ansdell M.,  Rosotti G.,  Hughes A.,  Armitage P.,  Lodato G.,   Williams J.,  2022, \mn@doi [arXiv e-prints] {10.48550/arXiv.2203.09930}, p. arXiv:2203.09930

\bibitem[\protect\citeauthoryear{Michel, van~der Marel  \& Matthews}{Michel et~al.}{2021}]{2021ApJ...921...72M}
Michel A.,  van~der Marel N.,   Matthews B.~C.,  2021, \mn@doi [\apj] {10.3847/1538-4357/ac1bbb}, 921, 72

\bibitem[\protect\citeauthoryear{{Miotello} et~al.,}{{Miotello} et~al.}{2017}]{2017A&A...599A.113M}
{Miotello} A.,  et~al., 2017, \mn@doi [\aap] {10.1051/0004-6361/201629556}, \href {https://ui.adsabs.harvard.edu/abs/2017A&A...599A.113M} {599, A113}

\bibitem[\protect\citeauthoryear{{Morishima}}{{Morishima}}{2012}]{2012MNRAS.420.2851M}
{Morishima} R.,  2012, \mn@doi [\mnras] {10.1111/j.1365-2966.2011.19940.x}, \href {https://ui.adsabs.harvard.edu/abs/2012MNRAS.420.2851M} {420, 2851}

\bibitem[\protect\citeauthoryear{{Najita} \& {Bergin}}{{Najita} \& {Bergin}}{2018}]{2018ApJ...864..168N}
{Najita} J.~R.,  {Bergin} E.~A.,  2018, \mn@doi [\apj] {10.3847/1538-4357/aad80c}, \href {https://ui.adsabs.harvard.edu/abs/2018ApJ...864..168N} {864, 168}

\bibitem[\protect\citeauthoryear{Natta, Testi, Alcal{\'{a}}, Rigliaco, Covino, Stelzer  \& D'Elia}{Natta et~al.}{2014}]{2014A&A...569A...5N}
Natta A.,  Testi L.,  Alcal{\'{a}} J.,  Rigliaco E.,  Covino E.,  Stelzer B.,   D'Elia V.,  2014, \mn@doi [\aap] {10.1051/0004-6361/201424136}, 569, A5

\bibitem[\protect\citeauthoryear{Nelson, Gressel  \& Umurhan}{Nelson et~al.}{2013}]{2013MNRAS.435.2610N}
Nelson R.~P.,  Gressel O.,   Umurhan O.~M.,  2013, \mn@doi [\mnras] {10.1093/mnras/stt1475}, 435, 2610

\bibitem[\protect\citeauthoryear{{\"{O}}berg et~al.,}{{\"{O}}berg et~al.}{2021a}]{2021AJ....161...38O}
{\"{O}}berg K.~I.,  et~al., 2021a, \mn@doi [\aj] {10.3847/1538-3881/abc74d}, 161, 38

\bibitem[\protect\citeauthoryear{{{\"O}berg} et~al.,}{{{\"O}berg} et~al.}{2021b}]{2021ApJS..257....1O}
{{\"O}berg} K.~I.,  et~al., 2021b, \mn@doi [\apjs] {10.3847/1538-4365/ac1432}, \href {https://ui.adsabs.harvard.edu/abs/2021ApJS..257....1O} {257, 1}

\bibitem[\protect\citeauthoryear{Otter, Ginsburg, Ballering, Bally, Eisner, Goddi, Plambeck  \& Wright}{Otter et~al.}{2021}]{2021ApJ...923..221O}
Otter J.,  Ginsburg A.,  Ballering N.~P.,  Bally J.,  Eisner J.,  Goddi C.,  Plambeck R.,   Wright M.,  2021, \mn@doi [\apj] {10.3847/1538-4357/ac29c2}, 923, 221

\bibitem[\protect\citeauthoryear{{Pascucci}, {Cabrit}, {Edwards}, {Gorti}, {Gressel}  \& {Suzuki}}{{Pascucci} et~al.}{2023}]{2023ASPC..534..567P}
{Pascucci} I.,  {Cabrit} S.,  {Edwards} S.,  {Gorti} U.,  {Gressel} O.,   {Suzuki} T.~K.,  2023, in {Inutsuka} S.,  {Aikawa} Y.,  {Muto} T.,  {Tomida} K.,   {Tamura} M.,  eds,  Astronomical Society of the Pacific Conference Series Vol. 534, Astronomical Society of the Pacific Conference Series. p.~567 (\mn@eprint {arXiv} {2203.10068}), \mn@doi{10.48550/arXiv.2203.10068}

\bibitem[\protect\citeauthoryear{Pegues et~al.,}{Pegues et~al.}{2021}]{2021ApJ...911..150P}
Pegues J.,  et~al., 2021, \mn@doi [\apj] {10.3847/1538-4357/abe870}, 911, 150

\bibitem[\protect\citeauthoryear{{Pinilla}, {Flock}, {Ovelar}  \& {Birnstiel}}{{Pinilla} et~al.}{2016}]{2016A&A...596A..81P}
{Pinilla} P.,  {Flock} M.,  {Ovelar} M. d.~J.,   {Birnstiel} T.,  2016, \mn@doi [\aap] {10.1051/0004-6361/201628441}, \href {https://ui.adsabs.harvard.edu/abs/2016A&A...596A..81P} {596, A81}

\bibitem[\protect\citeauthoryear{{Reg{\'a}ly}, {Juh{\'a}sz}, {S{\'a}ndor}  \& {Dullemond}}{{Reg{\'a}ly} et~al.}{2012}]{2012MNRAS.419.1701R}
{Reg{\'a}ly} Z.,  {Juh{\'a}sz} A.,  {S{\'a}ndor} Z.,   {Dullemond} C.~P.,  2012, \mn@doi [\mnras] {10.1111/j.1365-2966.2011.19834.x}, \href {https://ui.adsabs.harvard.edu/abs/2012MNRAS.419.1701R} {419, 1701}

\bibitem[\protect\citeauthoryear{Ribas, Bouy  \& Mer{\'\i}n}{Ribas et~al.}{2015}]{2015A&A...576A..52R}
Ribas {\'{A}}.,  Bouy H.,   Mer{\'\i}n B.,  2015, \mn@doi [\aap] {10.1051/0004-6361/201424846}, 576, A52

\bibitem[\protect\citeauthoryear{Richert, Getman, Feigelson, Kuhn, Broos, Povich, Bate  \& Garmire}{Richert et~al.}{2018}]{2018MNRAS.477.5191R}
Richert A.,  Getman K.,  Feigelson E.,  Kuhn M.,  Broos P.,  Povich M.,  Bate M.,   Garmire G.,  2018, \mn@doi [\mnras] {10.1093/mnras/sty949}, 477, 5191

\bibitem[\protect\citeauthoryear{{Rosotti}}{{Rosotti}}{2023}]{2023NewAR..9601674R}
{Rosotti} G.~P.,  2023, \mn@doi [\nar] {10.1016/j.newar.2023.101674}, \href {https://ui.adsabs.harvard.edu/abs/2023NewAR..9601674R} {96, 101674}

\bibitem[\protect\citeauthoryear{Rosotti, Tazzari, Booth, Testi, Lodato  \& Clarke}{Rosotti et~al.}{2019}]{2019MNRAS.486.4829R}
Rosotti G.~P.,  Tazzari M.,  Booth R.~A.,  Testi L.,  Lodato G.,   Clarke C.,  2019, \mn@doi [\mnras] {10.1093/mnras/stz1190}, 486, 4829

\bibitem[\protect\citeauthoryear{Rugel, Fedele  \& Herczeg}{Rugel et~al.}{2018}]{2018A&A...609A..70R}
Rugel M.,  Fedele D.,   Herczeg G.,  2018, \mn@doi [\aap] {10.1051/0004-6361/201630111}, 609, A70

\bibitem[\protect\citeauthoryear{Sanchis et~al.,}{Sanchis et~al.}{2021}]{2021A&A...649A..19S}
Sanchis E.,  et~al., 2021, \mn@doi [\aap] {10.1051/0004-6361/202039733}, 649, A19

\bibitem[\protect\citeauthoryear{{Schwarz}, {Bergin}, {Cleeves}, {Blake}, {Zhang}, {{\"O}berg}, {van Dishoeck}  \& {Qi}}{{Schwarz} et~al.}{2016}]{2016ApJ...823...91S}
{Schwarz} K.~R.,  {Bergin} E.~A.,  {Cleeves} L.~I.,  {Blake} G.~A.,  {Zhang} K.,  {{\"O}berg} K.~I.,  {van Dishoeck} E.~F.,   {Qi} C.,  2016, \mn@doi [\apj] {10.3847/0004-637X/823/2/91}, \href {https://ui.adsabs.harvard.edu/abs/2016ApJ...823...91S} {823, 91}

\bibitem[\protect\citeauthoryear{{Shakura} \& {Sunyaev}}{{Shakura} \& {Sunyaev}}{1973}]{1973A&A....24..337S}
{Shakura} N.~I.,  {Sunyaev} R.~A.,  1973, \aap, \href {https://ui.adsabs.harvard.edu/abs/1973A&A....24..337S} {24, 337}

\bibitem[\protect\citeauthoryear{{Simon}, {Bai}, {Armitage}, {Stone}  \& {Beckwith}}{{Simon} et~al.}{2013}]{2013ApJ...775...73S}
{Simon} J.~B.,  {Bai} X.-N.,  {Armitage} P.~J.,  {Stone} J.~M.,   {Beckwith} K.,  2013, \mn@doi [\apj] {10.1088/0004-637X/775/1/73}, \href {https://ui.adsabs.harvard.edu/abs/2013ApJ...775...73S} {775, 73}

\bibitem[\protect\citeauthoryear{{Simon}, {Bai}, {Flaherty}  \& {Hughes}}{{Simon} et~al.}{2018}]{2018ApJ...865...10S}
{Simon} J.~B.,  {Bai} X.-N.,  {Flaherty} K.~M.,   {Hughes} A.~M.,  2018, \mn@doi [\apj] {10.3847/1538-4357/aad86d}, \href {https://ui.adsabs.harvard.edu/abs/2018ApJ...865...10S} {865, 10}

\bibitem[\protect\citeauthoryear{Sturm, McClure, Harsono, Facchini, Long, Kama, Bergin  \& van Dishoeck}{Sturm et~al.}{2022}]{2022A&A...660A.126S}
Sturm J.,  McClure M.,  Harsono D.,  Facchini S.,  Long F.,  Kama M.,  Bergin E.,   van Dishoeck E.,  2022, \mn@doi [\aap] {10.1051/0004-6361/202141860}, 660, A126

\bibitem[\protect\citeauthoryear{Suzuki, Ogihara, Morbidelli, Crida  \& Guillot}{Suzuki et~al.}{2016}]{2016A&A...596A..74S}
Suzuki T.~K.,  Ogihara M.,  Morbidelli A.,  Crida A.,   Guillot T.,  2016, \mn@doi [\aap] {10.1051/0004-6361/201628955}, 596, A74

\bibitem[\protect\citeauthoryear{{Tabone} et~al.,}{{Tabone} et~al.}{2020}]{2020A&A...640A..82T}
{Tabone} B.,  et~al., 2020, \mn@doi [\aap] {10.1051/0004-6361/201834377}, \href {https://ui.adsabs.harvard.edu/abs/2020A&A...640A..82T} {640, A82}

\bibitem[\protect\citeauthoryear{Tabone, Rosotti, Lodato, Armitage, Cridland  \& van Dishoeck}{Tabone et~al.}{2022a}]{2022MNRAS.512L..74T}
Tabone B.,  Rosotti G.,  Lodato G.,  Armitage P.,  Cridland A.,   van Dishoeck E.,  2022a, \mn@doi [\mnras] {10.1093/mnrasl/slab124}, 512, L74

\bibitem[\protect\citeauthoryear{{Tabone}, {Rosotti}, {Cridland}, {Armitage}  \& {Lodato}}{{Tabone} et~al.}{2022b}]{2022MNRAS.512.2290T}
{Tabone} B.,  {Rosotti} G.~P.,  {Cridland} A.~J.,  {Armitage} P.~J.,   {Lodato} G.,  2022b, \mn@doi [\mnras] {10.1093/mnras/stab3442}, \href {https://ui.adsabs.harvard.edu/abs/2022MNRAS.512.2290T} {512, 2290}

\bibitem[\protect\citeauthoryear{Teague et~al.,}{Teague et~al.}{2018}]{2018ApJ...864..133T}
Teague R.,  et~al., 2018, \mn@doi [\apj] {10.3847/1538-4357/aad80e}, 864, 133

\bibitem[\protect\citeauthoryear{{The Astropy Collaboration} et~al.,}{{The Astropy Collaboration} et~al.}{2018}]{astropy}
{The Astropy Collaboration} et~al., 2018, \mn@doi [\\aj] {10.3847/1538-3881/aabc4f}, \href {https://ui.adsabs.harvard.edu/abs/2018AJ....156..123T} {156, 123}

\bibitem[\protect\citeauthoryear{Tielens \& Hollenbach}{Tielens \& Hollenbach}{1985}]{1985ApJ...291..722T}
Tielens A.,  Hollenbach D.,  1985, \mn@doi [\apj] {10.1086/163111}, 291, 722

\bibitem[\protect\citeauthoryear{{Toci}, {Lodato}, {Livio}, {Rosotti}  \& {Trapman}}{{Toci} et~al.}{2023}]{2023MNRAS.518L..69T}
{Toci} C.,  {Lodato} G.,  {Livio} F.~G.,  {Rosotti} G.,   {Trapman} L.,  2023, \mn@doi [\mnras] {10.1093/mnrasl/slac137}, \href {https://ui.adsabs.harvard.edu/abs/2023MNRAS.518L..69T} {518, L69}

\bibitem[\protect\citeauthoryear{Trapman, Facchini, Hogerheijde, van Dishoeck  \& Bruderer}{Trapman et~al.}{2019}]{2019A&A...629A..79T}
Trapman L.,  Facchini S.,  Hogerheijde M.,  van Dishoeck E.,   Bruderer S.,  2019, \mn@doi [\aap] {10.1051/0004-6361/201834723}, 629, A79

\bibitem[\protect\citeauthoryear{Trapman, Rosotti, Bosman, Hogerheijde  \& van Dishoeck}{Trapman et~al.}{2020}]{2020A&A...640A...5T}
Trapman L.,  Rosotti G.,  Bosman A.,  Hogerheijde M.,   van Dishoeck E.,  2020, \mn@doi [\aap] {10.1051/0004-6361/202037673}, 640, A5

\bibitem[\protect\citeauthoryear{{Trapman}, {Tabone}, {Rosotti}  \& {Zhang}}{{Trapman} et~al.}{2022}]{2022ApJ...926...61T}
{Trapman} L.,  {Tabone} B.,  {Rosotti} G.,   {Zhang} K.,  2022, \mn@doi [\apj] {10.3847/1538-4357/ac3ed5}, \href {https://ui.adsabs.harvard.edu/abs/2022ApJ...926...61T} {926, 61}

\bibitem[\protect\citeauthoryear{{Trapman}, {Rosotti}, {Zhang}  \& {Tabone}}{{Trapman} et~al.}{2023}]{2023ApJ...954...41T}
{Trapman} L.,  {Rosotti} G.,  {Zhang} K.,   {Tabone} B.,  2023, \mn@doi [\apj] {10.3847/1538-4357/ace7d1}, \href {https://ui.adsabs.harvard.edu/abs/2023ApJ...954...41T} {954, 41}

\bibitem[\protect\citeauthoryear{Umebayashi \& Nakano}{Umebayashi \& Nakano}{1981}]{1981PASJ...33..617U}
Umebayashi T.,  Nakano T.,  1981, \pasj, 33, 617

\bibitem[\protect\citeauthoryear{Umebayashi \& Nakano}{Umebayashi \& Nakano}{2009}]{2009ApJ...690...69U}
Umebayashi T.,  Nakano T.,  2009, \mn@doi [\apj] {10.1088/0004-637X/690/1/69}, 690, 69

\bibitem[\protect\citeauthoryear{Urpin}{Urpin}{2003}]{2003A&A...404..397U}
Urpin V.,  2003, \mn@doi [\aap] {10.1051/0004-6361:20030513}, 404, 397

\bibitem[\protect\citeauthoryear{Urpin \& Brandenburg}{Urpin \& Brandenburg}{1998}]{1998MNRAS.294..399U}
Urpin V.,  Brandenburg A.,  1998, \mn@doi [\mnras] {10.1046/j.1365-8711.1998.01118.x}, 294, 399

\bibitem[\protect\citeauthoryear{Venuti et~al.,}{Venuti et~al.}{2019}]{2019A&A...632A..46V}
Venuti L.,  et~al., 2019, \mn@doi [\aap] {10.1051/0004-6361/201935745}, 632, A46

\bibitem[\protect\citeauthoryear{Vincke \& Pfalzner}{Vincke \& Pfalzner}{2016}]{2016ApJ...828...48V}
Vincke K.,  Pfalzner S.,  2016, \mn@doi [\apj] {10.3847/0004-637X/828/1/48}, 828, 48

\bibitem[\protect\citeauthoryear{Virtanen et~al.,}{Virtanen et~al.}{2020}]{scipy2020}
Virtanen P.,  et~al., 2020, \mn@doi [Nature Methods] {10.1038/s41592-019-0686-2}, 17, 261

\bibitem[\protect\citeauthoryear{Weder, Mordasini  \& Emsenhuber}{Weder et~al.}{2023}]{2023A&A...674A.165W}
Weder J.,  Mordasini C.,   Emsenhuber A.,  2023, \mn@doi [\aap] {10.1051/0004-6361/202243453}, 674, A165

\bibitem[\protect\citeauthoryear{{W}es {M}c{K}inney}{{W}es {M}c{K}inney}{2010}]{pandas2010}
{W}es {M}c{K}inney 2010, in {S}t\'efan van~der {W}alt {J}arrod {M}illman eds, {P}roceedings of the 9th {P}ython in {S}cience {C}onference. pp 56 -- 61, \mn@doi{10.25080/Majora-92bf1922-00a}

\bibitem[\protect\citeauthoryear{Winter, Clarke, Rosotti, Ih, Facchini  \& Haworth}{Winter et~al.}{2018}]{2018MNRAS.478.2700W}
Winter A.,  Clarke C.,  Rosotti G.,  Ih J.,  Facchini S.,   Haworth T.,  2018, \mn@doi [\mnras] {10.1093/mnras/sty984}, 478, 2700

\bibitem[\protect\citeauthoryear{Yamato et~al.,}{Yamato et~al.}{2023}]{2023ApJ...951...11Y}
Yamato Y.,  et~al., 2023, \mn@doi [\apj] {10.3847/1538-4357/accd71}, 951, 11

\bibitem[\protect\citeauthoryear{Yang \& Bai}{Yang \& Bai}{2021}]{2021ApJ...922..201Y}
Yang H.,  Bai X.-N.,  2021, \mn@doi [\apj] {10.3847/1538-4357/ac250a}, 922, 201

\bibitem[\protect\citeauthoryear{Yen et~al.,}{Yen et~al.}{2014}]{2014ApJ...793....1Y}
Yen H.-W.,  et~al., 2014, \mn@doi [\apj] {10.1088/0004-637X/793/1/1}, 793, 1

\bibitem[\protect\citeauthoryear{Yu, Teague, Bae  \& {\"{O}}berg}{Yu et~al.}{2021}]{2021ApJ...920L..33Y}
Yu H.,  Teague R.,  Bae J.,   {\"{O}}berg K.,  2021, \mn@doi [\apjl] {10.3847/2041-8213/ac283e}, 920, L33

\bibitem[\protect\citeauthoryear{{Zagaria}, {Rosotti}, {Clarke}  \& {Tabone}}{{Zagaria} et~al.}{2022}]{2022MNRAS.514.1088Z}
{Zagaria} F.,  {Rosotti} G.~P.,  {Clarke} C.~J.,   {Tabone} B.,  2022, \mn@doi [\mnras] {10.1093/mnras/stac1461}, \href {https://ui.adsabs.harvard.edu/abs/2022MNRAS.514.1088Z} {514, 1088}

\bibitem[\protect\citeauthoryear{Zagaria, Facchini, Miotello, Manara, Toci  \& Clarke}{Zagaria et~al.}{2023}]{2023A&A...672L..15Z}
Zagaria F.,  Facchini S.,  Miotello A.,  Manara C.~F.,  Toci C.,   Clarke C.~J.,  2023, \mn@doi [\aap] {10.1051/0004-6361/202346164}, 672, L15

\bibitem[\protect\citeauthoryear{Zhang, Bergin, Blake, Cleeves, Hogerheijde, Salinas  \& Schwarz}{Zhang et~al.}{2016}]{2016ApJ...818L..16Z}
Zhang K.,  Bergin E.~A.,  Blake G.~A.,  Cleeves L.~I.,  Hogerheijde M.,  Salinas V.,   Schwarz K.~R.,  2016, \mn@doi [\apjl] {10.3847/2041-8205/818/1/L16}, 818, L16

\bibitem[\protect\citeauthoryear{{Zormpas}, {Birnstiel}, {Rosotti}  \& {Andrews}}{{Zormpas} et~al.}{2022}]{2022A&A...661A..66Z}
{Zormpas} A.,  {Birnstiel} T.,  {Rosotti} G.~P.,   {Andrews} S.~M.,  2022, \mn@doi [\aap] {10.1051/0004-6361/202142046}, \href {https://ui.adsabs.harvard.edu/abs/2022A&A...661A..66Z} {661, A66}

\bibitem[\protect\citeauthoryear{{de Valon}, {Dougados}, {Cabrit}, {Louvet}, {Zapata}  \& {Mardones}}{{de Valon} et~al.}{2020}]{2020A&A...634L..12D}
{de Valon} A.,  {Dougados} C.,  {Cabrit} S.,  {Louvet} F.,  {Zapata} L.~A.,   {Mardones} D.,  2020, \mn@doi [\aap] {10.1051/0004-6361/201936950}, \href {https://ui.adsabs.harvard.edu/abs/2020A&A...634L..12D} {634, L12}

\bibitem[\protect\citeauthoryear{pandas~development team}{pandas~development team}{2020}]{pandas2020}
pandas~development team T.,  2020, pandas-dev/pandas: Pandas, \mn@doi{10.5281/zenodo.3509134}, \url {https://doi.org/10.5281/zenodo.3509134}

\bibitem[\protect\citeauthoryear{van Dishoeck \& Black}{van Dishoeck \& Black}{1988}]{1988ApJ...334..771V}
van Dishoeck E.~F.,  Black J.~H.,  1988, \mn@doi [\apj] {10.1086/166877}, 334, 771

\bibitem[\protect\citeauthoryear{van~der Marel et~al.,}{van~der Marel et~al.}{2013}]{2013Sci...340.1199V}
van~der Marel N.,  et~al., 2013, \mn@doi [Science] {10.1126/science.1236770}, 340, 1199

\bibitem[\protect\citeauthoryear{van~der Walt, Colbert  \& Varoquaux}{van~der Walt et~al.}{2011}]{numpy}
van~der Walt S.,  Colbert S.~C.,   Varoquaux G.,  2011, \mn@doi [Computing in Science & Engineering] {10.1109/MCSE.2011.37}, 13, 22

\bibitem[\protect\citeauthoryear{van't Hoff et~al.,}{van't Hoff et~al.}{2023}]{2023ApJ...951...10V}
van't Hoff M.~L.,  et~al., 2023, \mn@doi [\apj] {10.3847/1538-4357/accf87}, 951, 10

\makeatother
\end{thebibliography}


\appendix
\section{Code testing}\label{sec:appendix_test}
\begin{figure}
    \centering
    \includegraphics[width=0.44\textwidth]{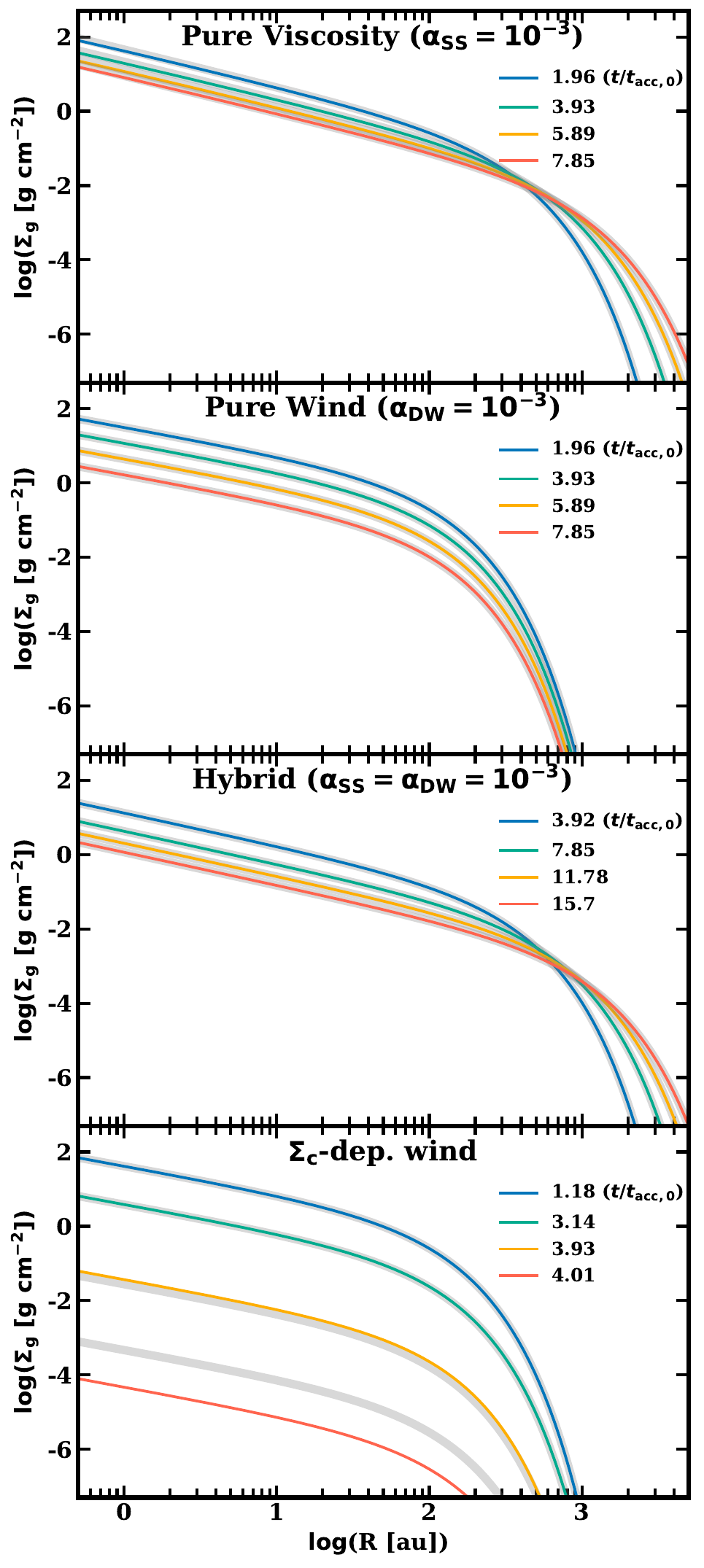}
    \caption{Comparison between the analytical and numerical solutions on the gas surface density for different scenarios. Numerical solutions are shown in polychromic lines for different times normalised to the initial accretion time-scale and analytical solutions are shown in monochronic lines.}
    \label{fig:test}
\end{figure}

Our numerical results (coloured lines) are plotted over the analytical solutions (indicated by grey shades at the corresponding time), which are normalized to the initial accretion time-scale $t_\mathrm{acc,0}=R_{c,0}/(3\epsilon_c c_\mathrm{s,c}\Tilde{\alpha}(t=0))$. $\Tilde{\alpha}(t=0)$ is the summation of $\alpha_\mathrm{DW}$ and $\alpha_\mathrm{SS}$ at $t=0$. $\epsilon_c$ and $c_\mathrm{s,c}$ are the aspect ratio ($H/R$) and sound speed at the initial characteristic radius $R_{c,0}$, respectively. Here, we fix the initial $\alpha_\mathrm{DW}$ or $\alpha_\mathrm{SS}$ to be $10^{-3}$. In cases where both effects are considered, the same value ($10^{-3}$) is assigned to each, giving rise to $\Tilde{\alpha}(t=0)=2\times10^{-3}$. Figure \ref{fig:test} shows that numerical results match well with the analytical solutions for the pure wind and the hybrid cases, but are a little off for the pure viscosity case and the $\Sigma_c$-dependent case in the later evolutionary stage. We attribute the deviation in the former to the zero-torque boundary condition imposed in the inner boundary. The latter arises from the numerical discretization and is further complicated by the dependence of $\alpha_\mathrm{DW}$ on the disc mass computed from the surface density ($\alpha_\mathrm{DW}\propto M_d(t)^{-\omega}$). The variation of $\Sigma_\mathrm{g}$ can result in changes in $\alpha_\mathrm{DW}$ and these quantities jointly determine accretion rates driven by viscosity and MHD winds, which in turn alter the disc mass and hence the surface density profile. However, even though the relative difference in the surface density between the numerical and the analytical solution looks large, the absolute difference is negligible, as only $10^{-5}$ of the initial gas disc mass remains at $t=4~t_\mathrm{acc,0}$. 

\section{Measurements of the characteristic radius \texorpdfstring{$\boldsymbol{R_c}$}{Rc}}\label{sec:appendix_r_c}

The two-fold physical meaning of $R_c$: the \say{cutoff} radius, beyond which the surface density drops exponentially; and the radius enclosing $63$ per cent of the total disc mass, inspires us to characterise it from two approaches. First, we measure $d\log\Sigma_g/d\log R$ for every two adjacent cells and then compute the distribution of these slopes, which is further used to calculate cumulative frequency-weighted slopes by varying the fraction of slopes included in the calculation. We use the initial surface density, where the $R_{c,0}$ is determined, to calibrate the threshold fraction. We also make sure $R_c$ is always beyond any gas substructures present in the profile. The characteristic radius $R_c$ evaluated in this way is denoted as $R_\mathrm{c,exp}$. Second, we measure the radius that encloses $63$ per cent of total disc mass and denote it as $R_\mathrm{63}$.
 
\par For each model, we measure $R_c$ every $0.5~$Myr by both approaches and compare them. The relative differences between two $R_c$ are within $30$ per cent for more than $70$ per cent measured disc sizes. The remaining $\sim 30$ per cent disc sizes are mainly ($\sim 94$ per cent) from discs falling in Category A (see Section \ref{sec:para_exp_surf_cat}), especially for those with a large $\alpha_\mathrm{SS,out}=10^{-2}$. The small $\alpha_\mathrm{DW,dz}$ enhances mass accumulation in the intermediate disc while the large $\alpha_\mathrm{SS,out}$ facilitates the disc expansion in the outer disc, enhancing the disparity in the surface density around $R_\mathrm{dz,out}$ and pushing $R_\mathrm{63}$ to a smaller radius than its initial location. When the jump in the surface density is smoothed by the viscosity at later times, $R_\mathrm{63}$ returns to be comparable to $R_\mathrm{c,exp}$ (relative differences $<30$ per cent). Hence, in this study, we use $R_\mathrm{c,exp}$ as the characteristic radius $R_c$.

\section{Distribution of measured \texorpdfstring{$\boldsymbol{{}^{12}\mathrm{CO}}$}{12CO} disc size}\label{sec:appendix_size}
We collated gas disc sizes traced by ${}^{12}\mathrm{CO}$ from previous studies \citep{2017ApJ...851...85B, 2018ApJ...859...21A, 2019A&A...626L...2F, 2021A&A...649A..19S, 2021ApJS..257....3L, 2021ApJ...911..150P, 2021A&A...645A.139K, 2021ApJ...920L..33Y, 2021MNRAS.507.3789C, 2022ApJ...931....6L, 2022ApJ...932..114L, 2023MNRAS.522.2611A, 2023ApJ...948...60L}. We ignore the differences in disc sizes characterised by different rotational transitions, ${}^{12}\mathrm{CO}$ (2-1) and ${}^{12}\mathrm{CO}$ (3-2), as they tend to be less than $10$ per cent \citep{2019A&A...629A..79T}. These sizes are measured in two approaches. When the measurement is directly performed in the image plane, a disc size enclosing a certain fraction (commonly $68$ or $90$ per cent) of the total flux density is either obtained from an increasing elliptical aperture \citep[e.g.,][]{2018ApJ...859...21A}, or an azimuthally averaged radial intensity profile of the disc \citep[e.g.,][]{2021A&A...645A.139K, 2022ApJ...931....6L}. The other method first requires an input for the visibility modelling, and then measures the disc size from the modelled image plane following methods mentioned above. Commonly used models for visibilities from previous studies are Gaussian, Nuker and power-law models \citep[e.g.,][]{2017ApJ...851...85B, 2021A&A...649A..19S}.

\par
\begin{figure}
    \centering
    \includegraphics[width=0.49\textwidth]{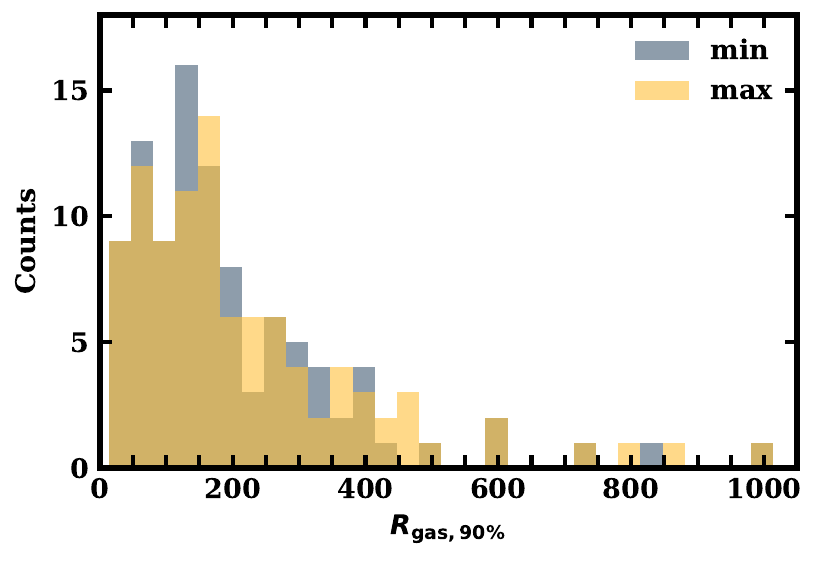}
    \caption{Histogram of gas disc sizes $R_\mathrm{CO,90}$ traced by ${}^{12}\mathrm{CO}$ (2-1) or ${}^{12}\mathrm{CO}$ (3-2) from recent observations. For discs that are measured multiple times, maximum measurements (yellow) and minimum measurements (grey) are taken respectively and presented in two separate distributions.}
    \label{fig:disc_size}
\end{figure}

 We define the size as a radius enclosing $90$ per cent of the total flux density ($R_\mathrm{CO,90}$). For literature that uses $68$ per cent ($R_\mathrm{CO,68}$) instead, we simply assume discs are Gaussian and convert $R_\mathrm{CO,68}$ to $R_\mathrm{CO,90}$ by multiplying a factor of $1.42$ \citep{2021A&A...649A..19S}. If discs are explicitly denoted as non-Gaussian in previous studies, we adopt the $R_\mathrm{CO,68}$ directly. For example, some discs in \citet{2021A&A...649A..19S} are modelled with Nuker profiles and only $R_\mathrm{CO,68}$ is provided. Discs measured in the visibility plane and modelled using a non-Gaussian function, specifically a power-law function in this instance \citep{2017ApJ...851...85B}, have their sizes as reported in the literature here. The largest and smallest measurements are taken separately for discs that have been measured multiple times and plotted in two distributions in Figure \ref{fig:disc_size}. 

\section{Distributions of varying parameters in the population synthesis}\label{sec:appendix_pop}
\begin{figure*}
    \centering
    \includegraphics[width=\textwidth]{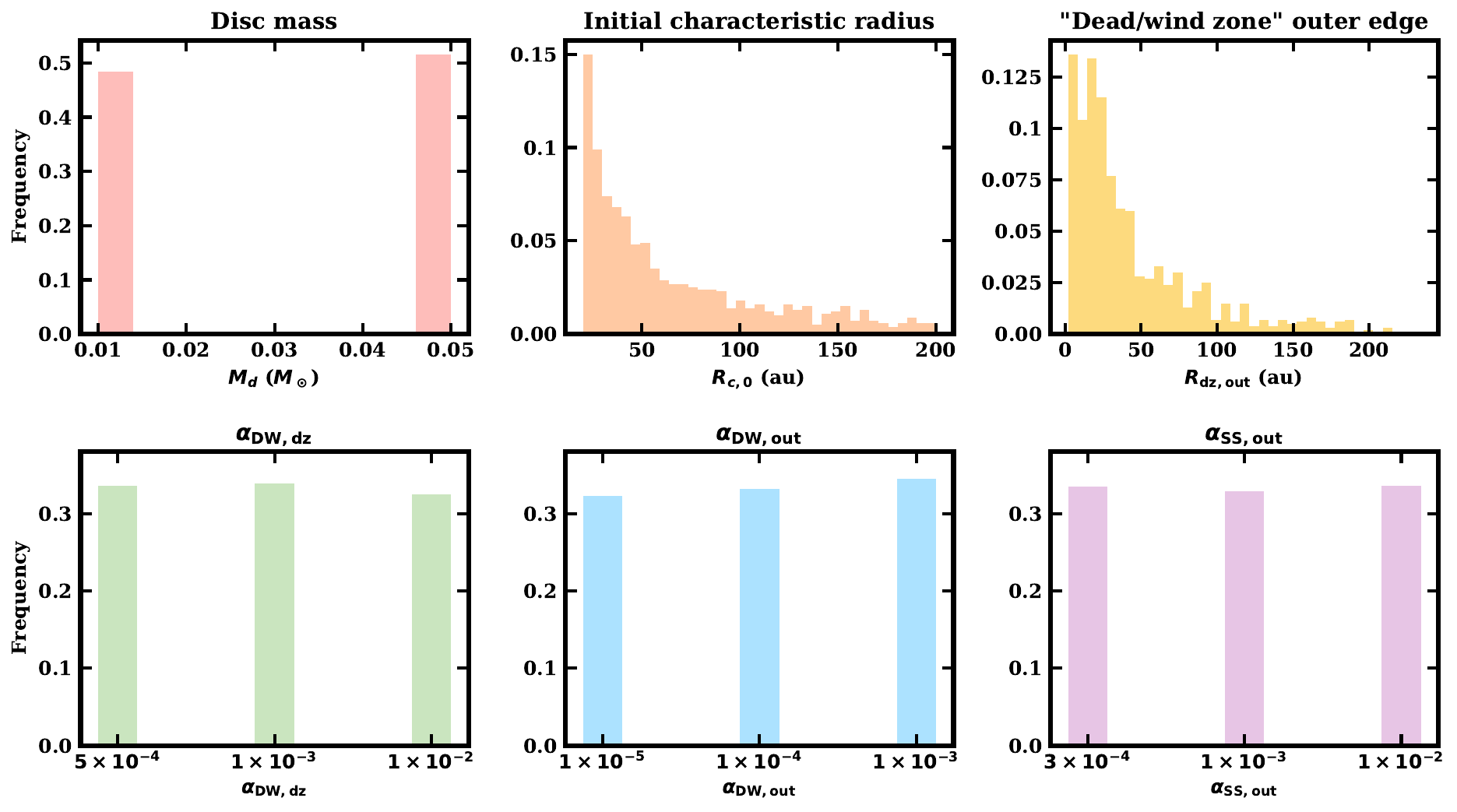}
    \caption{Distributions of initial disc masses $M_d$ (upper left panel), initial characteristic radii $R_\mathrm{c,0}$ (upper middle panel), ``dead/wind zone'' outer edges $R_\mathrm{dz,out}$ (upper right panel), $\alpha_\mathrm{DW,dz}$ (lower left panel), $\alpha_\mathrm{DW,out}$ (lower middle panel) and $\alpha_\mathrm{SS,out}$ (lower left panel) drawn for the first and second populations.}
    \label{fig:pop_syn_distri}
\end{figure*}

Figure \ref{fig:pop_syn_distri} shows the distributions of parameters, including initial disc masses $M_d$, initial characteristic radii $R_{c,0}$, \say{dead/wind zone} outer edges $R_\mathrm{dz,out}$, $\alpha_\mathrm{DW,dz}$, $\alpha_\mathrm{DW,out}$ and $\alpha_\mathrm{SS,out}$ for 1000 samples in the first and second populations described in Section \ref{sec:disc_pop}.
\section{List of simulations}\label{sec:appendix_table}
A full list of 92 models carried out in this study and their results on disc lifetimes, cumulative mass-loss fractions by wind extraction, stellar accretion and photoevaporation.

\onecolumn
\begin{center}
\renewcommand*{\arraystretch}{1.1}
\begin{longtable}{cccccccccc}
\caption{Summary of simulations that we have studied in Section \ref{sec:explor}. Column 2-6 list parameters for each simulation. Column 7 provides lifetimes of discs (see Section \ref{sec:para_exp_life_fixed}). Column 8-10 give the cumulative mass-loss fractions of wind extraction, stellar accretion and internal photoevaporation.} \label{tb:parameters} \\
\toprule \toprule
No. & \multicolumn{3}{c}{$\alpha$}                     & \multicolumn{2}{c}{Radius (au)} & \multicolumn{1}{c}{Lifetime}& \multicolumn{3}{c}{Mass loss fraction}\\
\cmidrule(r){2-4}
\cmidrule(r){5-6}
\cmidrule(r){8-10}
& $\alpha_\mathrm{DW, dz}$ & $\alpha_\mathrm{DW, out}$ & $\alpha_\mathrm{SS, out}$ & $R_\mathrm{c,0}$ & $R_\mathrm{dz,out}$ &  (Myr) &  Wind & Acc. & PE \\ 
(1) & (2) & (3) & (4) & (5) & (6) & (7) &  (8) & (9) & (10)\\
\addlinespace[0.5ex]
\hline
\addlinespace[1ex]
\endfirsthead

\bottomrule
\endfoot

\bottomrule
\endlastfoot

\toprule \toprule
& \multicolumn{3}{c}{$\alpha$}                                    & \multicolumn{2}{c}{Radius (au)} & \multicolumn{1}{c}{Lifetime} & \multicolumn{3}{c}{Mass loss fraction}\\
\cmidrule(r){2-4}
\cmidrule(r){5-6}
\cmidrule(r){8-10}
& $\alpha_\mathrm{DW, dz}$ & $\alpha_\mathrm{DW, out}$ & $\alpha_\mathrm{SS, out}$ & $R_\mathrm{c,0}$ & $R_\mathrm{dzo}$ & (Myr)  & Wind & Acc & PE  \\
(1) & (2) & (3) & (4) & (5) & (6) & (7) & (8) & (10) & (11)\\
\addlinespace[0.5ex]
\hline
\addlinespace[1ex]
\endhead
visc. & $1 \times 10^{-3}$ & $1 \times 10^{-3}$ & $1 \times 10^{-3}$ & 60 & -- & >12 & -- & 0.84 & 0.16 \\
wind & $1 \times 10^{-3}$ & $1 \times 10^{-3}$ & $1 \times 10^{-3}$ & 60 & -- & $\sim 5.1$ & 0.87 & 0.13 & -- \\  
\midrule
1 & $5 \times 10^{-4}$ & $1 \times 10^{-5}$ & $3 \times 10^{-4}$ & 60 & 30 & 10.37 & 0.49 & 0.20 & 0.31 \\
2 & $5 \times 10^{-4}$ & $1 \times 10^{-5}$ & $1 \times 10^{-3}$ & 60 & 30 & 7.79 & 0.55 & 0.22 & 0.23 \\
3 & $5 \times 10^{-4}$ & $1 \times 10^{-5}$ & $1 \times 10^{-2}$ & 60 & 30 & 4.13 & 0.63 & 0.26 & 0.11 \\
4 & $5 \times 10^{-4}$ & $1 \times 10^{-4}$ & $3 \times 10^{-4}$ & 60 & 30 & 10.36 & 0.53 & 0.20 & 0.27 \\
5 & $5 \times 10^{-4}$ & $1 \times 10^{-4}$ & $1 \times 10^{-3}$ & 60 & 30 & 7.68 & 0.57 & 0.23 & 0.21 \\
6 & $5 \times 10^{-4}$ & $1 \times 10^{-4}$ & $1 \times 10^{-2}$ & 60 & 30 & 4.11 & 0.63 & 0.26 & 0.11 \\
7 & $5 \times 10^{-4}$ & $1 \times 10^{-3}$ & $3 \times 10^{-4}$ & 60 & 30 & 5.87 & 0.71 & 0.23 & 0.06 \\
8 & $5 \times 10^{-4}$ & $1 \times 10^{-3}$ & $1 \times 10^{-3}$ & 60 & 30 & 6.06 & 0.67 & 0.23 & 0.10 \\
9 & $5 \times 10^{-4}$ & $1 \times 10^{-3}$ & $1 \times 10^{-2}$ & 60 & 30 & 3.88 & 0.65 & 0.26 & 0.09 \\
10 & $1 \times 10^{-3}$ & $1 \times 10^{-5}$ & $3 \times 10^{-4}$ & 60 & 30 & 8.93 & 0.52 & 0.18 & 0.30 \\
11 & $1 \times 10^{-3}$ & $1 \times 10^{-5}$ & $1 \times 10^{-3}$ & 60 & 30 & 6.72 & 0.59 & 0.20 & 0.21 \\
12 & $1 \times 10^{-3}$ & $1 \times 10^{-5}$ & $1 \times 10^{-2}$ & 60 & 30 & 3.52 & 0.66 & 0.23 & 0.11 \\
13 & $1 \times 10^{-3}$ & $1 \times 10^{-4}$ & $3 \times 10^{-4}$ & 60 & 30 & 9.00 & 0.56 & 0.18 & 0.26 \\
14 & $1 \times 10^{-3}$ & $1 \times 10^{-4}$ & $1 \times 10^{-3}$ & 60 & 30 & 6.64 & 0.60 & 0.20 & 0.20 \\
15 & $1 \times 10^{-3}$ & $1 \times 10^{-4}$ & $1 \times 10^{-2}$ & 60 & 30 & 3.49 & 0.66 & 0.23 & 0.11 \\
16 & $1 \times 10^{-3}$ & $1 \times 10^{-3}$ & $3 \times 10^{-4}$ & 60 & 30 & 5.03 & 0.74 & 0.21 & 0.05 \\
17 & $1 \times 10^{-3}$ & $1 \times 10^{-3}$ & $1 \times 10^{-3}$ & 60 & 30 & 5.22 & 0.71 & 0.21 & 0.08 \\
18 & $1 \times 10^{-3}$ & $1 \times 10^{-3}$ & $1 \times 10^{-2}$ & 60 & 30 & 3.28 & 0.68 & 0.23 & 0.09 \\
19 & $1 \times 10^{-2}$ & $1 \times 10^{-5}$ & $3 \times 10^{-4}$ & 60 & 30 & 7.27 & 0.57 & 0.16 & 0.27 \\
20 & $1 \times 10^{-2}$ & $1 \times 10^{-5}$ & $1 \times 10^{-3}$ & 60 & 30 & 5.51 & 0.63 & 0.18 & 0.19 \\
21 & $1 \times 10^{-2}$ & $1 \times 10^{-5}$ & $1 \times 10^{-2}$ & 60 & 30 & 2.82 & 0.70 & 0.20 & 0.10 \\
22 & $1 \times 10^{-2}$ & $1 \times 10^{-4}$ & $3 \times 10^{-4}$ & 60 & 30 & 7.41 & 0.60 & 0.16 & 0.24 \\
23 & $1 \times 10^{-2}$ & $1 \times 10^{-4}$ & $1 \times 10^{-3}$ & 60 & 30 & 5.44 & 0.64 & 0.18 & 0.18 \\
24 & $1 \times 10^{-2}$ & $1 \times 10^{-4}$ & $1 \times 10^{-2}$ & 60 & 30 & 2.79 & 0.71 & 0.20 & 0.09 \\
25 & $1 \times 10^{-2}$ & $1 \times 10^{-3}$ & $3 \times 10^{-4}$ & 60 & 30 & 4.11 & 0.77 & 0.19 & 0.04 \\
26 & $1 \times 10^{-2}$ & $1 \times 10^{-3}$ & $1 \times 10^{-3}$ & 60 & 30 & 4.24 & 0.74 & 0.19 & 0.07 \\
27 & $1 \times 10^{-2}$ & $1 \times 10^{-3}$ & $1 \times 10^{-2}$ & 60 & 30 & 2.57 & 0.72 & 0.20 & 0.08 \\

\midrule
28 & $5 \times 10^{-4}$ & $1 \times 10^{-5}$ & $3 \times 10^{-4}$ & 60 & 75 & 7.89 & 0.59 & 0.21 & 0.20 \\
29 & $5 \times 10^{-4}$ & $1 \times 10^{-5}$ & $3 \times 10^{-4}$ & 60 & 135 & 8.02 & 0.65 & 0.22 & 0.13 \\
30 & $5 \times 10^{-4}$ & $1 \times 10^{-5}$ & $3 \times 10^{-4}$ & 120 & 30 & >12.00 & 0.59 & 0.22 & 0.19 \\
31 & $5 \times 10^{-4}$ & $1 \times 10^{-5}$ & $3 \times 10^{-4}$ & 120 & 75 & 11.60 & 0.48 & 0.15 & 0.37 \\
32 & $5 \times 10^{-4}$ & $1 \times 10^{-5}$ & $3 \times 10^{-4}$ & 120 & 135 & 11.88 & 0.56 & 0.16 & 0.28 \\
33 & $5 \times 10^{-4}$ & $1 \times 10^{-5}$ & $1 \times 10^{-3}$ & 60 & 75 & 6.71 & 0.63 & 0.22 & 0.15 \\
34 & $5 \times 10^{-4}$ & $1 \times 10^{-5}$ & $1 \times 10^{-3}$ & 60 & 135 & 7.51 & 0.67 & 0.22 & 0.11 \\
35 & $5 \times 10^{-4}$ & $1 \times 10^{-5}$ & $1 \times 10^{-3}$ & 120 & 30 & 11.16 & 0.47 & 0.18 & 0.35 \\
36 & $5 \times 10^{-4}$ & $1 \times 10^{-5}$ & $1 \times 10^{-3}$ & 120 & 75 & 9.79 & 0.55 & 0.17 & 0.28 \\
37 & $5 \times 10^{-4}$ & $1 \times 10^{-5}$ & $1 \times 10^{-3}$ & 120 & 135 & 10.89 & 0.61 & 0.17 & 0.22 \\
38 & $5 \times 10^{-4}$ & $1 \times 10^{-5}$ & $1 \times 10^{-2}$ & 60 & 75 & 4.96 & 0.68 & 0.24 & 0.08 \\
39 & $5 \times 10^{-4}$ & $1 \times 10^{-5}$ & $1 \times 10^{-2}$ & 60 & 135 & 6.48 & 0.70 & 0.23 & 0.07 \\
40 & $5 \times 10^{-4}$ & $1 \times 10^{-5}$ & $1 \times 10^{-2}$ & 120 & 30 & 5.96 & 0.59 & 0.23 & 0.18 \\
41 & $5 \times 10^{-4}$ & $1 \times 10^{-5}$ & $1 \times 10^{-2}$ & 120 & 75 & 6.47 & 0.66 & 0.21 & 0.13 \\
42 & $5 \times 10^{-4}$ & $1 \times 10^{-5}$ & $1 \times 10^{-2}$ & 120 & 135 & 8.62 & 0.69 & 0.20 & 0.11 \\
43 & $5 \times 10^{-4}$ & $1 \times 10^{-4}$ & $1 \times 10^{-2}$ & 60 & 75 & 4.95 & 0.68 & 0.24 & 0.08 \\
44 & $5 \times 10^{-4}$ & $1 \times 10^{-4}$ & $1 \times 10^{-2}$ & 60 & 135 & 6.47 & 0.70 & 0.23 & 0.07 \\
45 & $5 \times 10^{-4}$ & $1 \times 10^{-4}$ & $1 \times 10^{-2}$ & 120 & 30 & 5.91 & 0.60 & 0.23 & 0.17 \\
46 & $5 \times 10^{-4}$ & $1 \times 10^{-4}$ & $1 \times 10^{-2}$ & 120 & 75 & 6.46 & 0.66 & 0.21 & 0.13 \\
47 & $5 \times 10^{-4}$ & $1 \times 10^{-4}$ & $1 \times 10^{-2}$ & 120 & 135 & 8.61 & 0.69 & 0.19 & 0.11 \\
48 & $1 \times 10^{-3}$ & $1 \times 10^{-5}$ & $3 \times 10^{-4}$ & 60 & 75 & 5.19 & 0.64 & 0.19 & 0.17 \\
49 & $1 \times 10^{-3}$ & $1 \times 10^{-5}$ & $3 \times 10^{-4}$ & 60 & 135 & 4.37 & 0.71 & 0.20 & 0.09 \\
50 & $1 \times 10^{-3}$ & $1 \times 10^{-5}$ & $3 \times 10^{-4}$ & 120 & 30 & >12.00 & 0.53 & 0.17 & 0.30 \\
51 & $1 \times 10^{-3}$ & $1 \times 10^{-5}$ & $3 \times 10^{-4}$ & 120 & 75 & 8.65 & 0.52 & 0.14 & 0.34 \\
52 & $1 \times 10^{-3}$ & $1 \times 10^{-5}$ & $3 \times 10^{-4}$ & 120 & 135 & 7.34 & 0.61 & 0.15 & 0.24 \\
53 & $1 \times 10^{-3}$ & $1 \times 10^{-5}$ & $1 \times 10^{-3}$ & 60 & 75 & 4.22 & 0.67 & 0.20 & 0.13 \\
54 & $1 \times 10^{-3}$ & $1 \times 10^{-5}$ & $1 \times 10^{-3}$ & 60 & 135 & 4.01 & 0.72 & 0.21 & 0.07 \\
55 & $1 \times 10^{-3}$ & $1 \times 10^{-5}$ & $1 \times 10^{-3}$ & 120 & 30 & 9.94 & 0.50 & 0.16 & 0.34 \\
56 & $1 \times 10^{-3}$ & $1 \times 10^{-5}$ & $1 \times 10^{-3}$ & 120 & 75 & 7.21 & 0.59 & 0.16 & 0.25 \\
57 & $1 \times 10^{-3}$ & $1 \times 10^{-5}$ & $1 \times 10^{-3}$ & 120 & 135 & 6.50 & 0.66 & 0.16 & 0.18 \\
58 & $1 \times 10^{-3}$ & $1 \times 10^{-4}$ & $1 \times 10^{-2}$ & 60 & 75 & 2.77 & 0.72 & 0.22 & 0.06 \\
59 & $1 \times 10^{-3}$ & $1 \times 10^{-4}$ & $1 \times 10^{-2}$ & 60 & 135 & 3.36 & 0.74 & 0.21 & 0.05 \\
60 & $1 \times 10^{-3}$ & $1 \times 10^{-4}$ & $1 \times 10^{-2}$ & 120 & 30 & 5.24 & 0.63 & 0.21 & 0.16 \\
61 & $1 \times 10^{-3}$ & $1 \times 10^{-4}$ & $1 \times 10^{-2}$ & 120 & 75 & 4.19 & 0.69 & 0.19 & 0.12 \\
62 & $1 \times 10^{-3}$ & $1 \times 10^{-4}$ & $1 \times 10^{-2}$ & 120 & 135 & 4.79 & 0.73 & 0.18 & 0.09 \\
63 & $1 \times 10^{-2}$ & $1 \times 10^{-5}$ & $3 \times 10^{-4}$ & 60 & 75 & 3.00 & 0.69 & 0.18 & 0.13 \\
64 & $1 \times 10^{-2}$ & $1 \times 10^{-5}$ & $3 \times 10^{-4}$ & 60 & 135 & 1.18 & 0.75 & 0.19 & 0.06 \\
65 & $1 \times 10^{-2}$ & $1 \times 10^{-5}$ & $3 \times 10^{-4}$ & 120 & 30 & 10.47 & 0.44 & 0.11 & 0.45 \\
66 & $1 \times 10^{-2}$ & $1 \times 10^{-5}$ & $3 \times 10^{-4}$ & 120 & 75 & 5.81 & 0.57 & 0.13 & 0.30 \\
67 & $1 \times 10^{-2}$ & $1 \times 10^{-5}$ & $3 \times 10^{-4}$ & 120 & 135 & 3.24 & 0.67 & 0.14 & 0.19 \\
68 & $1 \times 10^{-2}$ & $1 \times 10^{-5}$ & $1 \times 10^{-2}$ & 60 & 75 & 1.38 & 0.76 & 0.19 & 0.05 \\
69 & $1 \times 10^{-2}$ & $1 \times 10^{-5}$ & $1 \times 10^{-2}$ & 60 & 135 & 0.68 & 0.78 & 0.19 & 0.03 \\
70 & $1 \times 10^{-2}$ & $1 \times 10^{-5}$ & $1 \times 10^{-2}$ & 120 & 30 & 4.47 & 0.67 & 0.18 & 0.15 \\
71 & $1 \times 10^{-2}$ & $1 \times 10^{-5}$ & $1 \times 10^{-2}$ & 120 & 75 & 2.78 & 0.73 & 0.16 & 0.11 \\
72 & $1 \times 10^{-2}$ & $1 \times 10^{-5}$ & $1 \times 10^{-2}$ & 120 & 135 & 1.78 & 0.76 & 0.16 & 0.08 \\
73 & $1 \times 10^{-2}$ & $1 \times 10^{-4}$ & $1 \times 10^{-2}$ & 60 & 75 & 1.38 & 0.76 & 0.19 & 0.05 \\
74 & $1 \times 10^{-2}$ & $1 \times 10^{-4}$ & $1 \times 10^{-2}$ & 60 & 135 & 0.68 & 0.78 & 0.19 & 0.03 \\
75 & $1 \times 10^{-2}$ & $1 \times 10^{-4}$ & $1 \times 10^{-2}$ & 120 & 30 & 4.44 & 0.67 & 0.18 & 0.15 \\
76 & $1 \times 10^{-2}$ & $1 \times 10^{-4}$ & $1 \times 10^{-2}$ & 120 & 75 & 2.77 & 0.73 & 0.16 & 0.11 \\
77 & $1 \times 10^{-2}$ & $1 \times 10^{-4}$ & $1 \times 10^{-2}$ & 120 & 135 & 1.77 & 0.77 & 0.16 & 0.07 \\
78 & $1 \times 10^{-2}$ & $1 \times 10^{-3}$ & $3 \times 10^{-4}$ & 60 & 75 & 2.77 & 0.78 & 0.19 & 0.03 \\
79 & $1 \times 10^{-2}$ & $1 \times 10^{-3}$ & $3 \times 10^{-4}$ & 60 & 135 & 1.58 & 0.79 & 0.19 & 0.02 \\
80 & $1 \times 10^{-2}$ & $1 \times 10^{-3}$ & $3 \times 10^{-4}$ & 120 & 30 & 7.72 & 0.75 & 0.15 & 0.10 \\
81 & $1 \times 10^{-2}$ & $1 \times 10^{-3}$ & $3 \times 10^{-4}$ & 120 & 75 & 6.16 & 0.76 & 0.15 & 0.09 \\
82 & $1 \times 10^{-2}$ & $1 \times 10^{-3}$ & $3 \times 10^{-4}$ & 120 & 135 & 4.36 & 0.78 & 0.15 & 0.07 \\
83 & $1 \times 10^{-2}$ & $1 \times 10^{-3}$ & $1 \times 10^{-3}$ & 60 & 75 & 2.48 & 0.77 & 0.19 & 0.04 \\
84 & $1 \times 10^{-2}$ & $1 \times 10^{-3}$ & $1 \times 10^{-3}$ & 60 & 135 & 1.25 & 0.78 & 0.19 & 0.03 \\
85 & $1 \times 10^{-2}$ & $1 \times 10^{-3}$ & $1 \times 10^{-3}$ & 120 & 30 & 7.00 & 0.71 & 0.15 & 0.14 \\
86 & $1 \times 10^{-2}$ & $1 \times 10^{-3}$ & $1 \times 10^{-3}$ & 120 & 75 & 4.96 & 0.74 & 0.15 & 0.11 \\
87 & $1 \times 10^{-2}$ & $1 \times 10^{-3}$ & $1 \times 10^{-3}$ & 120 & 135 & 3.26 & 0.76 & 0.15 & 0.09 \\
88 & $1 \times 10^{-2}$ & $1 \times 10^{-3}$ & $1 \times 10^{-2}$ & 60 & 75 & 1.35 & 0.76 & 0.19 & 0.05 \\
89 & $1 \times 10^{-2}$ & $1 \times 10^{-3}$ & $1 \times 10^{-2}$ & 60 & 135 & 0.70 & 0.78 & 0.19 & 0.03 \\
90 & $1 \times 10^{-2}$ & $1 \times 10^{-3}$ & $1 \times 10^{-2}$ & 120 & 30 & 4.09 & 0.70 & 0.18 & 0.12 \\
91 & $1 \times 10^{-2}$ & $1 \times 10^{-3}$ & $1 \times 10^{-2}$ & 120 & 75 & 2.68 & 0.74 & 0.16 & 0.10 \\
92 & $1 \times 10^{-2}$ & $1 \times 10^{-3}$ & $1 \times 10^{-2}$ & 120 & 135 & 1.77 & 0.77 & 0.16 & 0.07 \\
\end{longtable}
\end{center}
\normalsize

\section{Visualisation of results from simulations}
\begin{figure*}
    \centering
    \includegraphics[width=\textwidth]{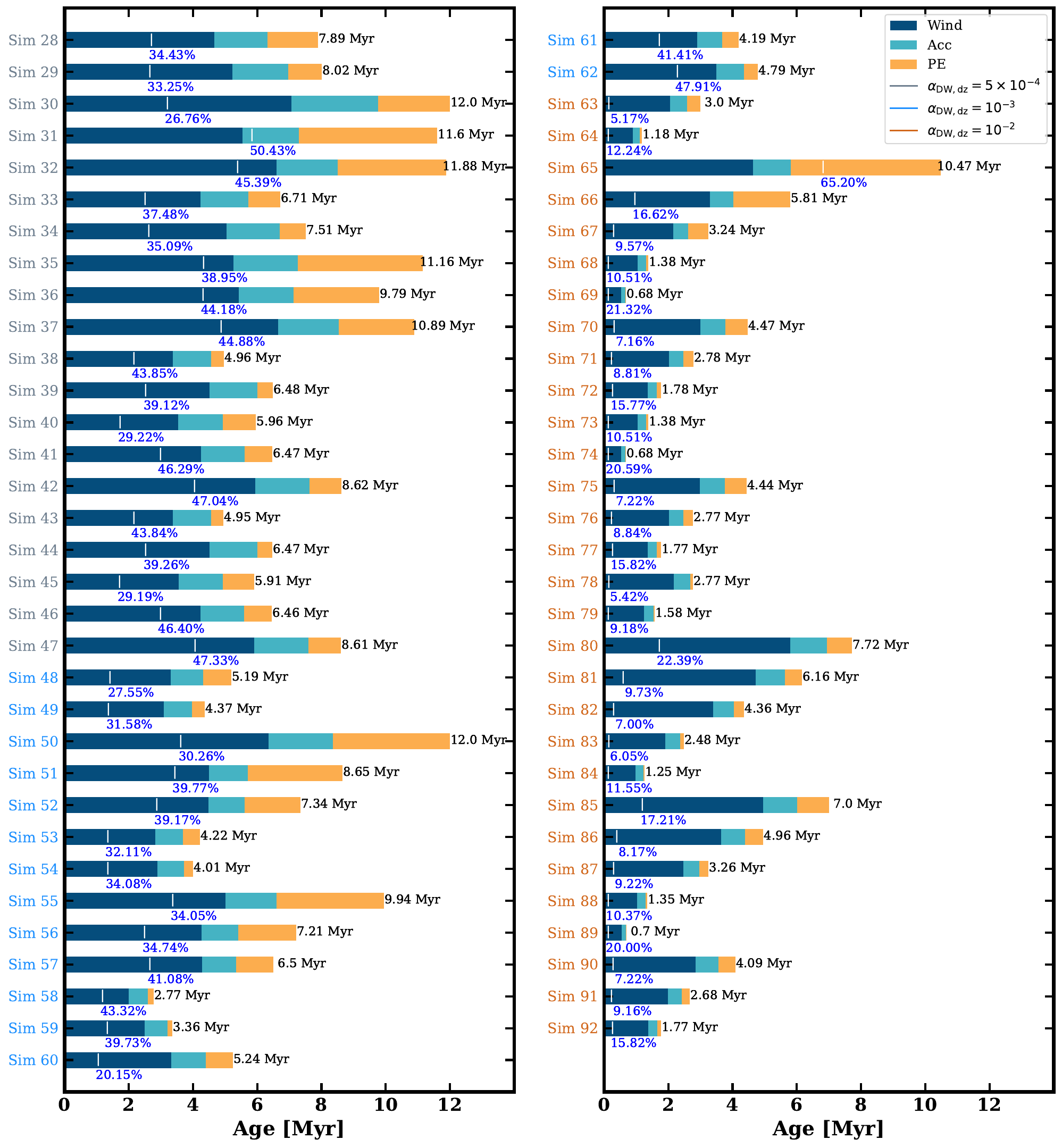}
    \caption{A similar diagram to Figure \ref{fig:cum_mass_loss} for simulations in the second group.}
    \label{fig:table_visual}
\end{figure*}

Figure \ref{fig:table_visual} is visualisation of data below the dividing line in Table \ref{tb:parameters}.


\bsp	
\label{lastpage}
\end{document}